\documentclass{aa}  

\usepackage{graphicx}
\usepackage{txfonts,textcomp}
\graphicspath{{./}{figures/}}

\usepackage{hyperref}
\hypersetup{
    colorlinks=true,
    linkcolor=blue,
    citecolor=blue,
    filecolor=magenta,      
    urlcolor=cyan,
    }

\begin{document}

   \title{The Blue Jay survey: Deep JWST spectroscopy for a representative sample of galaxies at cosmic noon}

   \titlerunning{The Blue Jay survey}
   \authorrunning{Belli et al.}

   \author{
        Sirio Belli\inst{\ref{unibo}} \and 
        Letizia Bugiani\inst{\ref{unibo}, \ref{inaf}} \and
        Minjung Park\inst{\ref{cfa},\ref{kavli},\ref{cavendish}} \and
        J. Trevor Mendel\inst{\ref{anu},\ref{arc}} \and
        Rebecca L. Davies\inst{\ref{arc},\ref{swinburne}} \and
        Amir H. Khoram\inst{\ref{unibo}, \ref{inaf}} \and
        Benjamin D. Johnson\inst{\ref{cfa}} \and
        Joel Leja\inst{\ref{psu1},\ref{psu2},\ref{psu3}} \and
        Sandro Tacchella\inst{\ref{kavli},\ref{cavendish}} \and
        Vanessa Brown\inst{\ref{colorado}} \and
        Charlie Conroy\inst{\ref{cfa}} \and 
        Razieh Emami\inst{\ref{cfa}} \and
        Yijia Li\inst{\ref{psu1},\ref{psu2}} \and
        Caterina Liboni\inst{\ref{unibo}, \ref{inaf}} \and 
        Gabriel Maheson\inst{\ref{kavli},\ref{cavendish}} \and
        Elijah P. Mathews\inst{\ref{psu1},\ref{psu2},\ref{psu3}} \and
        Rohan P. Naidu\inst{\ref{mit}} \and
        Erica J. Nelson\inst{\ref{colorado}} \and
        Bryan A. Terrazas\inst{\ref{oberlin}} \and
        Rainer Weinberger\inst{\ref{leibniz}}
        }
   \institute{
        \label{unibo} Dipartimento di Fisica e Astronomia, Università di Bologna, Bologna, Italy \and
        \label{inaf} INAF, Osservatorio di Astrofisica e Scienza dello Spazio, Bologna, Italy \and
        \label{cfa} Center for Astrophysics $|$ Harvard \& Smithsonian, Cambridge, MA, USA \and
        \label{kavli} Kavli Institute for Cosmology, University of Cambridge, Cambridge, UK \and
        \label{cavendish} Cavendish Laboratory, University of Cambridge, Cambridge, UK \and
        \label{anu} Research School of Astronomy and Astrophysics, Australian National University, Canberra, ACT, Australia \and
        \label{arc} ARC Centre of Excellence for All Sky Astrophysics in 3 Dimensions (ASTRO 3D) \and 
        \label{swinburne} Centre for Astrophysics and Supercomputing, Swinburne University of Technology, Hawthorn, Victoria, Australia \and
        \label{psu1} Department of Astronomy and Astrophysics, The Pennsylvania State University, University Park, PA, USA \and
        \label{psu2} Institute for Gravitation and the Cosmos, The Pennsylvania State University, University Park, PA, USA \and
        \label{psu3} Institute for Computational and Data Sciences, The Pennsylvania State University, University Park, PA, USA \and
        \label{colorado} Department for Astrophysical and Planetary Science, University of Colorado, Boulder, CO, USA \and
        \label{mit} MIT Kavli Institute for Astrophysics and Space Research, Cambridge, MA, USA \and        
        \label{oberlin} Department of Physics \& Astronomy, Oberlin College, Oberlin, OH, USA \and
        \label{leibniz} Leibniz Institute for Astrophysics, Potsdam, Germany
        }

   %\date{Received September 30, 20XX}

% \abstract{}{}{}{}{}
% 5 {} token are mandatory
 
  \abstract
  {We present the Blue Jay survey, a cycle 1 JWST program aimed at studying the stellar and gas content of galaxies at cosmic noon. The survey consists of deep spectroscopy for 153 targets observed over two pointings in the COSMOS field using the NIRSpec micro-shutter assembly (MSA). We employed the three medium-resolution gratings, G140M, G235M, and G395M, with exposure times of 13~hours, 3.2~hours, and 1.6~hours, respectively. We thus obtained full coverage of the 1 - 5~{\textmu}m range, corresponding to the entire rest-frame optical wavelength range, with a spectral resolution $R\sim1000$.
  Parallel observations provided deep multiband NIRCam imaging, which partially overlaps with the spectroscopic observations.
  The sample was carefully selected to provide a census of galaxies over the redshift range $1.7 < z < 3.5$ above a redshift-dependent minimum stellar mass that ranges from $10^{8.7}~M_\odot$ to $10^{9.3}~M_\odot$. The selection ensured that the Blue Jay sample is representative of the entire galaxy population at these redshifts, without strong biases in color, star formation rate, or other properties.
  The sizes of massive galaxies at these redshifts are comparable to or larger than the NIRSpec shutters, which requires custom strategies for designing and reducing the observations. Since the standard $A-B$ nod subtraction leads to flux self-subtraction for large galaxies, we constructed a master background from empty shutters and subtracted it from each of the science spectra. This in turn allowed us to use shorter slitlets consisting of only two shutters per galaxy instead of the usual three, with a substantial increase in the multiplexing of the NIRSpec MSA.
  In order to obtain self-consistent fits of spectroscopic and photometric data, we measured multiband photometry using space-based observations from the exact area in the sky where the NIRSpec 1D spectrum is extracted.
  The spectra contain a wealth of emission and absorption lines, including hydrogen lines (Balmer and Paschen series), helium lines, metal emission lines from ionized gas, metal stellar absorption lines, and resonant absorption lines that can be used to study neutral gas (Na~I~D and Ca~II H, K).  Spectroscopic redshifts are unambiguously measured for 93\% of the sample.
  The Blue Jay dataset, which we publicly release, represents the ideal sample for studying the stellar populations, neutral gas, and ionized gas in cosmic noon galaxies, including star formation histories, quenching, outflows, metallicity, and dust attenuation. 
  }


   \maketitle
   %\nolinenumbers

%%%%%%%%%%%%%%%%%%%%%%%%%%%%%%%%%%%%%%%%%%
\section{Introduction} 
\label{sec:intro}
%%%%%%%%%%%%%%%%%%%%%%%%%%%%%%%%%%%%%%%%%%

Cosmic noon is one of the most interesting and consequential phases in the history of the Universe, since about half of the current stellar mass was formed during this relatively short period, corresponding roughly to the redshift range $1 < z < 3$ \citep{madau14}. Cosmic noon is, by definition, the peak of the cosmic star formation history, but it is also the epoch when the cosmic black hole accretion rate is at its peak \citep{delvecchio14}, and when the population of massive galaxies becomes dominated by quiescent systems \citep{muzzin13, ilbert13}.

This redshift range, however, is also observationally challenging, because the rest-optical spectral features that contain most of the information about the physical properties of a galaxy are redshifted into the near-infrared, where the sensitivity of charge-coupled devices (CCDs) quickly declines, and ground-based observations are plagued by atmospheric absorption and emission. For sufficiently high redshifts, strong features in the far ultraviolet (UV) such as the Ly$\alpha$ emission line and/or the Lyman break enter the optical window and make it relatively easy to study star-forming galaxies \citep{steidel96}. At $z\sim1-3$, however, no strong spectral features fall in the observed optical window, and even measuring galaxy redshifts becomes difficult. For this reason, this epoch was dubbed the ``redshift desert'' \citep[e.g.,][]{stern99, steidel04}.

In the past two decades, the development of sensitive near-infrared detectors enabled the characterization of increasingly larger and more diverse samples of galaxies in the $1 < z < 3$ range, thus transforming this epoch from a hostile redshift desert to an exciting cosmic noon. A considerable amount of time on large ground-based telescopes has been devoted to spectroscopic surveys, which can be broadly divided into two categories according to whether they target the star-forming or the quiescent galaxy population.

The majority of spectroscopic observations at cosmic noon focus on star-forming galaxies, which are studied via the emission lines produced by ionized gas. The main challenge is represented by the effect of the Earth's atmosphere, which offers only limited wavelength ranges free from absorption (due to water vapor) and emission (due to OH molecules). As a result, it has been nearly impossible to observe a complete set of strong emission lines in the same galaxy. Nonetheless, rest-frame optical emission lines have been detected (and, in many cases, spatially resolved) for samples of hundreds or even thousands of galaxies at $1 < z < 3$ (\citealt{forster-schreiber09, steidel14, kriek15, wisnioski15, curti20}; see \citealt{forster-schreiber20} for a review). 

On the other hand, a subset of spectroscopic studies have focused on quiescent galaxies at $z>1$, which are gas-poor and lack strong emission lines; their study is thus mostly based on the detection of stellar absorption lines in the galaxy continuum emission.
These observations are made particularly challenging by the combination of atmospheric contamination and the requirement of extraordinarily long exposure times, which are needed in order to detect absorption lines against the faint stellar continuum \citep{kriek06, vandokkum09}. Additionally, the number density of quiescent galaxies quickly decreases with redshift, limiting the multiplexing capabilities of multi-object spectrographs on large telescopes. As a result, the samples of quiescent galaxies at cosmic noon that have been characterized spectroscopically are small, on the order of tens of galaxies \citep[e.g.][]{onodera12, vandesande13, bezanson13, belli14, belli17mosfire, carnall19, mendel20, kriek24}.

The advent of the \textit{James Webb} Space Telescope (JWST) enabled spectacular progress in spectroscopic studies of galaxies at $z>1$, in particular because of the exceptional capabilities of the NIRSpec instrument. The lack of atmospheric absorption and emission means that JWST spectra are more sensitive and can access a wider wavelength range than previous ground-based observations.
Moreover, while ground-based near-infrared spectrographs typically employ a high spectral resolution ($R \gtrsim 3000$) to isolate the contamination from atmospheric OH emission lines, JWST can employ intermediate ($R \sim 1000$) and even low ($R \sim 100$) spectral resolution, which minimizes detector noise. Finally, the absence of atmospheric seeing means that each high-redshift galaxy creates a much smaller footprint on the detector, enabling a substantially larger multiplexing and higher flexibility in target selection than ground-based instruments with the same field of view.
Starting with JWST cycle 1, the NIRSpec Multi-Object Spectroscopy (MOS) mode has been used by several programs to build samples of galaxies at cosmic noon, such as the CECILIA \citep{strom23}, AURORA \citep{shapley25}, and MARTA \citep{cataldi25} surveys, which mostly targeted emission-line galaxies, and SUSPENSE \citep{slob24}, which instead focused on absorption-line galaxies. Each survey differs in terms of observation depth, wavelength coverage and, crucially, sample selection. Observational programs are typically driven by a specific science goal and therefore tend to target a specific population of galaxies. This ensures the success of the primary goal of the survey, but it also produces a galaxy sample that is typically biased.

We present the Blue Jay survey, a cycle 1 program (GO 1810, PI: S. Belli) that collected deep NIRSpec spectra over the 1-5~{\textmu}m wavelength range for a sample of about 150 galaxies at $1.7 < z < 3.5$. Instead of focusing on a particular galaxy population, the Blue Jay survey was designed with the goal of probing a representative sample of galaxies uniformly spanning the target range in redshift and mass, regardless of other properties such as color or star formation rate. This can be considered as the extension to high redshifts of the strategy adopted by the SDSS survey at $z\sim0$ \citep{strauss02} and the LEGA-C survey at $z\sim1$ \citep{vanderwel16}. This design makes the Blue Jay sample ideal for studying a wide array of aspects related to galaxy evolution, and it ensures that the results are sufficiently general to be applied to the overall galaxy population, without being driven by a bias in the target selection.

We present the survey design, sample selection, and data reduction in the present work, while the results of spectral fits to the continuum and to individual emission lines are presented in a companion paper by Bugiani et al. (in prep.). 
Several results have already been obtained using the Blue Jay observations, focusing in particular on the massive end of the galaxy sample. The deep Blue Jay spectroscopy enabled the discovery of widespread neutral outflows \citep{davies24, liboni25}, which likely cause galaxy quenching \citep{belli24}; the detailed study of star formation histories \citep{park24} and ionized gas \citep{bugiani25} in quiescent galaxies; and the dust attenuation properties of galaxies at cosmic noon \citep{maheson25}.

We describe the sample selection and the JWST observations in Sect.~\ref{sec:design}, the NIRSpec data reduction in Sect.~\ref{sec:data_reduction}, and the self-consistent spectroscopic and photometric dataset in Sect.~\ref{sec:data}. In Sect.~\ref{sec:sample} we review the main properties of the Blue Jay sample. The key science objectives of the survey, together with a brief review of the results that have already been published, are listed in Sect.~\ref{sec:objectives}. Finally, we summarize the characteristics of the survey and briefly discuss its future developments in Sect.~\ref{sec:summary}.

%%%%%%%%%%%%%%%%%%%%%%%%%%%%%%%%%%%%%%%%%%
\section{Survey design and observations} 
\label{sec:design}
%%%%%%%%%%%%%%%%%%%%%%%%%%%%%%%%%%%%%%%%%%

\subsection{Guiding principles}

The main goal of the Blue Jay survey is to probe the full set of rest-optical absorption and emission lines for an unbiased representative sample of more than 100 galaxies at cosmic noon. To achieve this goal, we made the following choices:

\medskip
\noindent
$\bullet$ Given the high signal-to-noise ratio (S/N) on the continuum that is required to detect absorption lines, we adopted the medium-resolution gratings ($R\sim1000$). This resolution is sufficient to detect significant absorption and emission lines, and at the same time yields substantially higher sensitivity compared to the high-resolution gratings ($R\sim2700$), because the continuum flux is spread over fewer pixels and is thus affected by lower read-out noise, which is the dominant source of uncertainty. The primary disadvantage is that we are unable to resolve the kinematics below the instrumental resolution of $\sigma_\mathrm{inst} \sim 130$~km/s. To overcome this limitation, follow-up observations at higher resolution with the G235H grating were obtained in cycle 3 by program GO 5427 (PI: R. Davies).

\noindent
$\bullet$ We employed all three NIRSpec gratings with medium resolutions (G140M, G235M, G395M) to observe the wide wavelength range of 1 to 5~{\textmu}m so that all the spectral features of interest are captured by the NIRSpec spectra. This sets a lower limit on the redshift of the sample at $z\sim1.7$, because below this redshift the [O~II] emission line, the bluest among the strong nebular lines, lies outside the wavelength range that can be observed with the G140M/F100LP grating/filter combination. 

\noindent
$\bullet$ The sample selection must be performed with particular care in order to reduce all possible sources of bias. This requires a parent sample that is as complete as possible and well characterized at least in terms of redshift and stellar mass. Since Blue Jay was executed in the first half of cycle 1, the parent sample was necessarily based on pre-JWST data. Deep imaging in the near-infrared was only available up to a wavelength of 1.6~{\textmu}m, corresponding to the F160W filter on the \textit{Hubble} Space Telescope (HST). This sets an upper limit on the redshift range, because beyond $z\sim3.5$ the Balmer break is redshifted past F160W, making the parent catalog highly incomplete.

\noindent
$\bullet$ The only fields with sufficiently deep multiband imaging, including near-infrared data from HST, are those observed by the CANDELS survey. Among these, we selected the COSMOS field because it is an equatorial field and can therefore be observed from both hemispheres. This maximizes the potential for ground-based follow-up studies. 

\noindent
$\bullet$ To obtain a relatively large sample without sacrificing the exposure time, it is crucial to maximize the multiplexing capabilities of the NIRSpec micro-shutter assembly (MSA). Some programs achieve this by allowing spectral overlap among different sources \citep[e.g.][]{cataldi25, degraaff25}; this strategy can work for sparse emission lines but is not applicable in our case since overlapping continuum traces would be unusable. Instead, we opted for reducing the number of shutters allocated to each source from the standard three to just two. This choice reduces the availability of local background, but this is not an issue since we adopt a master background, as described in detail in Sect.~\ref{sec:data_reduction}.

\medskip
In this section, we describe how the parent catalog was constructed, explain how it was used to design the NIRSpec pointings, and describe the observational setup.

\subsection{Parent catalog}
\label{sec:parent_catalog}

The sample selection is based on the HST F160W imaging obtained by the CANDELS survey \citep{grogin11, koekemoer11}, adopting the source catalog for the COSMOS field produced by the 3D-HST team \citep[v4.1.5;][]{brammer12, skelton14, momcheva16}. This catalog, which we call the ``CANDELS/COSMOS'' catalog, includes 33879 sources. In the remainder of this work we adopt the IDs from this catalog to refer to individual sources.

First, we improved the position measurement of the sources in the CANDELS/COSMOS catalog in order to ensure a secure acquisition of the targets observed with NIRSpec.
Since the astrometric precision of the CANDELS mosaic in the COSMOS field does not meet the NIRSpec requirements, we made use of the newer mosaic released by the 3D-DASH team \citep{mowla22}, with a pixel scale of 100~milliarcsec. The mosaic extends over a wider field of view (targeted by the 3D-DASH survey); however, we focused only on the CANDELS/COSMOS footprint, where the mosaic includes the same data used for the original CANDELS mosaic. 
We used \texttt{forcepho} (B. D. Johnson et al. in prep.) to fit the surface brightness distribution for each of the 33879 sources in the CANDELS catalog. We then identified 134 \textit{Gaia} stars in the field that have well-measured proper motions in Data Release 3 \citep{gaia23}, and are well fit by \texttt{forcepho}. After correcting for proper motion, we detected a small shift between the DASH mosaic and the \textit{Gaia} astrometry ($\Delta \mathrm{RA} = -29$ mas, $\Delta \mathrm{dec} = 7$ mas). We applied this offset to the entire catalog of coordinates measured by \texttt{forcepho}, and adopted these as the final coordinates for the parent catalog. Comparing the final coordinates of the stars with those from the \textit{Gaia} catalog, we found a median absolute deviation of 12~mas on each coordinate.

\begin{figure}[tbp]
\includegraphics[width=0.45\textwidth]{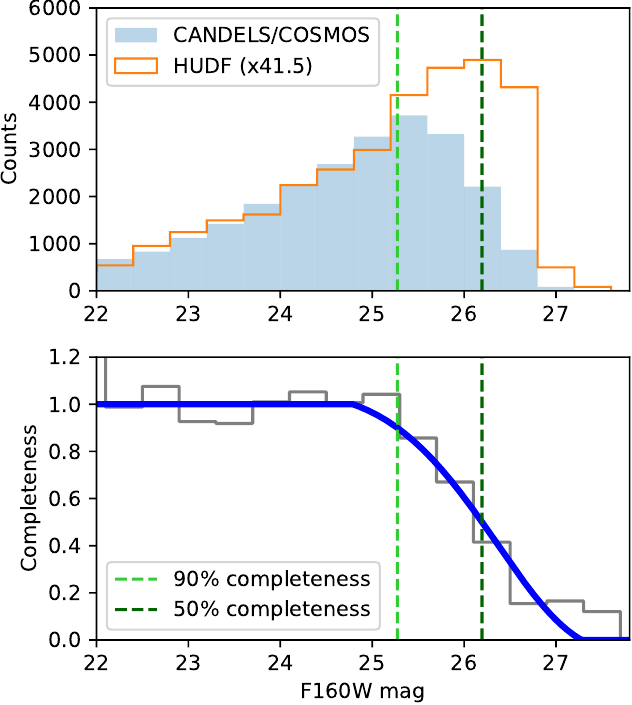}
\caption{Assessing the completeness of the CANDELS/COSMOS catalog. \textit{Top}: Source count for the CANDELS/COSMOS and the HUDF catalogs as a function of the F160W magnitude. The HUDF source counts have been normalized to account for the smaller area. \textit{Bottom}: Completeness of the catalog estimated by taking the ratio of the source counts shown in the top panel. The blue line represents a B-spline fit. The vertical dashed lines mark the 90\% and 50\% completeness limits.
\label{fig:completeness}}
\end{figure}

Second, we estimated the completeness of the F160W detections in the CANDELS/COSMOS catalog by comparing the source count to that from a substantially deeper catalog \citep[as done by, e.g.,][]{skelton14, tal14}. For this purpose we adopted the 3D-HST catalog for the GOODS-South field, where we selected all sources that lie within the small footprint of the \textit{Hubble} Ultra Deep Field (HUDF), which was observed with a considerably longer exposure time. The top panel of Fig.~\ref{fig:completeness} shows the number of sources flagged as ``usable'' and with S/N$>$5, as a function of F160W magnitude for the CANDELS/COSMOS catalog and the deeper HUDF catalog, after a normalization to account for the smaller HUDF area.
The two catalogs are very similar down to about $H\sim25$, beyond which the CANDELS/COSMOS source count begins to deviate due to incompleteness. The bottom panel shows how we estimated the completeness of CANDELS/COSMOS by taking the ratio of the source counts in the two catalogs (gray histogram), and then fitting a B-spline to obtain a smooth representation (blue curve). This lets us estimate that the CANDELS/COSMOS catalog is 90\% complete at $H_{90}=25.3$ and 50\% complete at $H_{50}=26.2$, in good agreement with the values found by \citet{skelton14} using slightly different methods.

Finally, we constructed the parent catalog by taking all objects in the CANDELS/COSMOS catalog that fulfill the following requirements:
\begin{itemize}
    \item Are flagged as ``usable'' (\texttt{use\_phot=1}) by \citet{skelton14}, which means that they are detected in F160W, are not identified as stars or too close to a bright star, are well exposed in F125W and F160W, and have meaningful fits to the photometry;
    \item Have a S/N in F160W larger than 5;
    \item Are brighter than $H_{50}=26.2$, to avoid probing galaxy populations that are mostly undetected in the CANDELS data.
\end{itemize}
The parent catalog defined in this way consists of 23755 sources.

\begin{figure}[tbp]
\includegraphics[width=0.47\textwidth]{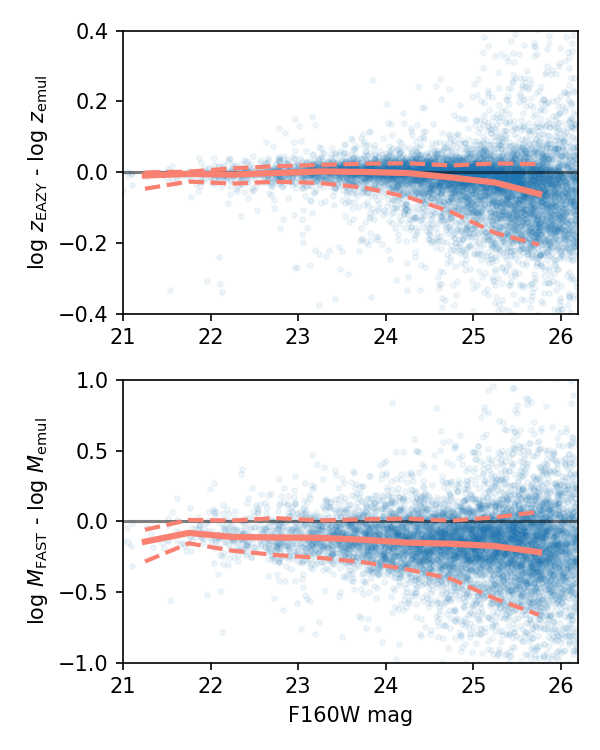}
\caption{Discrepancy in photometric redshift (\textit{top}) and stellar mass (\textit{bottom}) between the values in the 3D-HST catalog and those from the \texttt{Prospector} emulator as a function of magnitude for all galaxies in the parent catalog with redshift in the range $1.7 < z < 3.5$. In each panel, the thick line marks the running median, while the dashed lines show the interval containing the central 68\% of the distribution. The median stellar mass discrepancy is about -0.16 dex.
\label{fig:compare_mass}}
\end{figure}

\subsection{Physical properties of the parent sample}
\label{sec:parent_physical_prop}

The Blue Jay sample selection is based only on stellar mass and redshift, which can be derived from fits to the spectral energy distribution (SED) of each source. The 3D-HST catalog, in addition to the multiband HST CANDELS data, includes other ground- and space-based observations for a total of 27 broad bands and 17 medium bands, spanning the optical and near-infrared wavelength range. Using this photometric dataset, the 3D-HST team has derived photometric redshifts with the EAZY code \citep{brammer08} and stellar masses with FAST \citep{kriek09}; see \citet{skelton14} for details.

However, the assumptions used by the FAST fitting code, particularly the idealized exponentially declining star formation histories, are known to bias the resulting stellar masses \citep{leja19sed}.
This makes the FAST masses inconsistent with those measured using the \texttt{Prospector} code \citep{leja17, johnson21} assuming nonparametric star formation histories, as done for the Blue Jay survey (see \citealt{park24}; Bugiani et al. in prep.). Thus, to ensure self-consistency, we derived new stellar masses and photometric redshifts for the parent sample.

Running \texttt{Prospector} on the entire parent sample would require an extremely large amount of computational resources. Instead, we made use of an artificial neural network emulator as described in \citet{mathews23}. The emulator was trained on a subset of galaxies that are fit with \texttt{Prospector} using the FSPS library \citep{conroy09, conroy10}. The emulator is then able to produce an approximate posterior distribution for all the parameters in the \texttt{Prospector} model, given the observed photometry of a source, in a fraction of the computing time compared to directly running \texttt{Prospector}.

Figure~\ref{fig:compare_mass} shows a comparison of the emulator results with those present in the 3D-HST catalog. The photometric redshifts are approximately consistent, with an increasing scatter for fainter magnitudes. On the other hand, the stellar masses present a systematic offset, with FAST yielding masses that are $\sim0.16$~dex (median value) lower than those obtained with the emulator, with substantial scatter even at the bright end. These results are consistent with previous analysis of \texttt{Prospector} measurements compared to EAZY redshifts \citep{wang24} and FAST masses \citep{leja19sed}.
For the remainder of the article we adopted the emulator redshift and masses for the parent sample.

\begin{figure*}[htbp]
\sidecaption
\includegraphics[width=12cm]{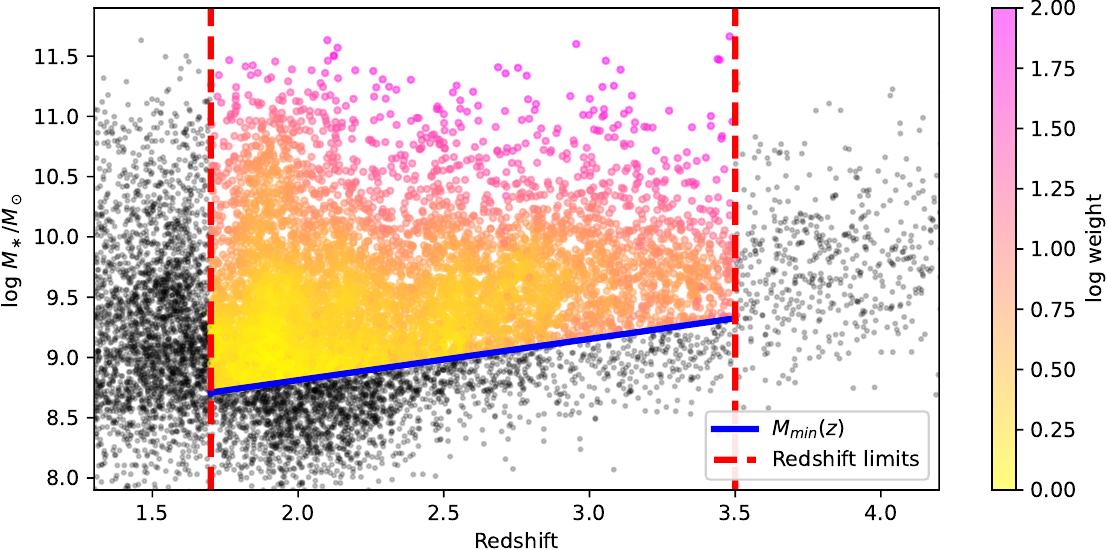}
\caption{Distribution of stellar masses and photometric redshifts for the parent sample. The redshift range of the Blue Jay survey, $1.7 < z < 3.5$ is shown in red, while the mass completeness limit is shown in blue. In the region defined by these boundaries, galaxies are color-coded by the spectroscopic weight assigned to them.
\label{fig:weights}}
\end{figure*}

\subsection{Spectroscopic weights}
\label{sec:weights}

Due to the large number of observable targets and the complex geometric requirements, NIRSpec pointings are determined automatically using an algorithm that maximizes the total weight of the selected targets. Such spectroscopic weights thus represent the relative probability of being selected for observations for each object in the parent catalog. Since our goal is to construct a spectroscopic sample that spans a wide range of properties, we chose to assign larger weights to galaxies that have values of stellar mass $M_\ast$ and redshift $z$ that are rare. In practice, this means setting the weight of each galaxy to be inversely proportional to the smoothed local density of potential targets in the $M_\ast$ versus $z$ plane, which is shown in Fig.~\ref{fig:weights}. We also divided the weight by the completeness function $f(H)$, shown in the bottom panel of Fig.~\ref{fig:completeness}, in order to account for similar targets that cannot be selected because they are missing from the parent sample (this correction is small for the majority of targets, and never exceeds 2 since we do not consider galaxies fainter than $H_{50}$).

Furthermore, we gave zero weight to galaxies outside the chosen redshift range $1.7 < z < 3.5$, and with masses below a completeness limit $M_\mathrm{min}$. We calculated this completeness limit following the method of \citet{pozzetti10}, as explained in detail in Appendix~\ref{sec:Mmin}. The resulting completeness limit as a function of redshift is $\log M_\mathrm{min}/M_\odot = 8.13 + 0.34 z$, and goes from $\log M_\mathrm{min}/M_\odot \sim 8.7$ at $z=1.7$ to $\log M_\mathrm{min}/M_\odot \sim 9.3$ at $z=3.5$.
We estimated a minor incompleteness for galaxies just above the limit, with $\log M_\mathrm{min} < \log M_\ast < \log M_\mathrm{min} + 0.36$, where we may miss about 15\% of the total population.

Finally, we scaled the weights so that the lowest values are around unity. The resulting weight distribution, shown in Fig.~\ref{fig:weights}, spans two orders of magnitude, reaching the largest values in the most massive systems. The sources in the parent sample that have a nonzero weight are 6327.

\subsection{Nod strategy}
\label{sec:nodding}

\begin{figure}[tbp]
\includegraphics[width=0.47\textwidth]{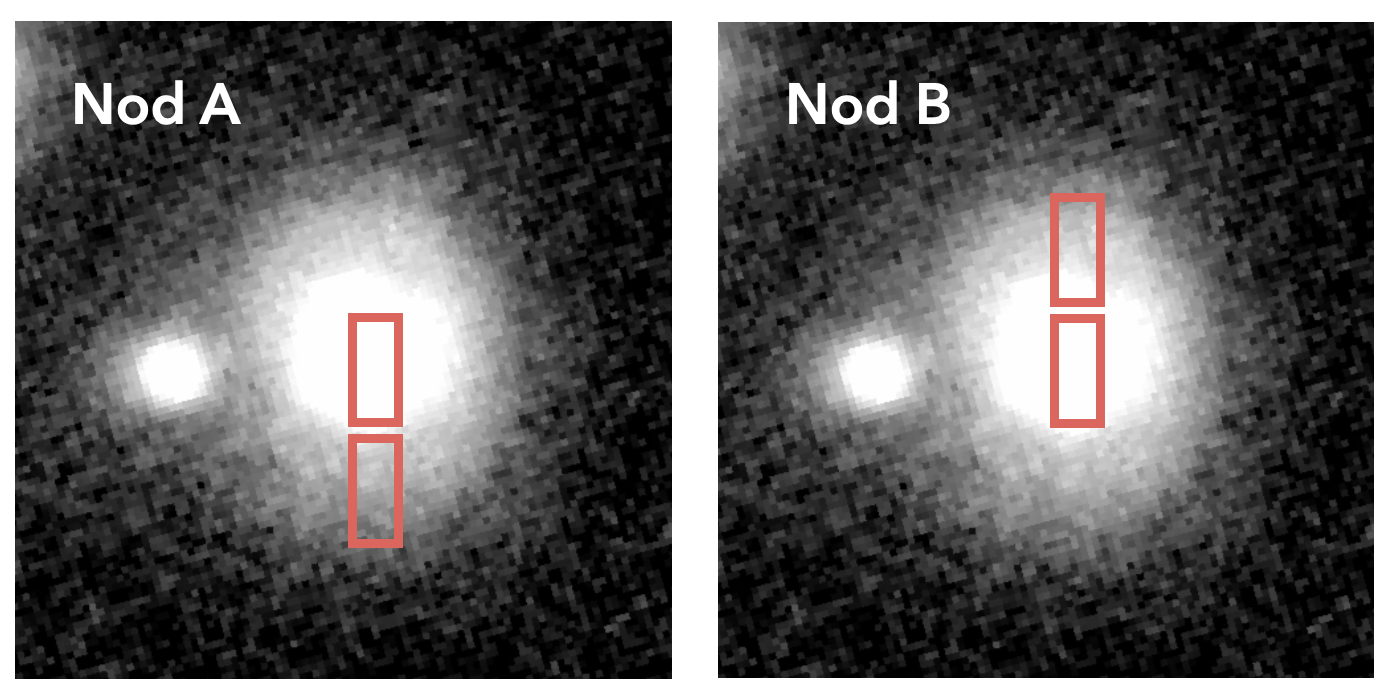}
\caption{Footprints of the NIRSpec shutters for galaxy 10592 ($z=1.8$, $\log M_\star/M_\odot=11.2$), illustrating the two-point nodding adopted by Blue Jay. Given the large size of the galaxy, it is not possible to derive a local background from shutters adjacent to the primary shutter -- instead, a master background must be used. The galaxy image is from NIRCam F200W.
\label{fig:shutters}}
\end{figure}

The MSA shutters are $0.2''\times0.46''$, or roughly $1.6\times3.8$ kpc, a size that is comparable to the typical effective size of galaxies at cosmic noon \citep[e.g.,][]{vanderwel14}. This means that the light from a galaxy will often contaminate the adjacent shutters --- a problem that affects in particular the most massive galaxies, as illustrated in Fig.~\ref{fig:shutters}. As a result, adjacent shutters cannot be used to derive a local background. This makes it impossible to adopt the default strategy for NIRSpec observations, which consists of employing three-shutter slitlets where one shutter is placed on the source and the other two shutters are used to measure the local background spectrum. Instead, we decided to perform a global background subtraction, using a high-S/N master background spectrum derived from empty shutters, as explained in detail in Sect.~\ref{sec:master_background}.

Since we did not make use of a local background, we did not necessarily need to employ the standard three-shutter slitlet. We adopted, instead, a two-shutter slitlet, which has the crucial advantage of increasing the multiplexing capabilities of the MSA, since we can allocate 50\% more galaxies using the same number of shutters. Moreover, shorter slitlets make the MSA design less prone to disruption by failed shutters (stuck open or stuck closed), leading to more efficient slitlet allocation.

We adopted a two-point nodding pattern, where the source is placed first on the upper shutter (``nod $A$'') and then in the lower shutter (``nod $B$''), as shown in Fig.~\ref{fig:shutters}. However, when reducing the data (as described in Sect.~\ref{sec:data_reduction}) we did not perform the usual $A-B$ subtraction, which in our case would lead to self-subtraction of the galaxy flux. Instead we treated the two nod positions as dithers, that is, we summed the flux obtained at the $A$ and $B$ positions. The main goal of dithering is to mitigate the impact of detector defects and flat fielding imperfections.

\begin{figure}[tbp]
\includegraphics[width=0.4\textwidth]{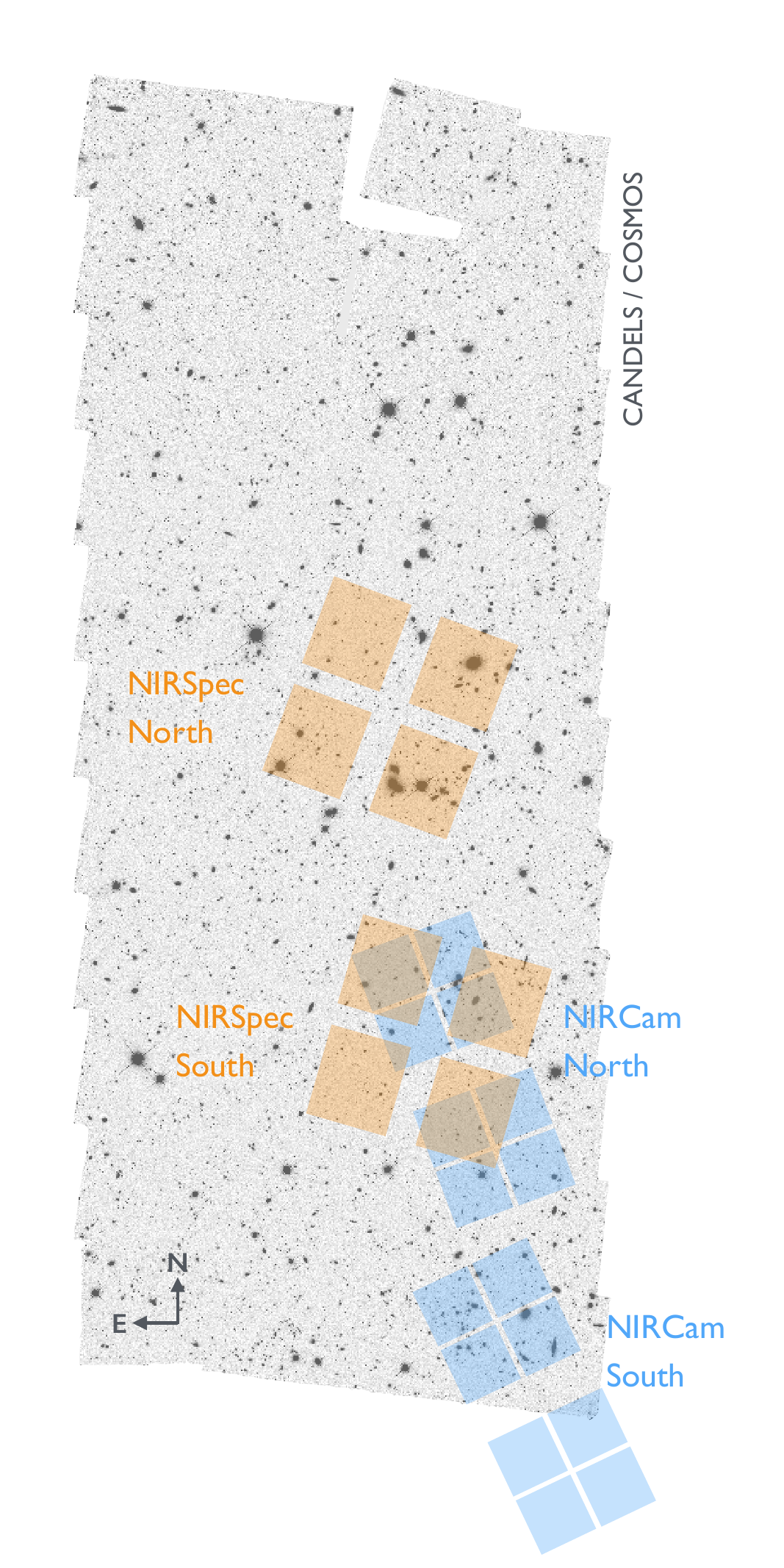}
\caption{Blue Jay NIRSpec pointings (in orange; each pointing consists of four MSA quadrants) and parallel NIRCam fields (in blue; each field consists of two separate arrays of $2\times2$ detectors). The NIRCam North field was observed in parallel to the NIRSpec North pointing. The background image is the HST F160W mosaic of the CANDELS/COSMOS field.
\label{fig:pointings}}
\end{figure}

\subsection{Configuration of NIRSpec MSA}
\label{sec:pointings}

We designed the NIRSpec pointings using the MSA Planning Tool available on the Astronomer's Proposal Tool software (version 2022.5.3). We made use of the spectroscopic weights described in Sect.~\ref{sec:weights} and adopted a two-shutter slitlet for each source. We did not allow the spectral traces to overlap on the detector, since this would make it impossible to study the continuum emission. However, we allowed science slitlets to be placed in areas affected by stuck open shutters. These are NIRSpec shutters that are stuck in the open position and cannot be closed. Each stuck open shutter produces a sky spectrum on the detector; when this overlaps with the spectrum of one of our targets, the science spectrum is contaminated.

\begin{table}[t]
\caption{NIRSpec MSA configurations.}
\label{tab:msa}
\centering
\begin{tabular}{lcc}
\hline \hline
 & North & South \\
 & Pointing & Pointing \\
\hline
\rule{0pt}{3ex}\textbf{Science} \\
\quad Slitlets            & 74$^a$        & 78$^b$ \\
\quad Shutters            & 180       & 191 \\
\rule{0pt}{3ex}\textbf{Background} \\
\quad Slitlets            & 38        & 38 \\
\quad Shutters            & 63        & 50 \\
\hline
\end{tabular}
\tablefoot{Slitlet and shutter counts include stuck open shutters.\\
$^a$ One slitlet composed of a single stuck open shutter (ID 22990).\\
$^b$ One slitlet contains two galaxies (IDs 11420 and 11451).}
\end{table}

We employed an ``unconstrained'' source centering criterion, meaning that the source will be allocated to a shutter even if the position in the parent catalog is exactly behind the shutter bar. This increases the efficiency of the MSA configuration, at the cost of potentially large slit losses. However, since our targets are typically much larger than the $0.07''$ shutter bar width, the slit losses do not depend strongly on the exact position of the source center.

With the chosen settings we were able to observe about 75 targets with a single MSA configuration; to reach the desired sample size of more than 100 galaxies we thus need two NIRSpec pointings.
We positioned the first pointing, dubbed South pointing, near an area densely populated by galaxies with high spectroscopic weights, corresponding to a known overdensity at $z\sim2.1$ \citep{spitler12}. Using the MSA Planning Tool, we searched for the optimal position (i.e., the one that maximizes the sum of the spectroscopic weights) using a large grid. We then used this position to start a new search with a finer grid, in an iterative fashion, until we reached a grid step size of 5 mas, which is smaller than the astrometric uncertainty. After we established the position of the South pointing, we repeated the procedure for the North pointing, which we placed in such a way that its parallel NIRCam imaging overlapped the South pointing, as illustrated in Fig.~\ref{fig:pointings}. This ensured that a fraction of the sample was observed by both NIRCam and NIRSpec. Moreover, the North pointing was sufficiently far from the $z\sim2.1$ overdensity to ensure that our sample probed a range of environments at each redshift.

After both pointings have been defined, we analyzed the resulting target sample by performing detailed SED fitting with \texttt{Prospector}. This is much more computationally demanding compared to the emulator fits described in Sect.~\ref{sec:parent_physical_prop}, but yields more robust results. We found that for a few galaxies the photometric redshift measured by \texttt{Prospector} is substantially different than the value predicted by the emulator. In some cases the new redshift is still within the survey limits, but for eight galaxies the new redshift is outside the $1.7 < z < 3.5$ range. We removed these targets from the parent sample and ran the MSA Planning Tool once more, using a very small grid to ensure that the majority of the targets are unchanged.

\begin{table*}[t]
\caption{Primary NIRSpec observations.}
\label{tab:nirspec}
\centering
\begin{tabular}{ccclcr}
\hline \hline
Grating/Filter & Integrations per exposure & $t_\mathrm{exposure}$ & Dithering pattern & N$_\mathrm{exposure}$ & $t_\mathrm{total}$ \\
\hline
G140M/F100LP & 4 & 97.2 min & $AA'BB'AA'BB'$ & 8 & 778 min \\
G235M/F170LP & 2 & 48.6 min & $AA'BB'$ & 4 & 194 min \\
G395M/F290LP & 1 & 24.3 min & $AA'BB'$ & 4 & 97 min \\
\hline
\end{tabular}
\tablefoot{Each integration consists of 20 groups in NRSIRS2 mode, corresponding to an integration time of 24.3 min. The dithering positions along the $y$ (cross-dispersion) direction were $A=+0\farcs529$, $A'=+0\farcs539$, $B=0\farcs000$, $B'=+0\farcs010$.}
\end{table*}

The resulting MSA configurations have a non-negligible fraction of detector area that is not illuminated, and we made manual changes to the MSA plan to take advantage of all the available space on the detector.
In some cases we found sources that are nominally outside an available shutter, but close enough to ensure that part of the light emitted by the source falls in the shutter nonetheless; we manually created slitlets for these targets. We also manually added four high-redshift ($z\sim 6$) targets, taken from a list of candidates identified with a dropout technique.
Finally, we systematically opened additional empty shutters to be used for the master background, until each row of the detector is illuminated, making sure to avoid any spectral overlap. We achieved this in two ways: first, we manually extended the galaxy slitlets by opening adjacent shutters, if they were available. We were able to do this for about one third of the slitlets, reaching a length of 3, 4, or 5 shutters.
Second, in the regions of the MSA that were still available we created new slits on empty parts of the sky, trying to uniformly sample the MSA quadrants.

Table~\ref{tab:msa} summarizes the number of shutters and slitlets in each of the two pointings. The total number of science slitlets is 152; since in one case two galaxies (both parts of the parent sample) fall in the same slitlet, the total number of galaxies observed with NIRSpec is 153. Due to the manual optimization of the MSA configurations, each science slitlet consists on average of 2.4 shutters. Additionally, there are 76 empty background slitlets, mostly consisting of one or two shutters, including those that are stuck open.

\begin{table}[t]
\caption{Parallel NIRCam observations.}
\label{tab:nircam}
\centering
\begin{tabular}{c c c}
\hline \hline
Filter & Total integrations & $t_\mathrm{total}$ \\
\hline
F090W   & 4      & 92 min \\
F115W   & 8      & 184 min \\
F150W   & 16     & 368 min \\
F200W   & 16     & 368 min \\
F277W   & 12     & 276 min \\
F356W   & 16     & 368 min \\
F444W   & 16     & 368 min \\
\hline
\end{tabular}
\tablefoot{Each integration consisted of seven groups in DEEP8 mode, corresponding to an integration time of 22.9 min.}
\end{table}

\subsection{Primary NIRSpec observations}

We observed the two NIRSpec pointings adopting a 2-point $AB$ nodding pattern, as explained in Sect.~\ref{sec:nodding}. We also performed parallel imaging observations with NIRCam, which benefits from sub-pixel dithering in order to improve the spatial resolution. This so-called compromise dither adds a second set of positions, $A'B'$, which are slightly offset from the $AB$ positions by $0.01''$ along the cross-dispersion direction. Thus, each NIRSpec observation must include at least four exposures, corresponding to the sequence $AA'BB'$.

We adopted the NRSIRS2 readout pattern with 20 groups per integration, corresponding to the maximum recommended integration time of about 1500 seconds. This is the optimal integration time, because it lowers the read-out noise as much as possible without being negatively impacted by cosmic rays. Using four exposures (because of the $AA'BB'$ pattern adopted) with just one integration each yields an optimal exposure time of about 97 minutes. To maximize the S/N and minimize the overheads, we set the exposure time for each grating to be a multiple of this value, by increasing the number of integrations and/or the number of exposures.

For each of the three gratings we determined the exposure time by requiring a S/N of 10 per $\AA$ for the most massive galaxies in the sample, to ensure that even fainter absorption lines can be robustly measured. Using the JWST exposure time calculator we found that this goal can be achieved in about 13 hours for G140M/F100LP; 3.2 hours for G235M/F170LP; and 1.6 hours for G395M/F290LP, as shown in Table~\ref{tab:nirspec}. The exposure times are progressively shorter at longer wavelengths due to the combination of brighter continuum (for massive, typically red galaxies) and higher NIRSpec efficiency.

The North and South pointings were observed, respectively, in November and December 2022, under JWST cycle 1 program GO 1810. The NIRSpec observations are available in the Mikulski Archive for Space Telescopes (MAST)\footnote{\url{http://doi.org/10.17909/t11e-xh37}} at the Space Telescope Science Institute.

\subsection{Parallel NIRCam imaging}

We obtained multiband NIRCam imaging in parallel to the NIRSpec observations. The CANDELS/COSMOS field was also observed with NIRCam by the PRIMER and COSMOS-Web programs, but the unusual depth of the Blue Jay observations offers the possibility of obtaining exquisite NIRCam imaging.
The parallel imaging observations include 41 out of 153 (27\%) Blue Jay spectroscopic targets, because the NIRCam fields partially overlap with the NIRSpec pointings by design, as explained in Sect.~\ref{sec:pointings}.

Table~\ref{tab:nircam} lists the seven broadband filters employed in the observations and the corresponding exposure times, which are partially constrained by the primary NIRSpec observations and by the NIRCam design, which allows simultaneous observations of the same field in a short-wavelength band and in a long-wavelength band. In four of the bands we reached an exposure time of about 6 hours each. For details about the NIRCam data, see \citet{conroy24}. The NIRCam parallel observations are publicly available on MAST\footnote{\url{https://doi.org/10.17909/vzzy-4w25}}. Additional medium-band imaging in the F182M and F410M filters obtained by the follow-up program GO 5427 will soon become available.

The position of the South NIRCam field was completely determined by the combination of the primary NIRSpec field location and the position angle assigned to our program. As shown in Fig.~\ref{fig:pointings}, the South NIRCam field falls partly outside the southwest corner of the CANDELS/COSMOS footprint. A nearby low-mass galaxy, Ark~227, happens to lie in that position: this coincidence enables a uniquely deep study of the outer stellar halo of a local galaxy, carried out by \citet{conroy24}.
In the two NIRCam fields we identified 6044 individual stars belonging to Ark~227, spanning the entire virial radius of the galaxy, up to 100~kpc from the galaxy center. The surface density distribution of the stars reveals the presence of ``accretion shelves,'' likely due to past minor merger events.

%%%%%%%%%%%%%%%%%%%%%%%%%%%%%%%%%%%%%%%%%%
\section{Data reduction} 
\label{sec:data_reduction}
%%%%%%%%%%%%%%%%%%%%%%%%%%%%%%%%%%%%%%%%%%

We downloaded the raw NIRSpec observations from MAST and reduced them with the JWST Calibration Pipeline \citep{jwst24} version 1.17.1, adopting version 1321 of the Calibration Reference Data System. In this section we describe the data reduction, with an emphasis on those steps in which our analysis deviates from the default Calibration Pipeline workflow.

\subsection{Slit definition}

The Calibration Pipeline relies on the description of each MSA configuration stored in the corresponding MSA metafile. Before running the pipeline, we applied a number of changes to the default MSA metafiles downloaded from MAST:
\begin{itemize}

\item Stuck open shutters are usually not included in the list of slitlets to be reduced. However, we manually added these shutters because they can be used for deriving the master background; moreover, in one case a science target (ID 22990) falls in a stuck open shutter.

\item In a few cases we manually opened a shutter when designing the MSA plan because a portion of a galaxy falls in it, despite the catalog position of the target being nominally outside the shutter area (as described in Sect.~\ref{sec:pointings}). We mark these shutters as containing a science target, and the relative coordinates were set to be formally inside the shutter footprint, but as close as possible to the catalog coordinates.

\item A small number of shutters did not open during the observations, as can be seen from the lack of background flux; these shutters were marked closed in the MSA metafile.

\item To each shutter not containing a science target we assigned a virtual source, with a unique ID, placed at the center of the shutter in the A nod position. This is purely for bookkeeping purposes and to facilitate the identification of stuck open shutters among the background slits.

\end{itemize}

\begin{figure*}[tbp]
\centering
\includegraphics[width=0.8\textwidth]{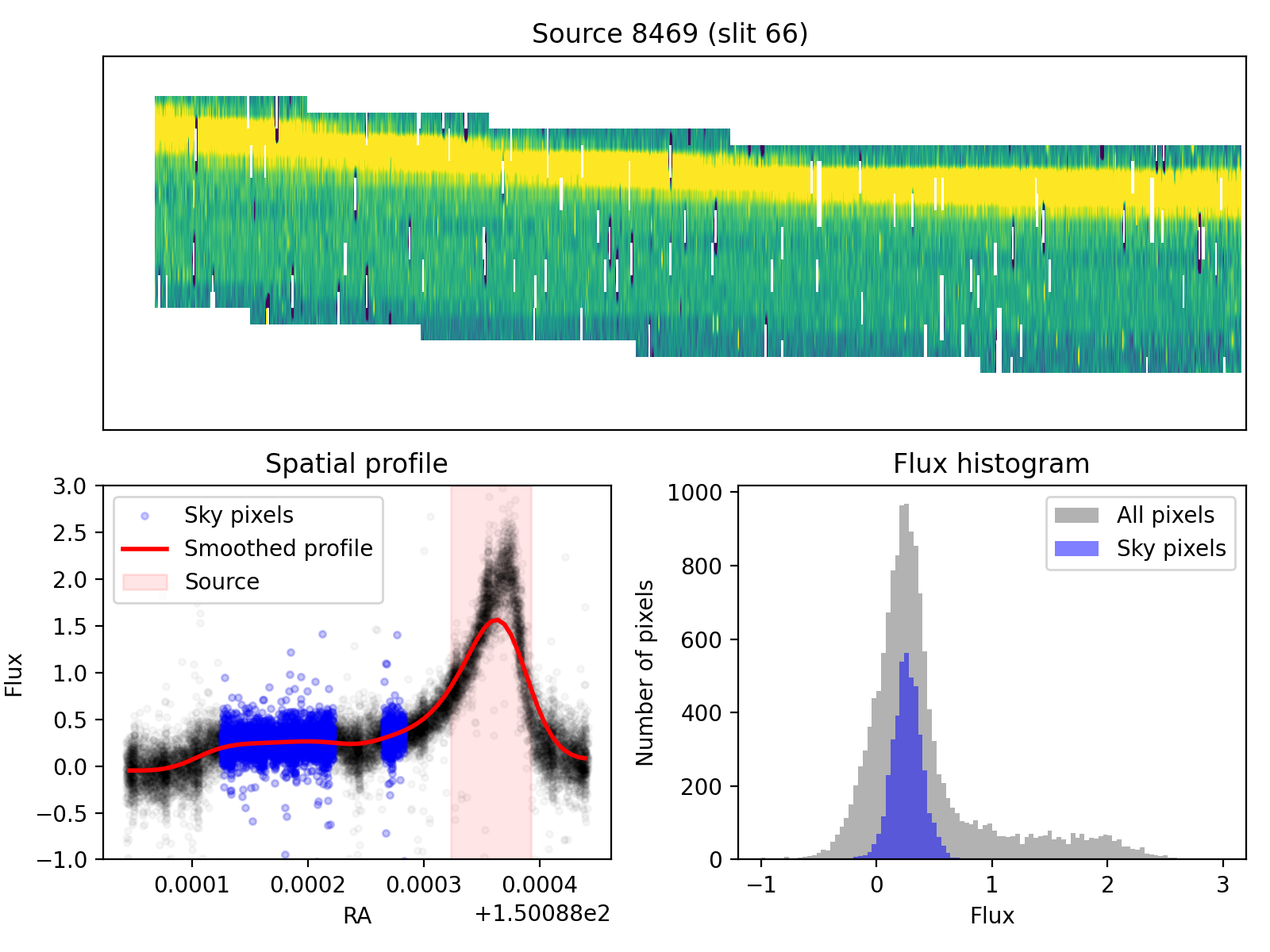}
\caption{\emph{Top:} Example of a calibrated, unrectified 2D spectrum (\texttt{cal} file) produced by the second stage of the pipeline. This is a two-shutter slitlet, with a compact galaxy in the top shutter. The shadow of the MSA bar between the two shutters is visible, despite the application of a correction by the pipeline, as the slightly darker region running along the horizontal direction, near the center of the slit, below the bright trace of the source. White vertical stripes are bad pixels (the aspect ratio of the figure is compressed along the horizontal direction in order to visualize the full spectrum). 
\emph{Bottom:} Selection of the sky pixels from the 2D spectrum. The spatial profile (\textit{left}) is used to identify source emission. Selected sky pixels, shown in blue, are those that are not affected by source emission and bar shadows. They represent fewer than half of the total pixels, and their fluxes typically have a normal distribution around a low positive value (\textit{right}).
\label{fig:background_selection}}
\end{figure*}

\subsection{Production of 2D spectra}
\label{sec:spec2d}

The uncalibrated data downloaded from MAST consist of successive, nondestructive reads of the charge in each pixel, and are stored in separate files for the NRS1 and NRS2 detectors.  
The first stage of the Calibration Pipeline performs detector-level corrections (linearity, dark current, bad pixel masking, etc.); identifies and masks those reads presenting an abrupt jump due to cosmic ray hits; and finally converts the reads to a single measurement of the count rate in each pixel. This yields the rate files that contain the traces from all the open MSA shutters.
We ran the first stage adopting the default parameters of the Calibration Pipeline and turning on the mitigation of large cosmic-ray events (so-called snowballs), which are known to affect near-infrared detectors.

One of the limitations of the pipeline is the inability to effectively mask all the bad pixels in the frame. We thus created a bad pixel mask for each detector and each pointing in the following way: we median-stacked an equal number of frames obtained with the G140M, G235M, and G395M gratings. Because each grating targets a different wavelength range, when stacking together the data we removed any sharp spectral features such as emission or absorption lines; however, the continuum traces are visible in the stack. We then identified pixels that are outliers compared to their neighbors along the dispersion direction. This method is similar to the \texttt{selfcal} algorithm that is an optional part of the JWST pipeline, but we applied it to the entire frame instead of the individual spectra. We found a number of bad pixels that is of the order of 1\% of the total.

Next, we measured the bias in each detector, which we defined as the median flux value of all the pixels that are not illuminated. We measured a small but nonzero bias, which varies from exposure to exposure. We subtracted the bias from each frame, and found that this improved the consistency of the master background measured from different frames, as discussed below.

After masking the bad pixels and correcting for the bias, we ran the second stage of the JWST pipeline, which includes the bulk of the data reduction processes, such as wavelength and flux calibration, resampling, and extraction of the 2D trace for each target. 
This yields unrectified, but calibrated, 2D spectra (\texttt{cal} files), an example of which is shown in the top panel of Fig.~\ref{fig:background_selection}, and rectified 2D spectra (\texttt{s2d} files).

We ran this stage twice: first, we adopted the standard $A-B$ subtraction, which has the advantage of removing background emission and contamination from stuck open shutters, but suffers from self-subtraction due to the large size of our targets; we used this only as a comparison dataset. We then ran the pipeline without $A-B$ subtraction and turning off any other type of background subtraction. We used this dataset to derive a master background, which we then used in the final steps of the data reduction.

\begin{figure*}[tbp]
\centering
\includegraphics[width=0.8\textwidth]{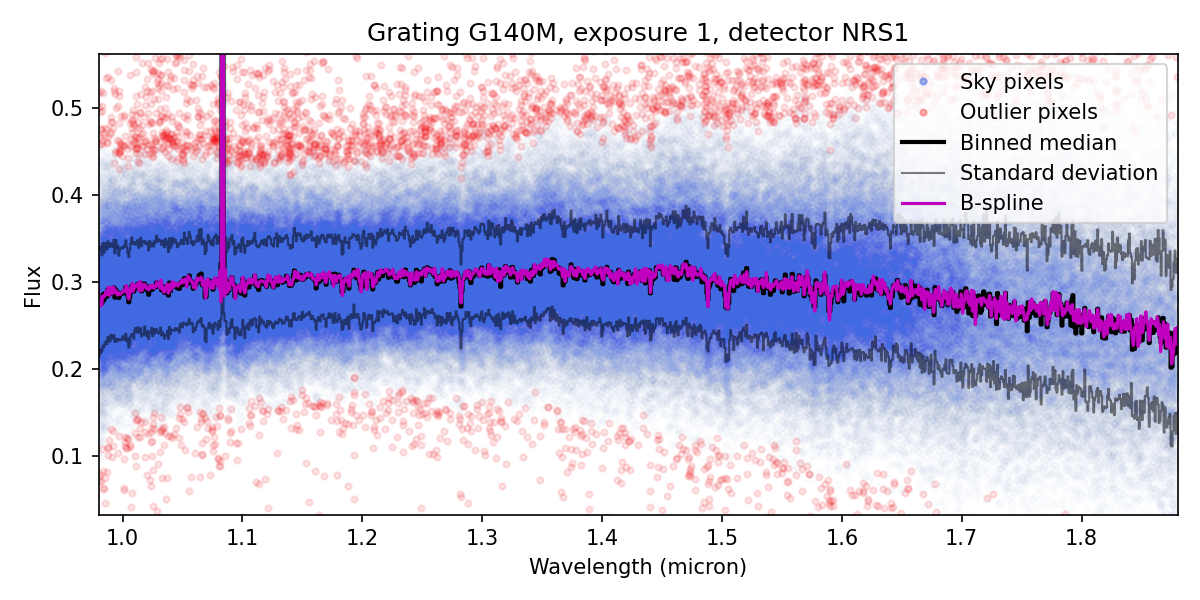}
\caption{Example of master background construction for the G140M grating. The flux is shown as a function of wavelength for more than 200,000 individual sky pixels that were selected from different slits for a given detector and exposure. After removing the outliers (red points), the master background is constructed by fitting a B-spline model (pink line) to the pixel fluxes. The running median (thick black line) and standard deviation (thin black lines) are shown for comparison. The strong emission line visible in the master background is He~I at 1.083~{\textmu}m.
\label{fig:master_background}}
\end{figure*}

\subsection{Master background}
\label{sec:master_background}

By design, the Blue Jay survey adopted a master background subtraction strategy to avoid the self-subtraction caused by the standard $A-B$ strategy applied to large target.
However, we note that the master background subtraction method is also optimal from the point of view of the S/N, because the master background spectrum can be determined with high accuracy by using a large number of slitlets, while in the $A-B$ subtraction the sky spectrum is determined using a small number of spatial pixels, and so its contribution to the final uncertainty can be non-negligible. For this reason the master background subtraction remains a valid choice even when the slitlets are longer, and is adopted by several surveys such as CECILIA \citep[see discussion in][]{rogers25} and RUBIES \citep{degraaff25}.

We constructed a master background spectrum for each combination of exposure, grating, detector, and pointing (i.e., for each \texttt{cal} file produced by the pipeline). When designing the MSA configuration, we purposefully added empty slitlets for measuring the background spectrum. Additionally, the science slitlets often include a spatial region that is free from emission from the source, either because the source is compact or because the slitlet was manually extended with additional shutters in the planning phase, as explained in Sect.~\ref{sec:pointings}. Thus, the first step is to identify all the ``sky'' pixels that are not contaminated by source emission, using the entire set of available 2D spectra.

The selection of sky pixels is illustrated in Fig.~\ref{fig:background_selection} for an example slitlet composed of two shutters: the upper shutter contains a source while the lower shutter is empty.
We started with the unrectified 2D spectrum (top panel), to avoid unnecessary resampling of the data; we then constructed the spatial profile along the slit (bottom left panel), smoothed it, and automatically identified the presence of sources by using a peak finder algorithm. 
Next, we excluded pixels that fall in the area of the detector that is not fully illuminated due to the shadow of the MSA bars. In principle, the pipeline applies a \texttt{barshadow} correction to the flux in each pixel to account for this effect, but the correction is imperfect and leaves residuals that are visible in the 2D spectrum. We thus excluded pixels with a \texttt{barshadow} correction larger than 20\%. 
The remaining pixels were selected as ``sky'' pixels, and are shown in blue in the bottom panels of Fig.~\ref{fig:background_selection}. Typically, these are fewer than half of the pixels in a slit; their histogram is a normal distribution around a low positive value. Slitlets containing large sources do not contribute any sky pixel.

\begin{figure*}[htbp]
\sidecaption
\includegraphics[width=12cm]{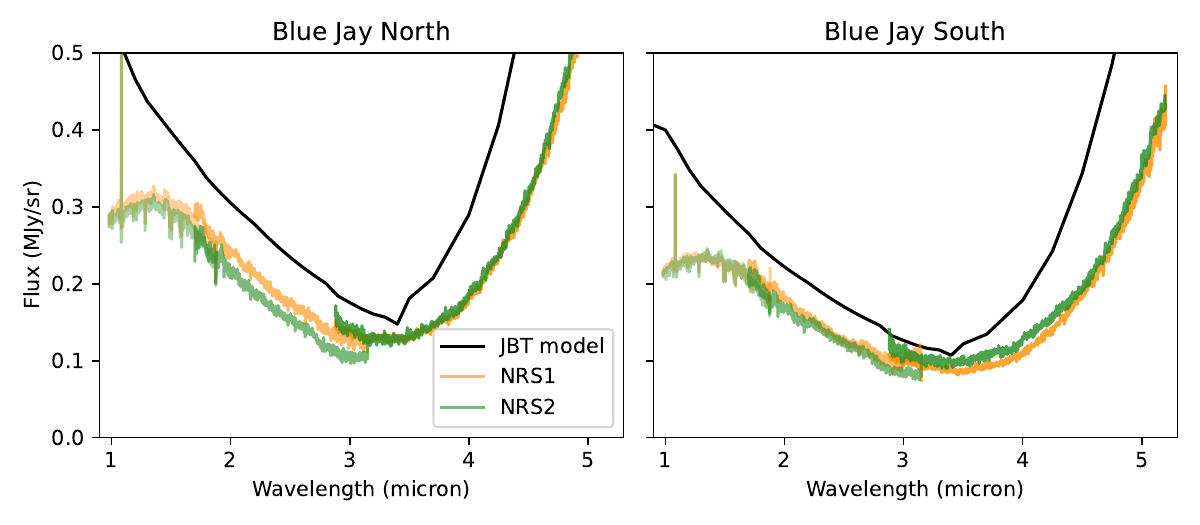}
\caption{Master background for the Blue Jay North (\textit{left}) and South (\textit{right}) pointings. The NRS1 and NRS2 master backgrounds are shown separately with different colors; for each detector, the three gratings are plotted with different levels of transparency. The black line is the JBT model generated for the coordinates and observation date of the two pointings.
\label{fig:masterbg-NS}}
\end{figure*}

\begin{figure*}[htbp]
\sidecaption
\includegraphics[width=12cm]{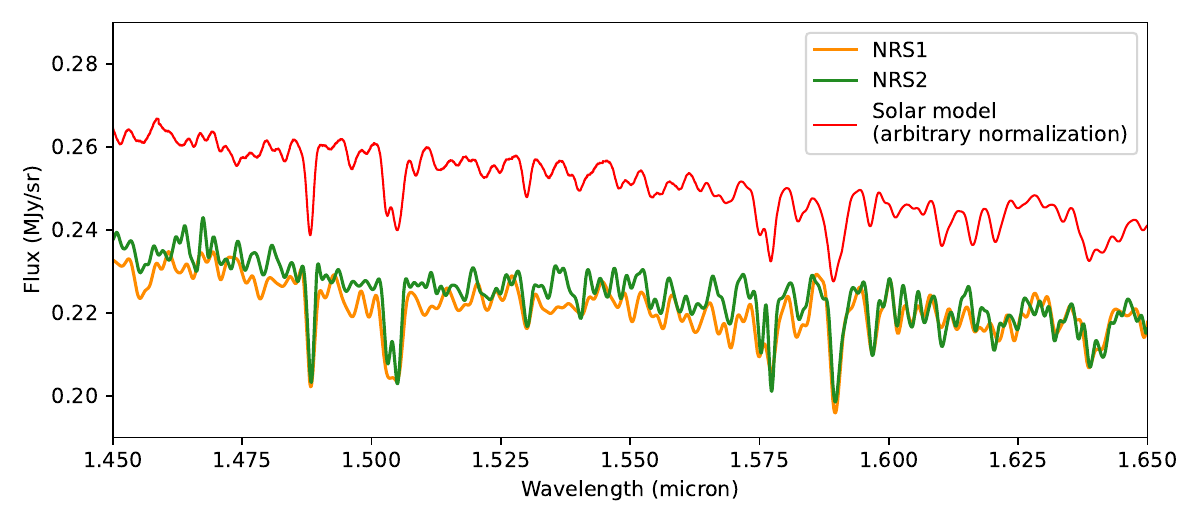}
\caption{Zoom-in on a small spectral region of the G140M grating, showing the master background in the South pointing for the NRS1 and NRS2 detectors. For comparison, we show the \citet{kurucz93} solar model with a $R\sim1000$ resolution, after having applied a continuum normalization derived from the JBT, and an arbitrary offset.
\label{fig:masterbg-solar}}
\end{figure*}

After selecting the sky pixels in all the available slitlets from a given \texttt{cal} file, we combined them to obtain a master background. We first excluded slits where the median sky flux is an outlier compared to the other slits; these are cases where the peak finder algorithm fails due to large and/or multiple sources in the slit. We also excluded pixels that have a large uncertainty, as calculated by the pipeline. The final selection of all sky pixels for a single detector includes a number of pixels of the order of $10^5$; an example is shown in Fig.~\ref{fig:master_background}.
After removing outlier pixels with a running sigma-clipping, we obtained the master background spectrum by fitting a B-spline to the sky pixel flux as a function of wavelength. For comparison, we also binned the sky pixels in wavelength and calculated the running median, finding a consistent result. We estimated the uncertainty by taking the standard deviation of the pixel flux values in each wavelength bin. We then compared this to the median of the pixel flux uncertainties provided by the pipeline. We found that the standard deviation is a factor 1.5-2 larger than the formal uncertainty, indicating that the pipeline underestimates the spectral error, as commonly reported \citep[e.g.,][]{maseda23}. We did not correct the spectral uncertainties at this point, because this correction is done by the so-called jitter parameter in the \texttt{Prospector} fitting, as discussed in \citet{park24} and Bugiani et al. (in prep.).

\begin{figure*}[htbp]
\sidecaption
\includegraphics[width=12cm]{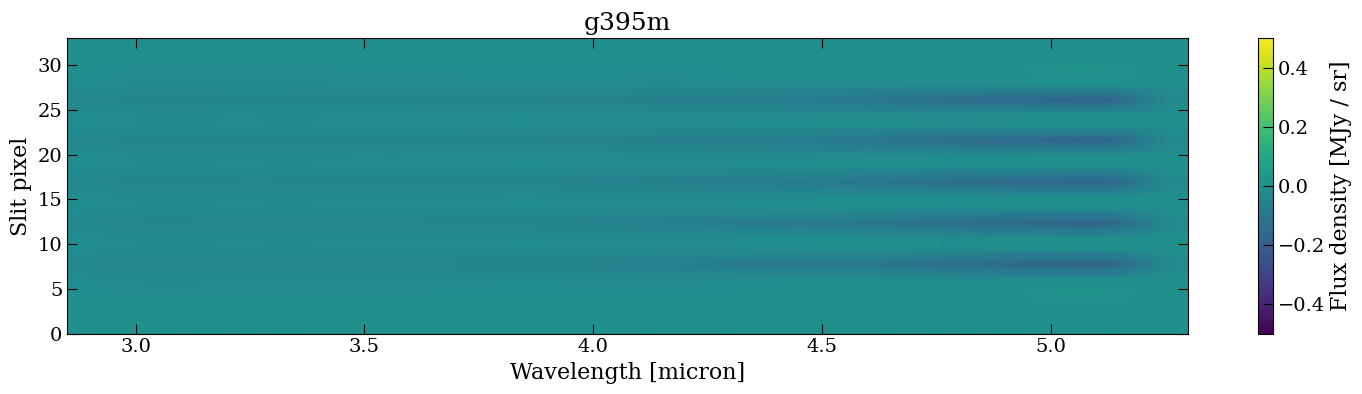}
\caption{Stack of all the pixels in empty shutters for the G395M grating, where the pixels have been binned according to their distance from the bar position. 
The residual background pattern is produced by insufficient bar-shadow attenuation of the master background spectrum. We correct for this effect by subtracting this stack from each 2D spectrum.
\label{fig:barshadow}}
\end{figure*}

We found that the variations in the master background spectrum from one exposure to another, or from one region of the detector to another, are small compared to the variations between the two detectors. We thus combined all the background spectra into a single master background for each grating and each detector. The master background spectra for the two Blue Jay pointings are shown in Fig.~\ref{fig:masterbg-NS}, separately for the NRS1 and NRS2 detectors. For each pointing and grating, the two detectors have slightly different background spectra, by up to 15\%; this could be due to small variations of throughput, bias, or other systematics. For this reason, we applied the master background subtraction separately for the two detectors instead of combining them together. We also noted a discrepancy in the background at 3{\textmu}m measured from the G235M and G395M filters, present in both the North and South pointings, which is likely due to imperfect flux calibration.

The master background spectra for the North and South pointings are clearly different in normalization and shape. The intensity of the 1.083~{\textmu}m He~I line, which is the only strong emission line in the sky spectrum, is also different in the two pointings. This is due to physical variations in the background, since the pointings were observed at different times of the year. We used the JWST Backgrounds Tool\footnote{\url{https://jwst-docs.stsci.edu/jwst-other-tools/jwst-backgrounds-tool}} (JBT) to estimate the expected background level given the coordinates and time of the year of the Blue Jay observations. While the overall shape of the JBT prediction is in broad agreement with the observed background, its normalization is about 20\% too high, and the discrepancy increases rapidly at the short wavelength end, as shown in Fig.~\ref{fig:masterbg-NS}. However, these are known issues with the JBT model \citep{rigby23}. The observed sky spectrum also contains several absorption lines, which are not modeled by the JBT. These are due to the zodiacal light, which is scattered sunlight and therefore has the same absorption lines that are found in the solar spectrum. To verify this, in Fig.~\ref{fig:masterbg-solar} we compare a small section of the G140M background spectrum with a solar model \citep{kurucz93}. The observed small-scale features clearly match the absorption lines in the solar spectrum. We thus concluded that our procedure is able to recover a robust measurement of the background spectrum.

\begin{figure*}[tbp]
\includegraphics[width=\textwidth]{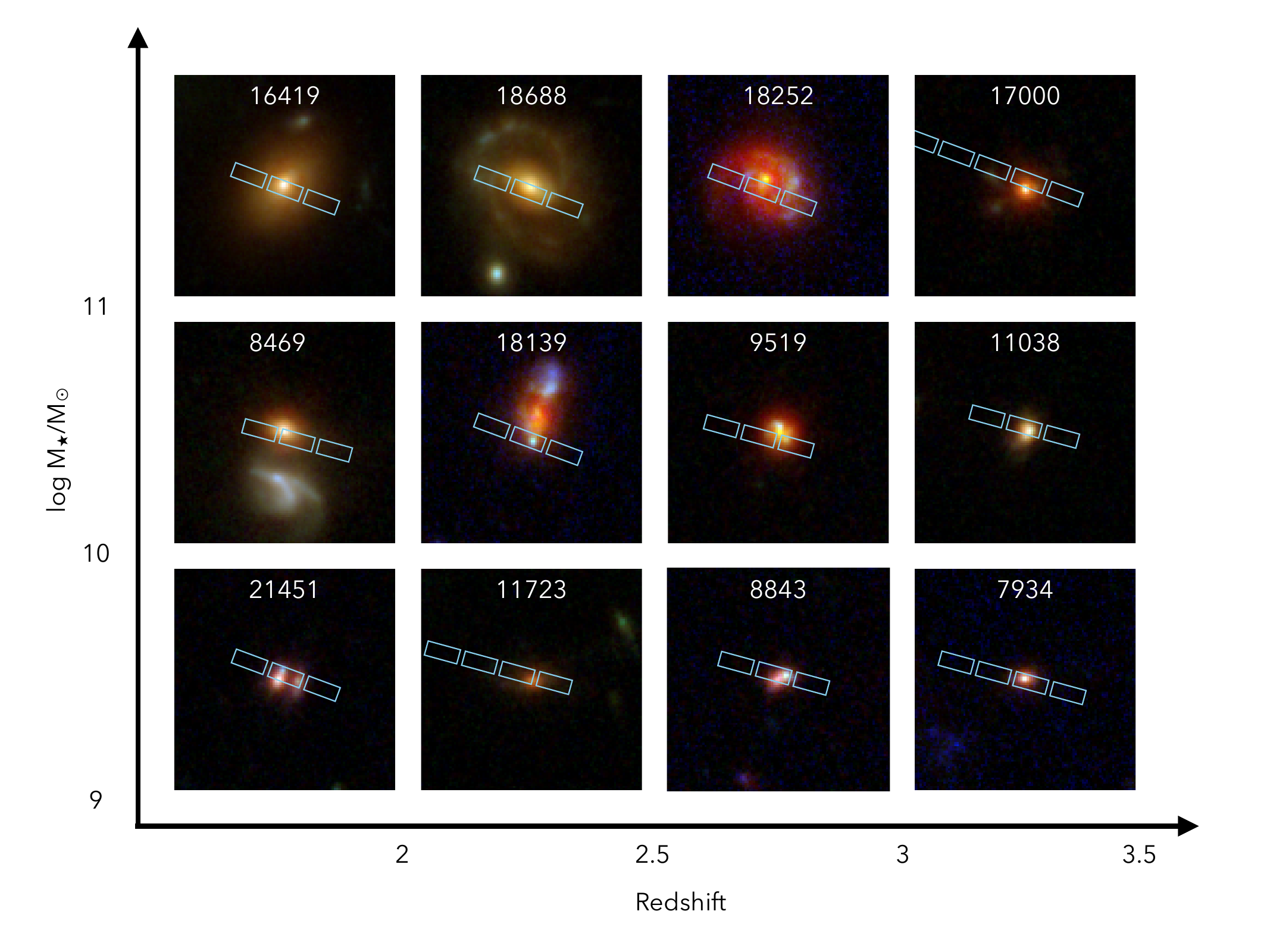}
\caption{Sample of Blue Jay spectroscopic targets spanning a range of redshifts and stellar masses. Cutouts are $3''$ on the side and are produced with the NIRCam F115W, F200W, and F444W images; north is up. The footprint of the NIRSpec microshutters is shown in each cutout. The sum of the $A$ and $B$ nod positions is shown so that two-shutter slitlets span three shutter positions on the sky.
\label{fig:cutouts}}
\end{figure*}

\subsection{Contamination masking and bar shadow correction}

We ran the second stage of the JWST pipeline for the third and last time, turning on the master background subtraction and providing the master background spectrum obtained with the method described above. Finally, we ran the third stage of the pipeline, which combines different exposures and gratings, and thus obtained fully calibrated, rectified, and background-subtracted 2D spectra. 

Some of the 2D spectra are clearly contaminated by the sky background produced by stuck open shutters, which can leave relatively bright traces on top of the science spectra. In other cases the contamination is due to some type of artifact that illuminates a small region of the detector (typically near the blue end). Usually, contamination is efficiently removed by the A-B subtraction, but since we adopted the master background subtraction we needed to manually mask all the affected pixels in each spectrum, separately for the A and B nods.

Another issue is that, in the empty slits, pixels affected by the bar shadow have fluxes centered on negative values, instead of zero. This effect is caused by the imperfect bar shadow correction, which the pipeline applies to the master background before subtracting it from the data. Since the pipeline does not sufficiently suppress the background spectrum in the pixels affected by the bar shadow, these pixels end up being over-subtracted. To correct for this issue, we constructed a 2D stack using all the empty shutters, and binning each pixel row according to their distance from the bar position. The result shows a clear pattern where the pixels that are fully illuminated have fluxes around zero, while those that are closer to the shutter bar have increasingly negative fluxes, particularly towards the red end of the NIRSpec range, where the background flux increases and the point spread function (PSF) becomes larger (see Fig.~\ref{fig:barshadow}). We subtracted this 2D background pattern from each science spectrum. This method substantially mitigates the imprint of the bar shadow on the 2D rectified spectra; however, many slitlets still feature flux residuals, either positive or negative, coinciding with the shutter bar position. This appears to be due to a variation in the shape and strength of the bar shadow with position on the MSA, which is not captured by our 2D stack. 

The imperfect bar shadow correction remains one of the limiting factors for the accuracy of the Blue Jay spectra. In order to improve on this, a more complex model of the bar shadow correction that is allowed to vary for each slitlet would be required \citep[see, e.g.][]{degraaff25, rogers25}.

%%%%%%%%%%%%%%%%%%%%%%%%%%%%%%%%%%%%%%%%%%
\section{Spectrophotometric dataset}
\label{sec:data}
%%%%%%%%%%%%%%%%%%%%%%%%%%%%%%%%%%%%%%%%%%

Figure~\ref{fig:cutouts} shows a select sample of Blue Jay targets as a function of their redshift and mass. The cutouts were made with JWST/NIRCam data (described below); the footprint of the microshutters that make up the MSA slitlet is also displayed. Most slitlets consist of two shutters, which however probe three positions when adding the $A$ and $B$ nods together.
In many cases the microshutters are small compared to the target, particularly for galaxies at intermediate and high masses. Moreover, the NIRSpec slitlet is often not centered on the galaxy, either because of geometrical constraints of the MSA (which consists of a fixed grid), or because the source center measured from the HST imaging, which was used for designing the observations, is biased towards blue, star-forming regions.
It is thus clear that, at these redshifts, NIRSpec MSA observations of massive galaxies are able to probe only a region, and not necessarily the central one, within the targeted galaxy. This may introduce a bias when analyzing spectroscopy together with broadband photometry, which is typically measured over the entire extent of the galaxy. We solved this problem by developing two different sets of photometric measurements: the ``source photometry'' was measured in an elliptical aperture encompassing the entire galaxy; the ``box photometry'' was measured in the small ``box'' aperture, which is used to extract the NIRSpec data, as explained in this section.

\subsection{Photometric data}
\label{sec:photometry}

All Blue Jay targets happen to be within the field observed by the PRIMER survey, a cycle 1 program (GO 1837; PI: J. Dunlop) that obtained multiband NIRCam observations in a subset of the COSMOS and UDS fields.
PRIMER imaging includes seven broadband filters (F090W, F115W, F150W, F200W, F277W, F356W, F444W) and one medium-band filter (F410M). 
These observations were taken after the Blue Jay survey was executed, so they were not used for the target selection.
We reduced the public imaging data and tied the astrometry to the \textit{Gaia} coordinates of reference stars.

We also adopted multiband HST imaging taken with ACS (F606W and F814W) and WFC3 (F125W, F140W, F160W) in the COSMOS field as part of the CANDELS survey \citep{grogin11, koekemoer11}. We downloaded the v7.0 mosaics of the HST imaging data for the PRIMER field from the Dawn JWST Archive\footnote{\url{https://dawn-cph.github.io/dja/}}. These mosaics are aligned with the DASH F160W mosaic and so we applied the same astrometric correction discussed in Sect.~\ref{sec:parent_catalog}.

The scientific analysis of the Blue Jay sample is thus based on HST and JWST photometry in 13 filters covering from 6000~\AA\ to 4.4~{\textmu}m, corresponding to a complete coverage of the rest-frame optical wavelength range for galaxies at cosmic noon, including parts of the rest-frame UV and near-infrared range.
This highly homogeneous, space-based dataset is crucial for accurately measuring the photometry in the small box aperture, as described below. 
Additional imaging in about 40 bands is available for the COSMOS field from several ground-based observations. We used these observations for characterizing the parent sample during the preparation of the Blue Jay survey (see Sect.~\ref{sec:parent_physical_prop}); however, we did not employ ground-based data for the scientific analysis because of its poor spatial resolution.

\begin{figure*}[htbp]
\centering
\includegraphics[width=\textwidth]{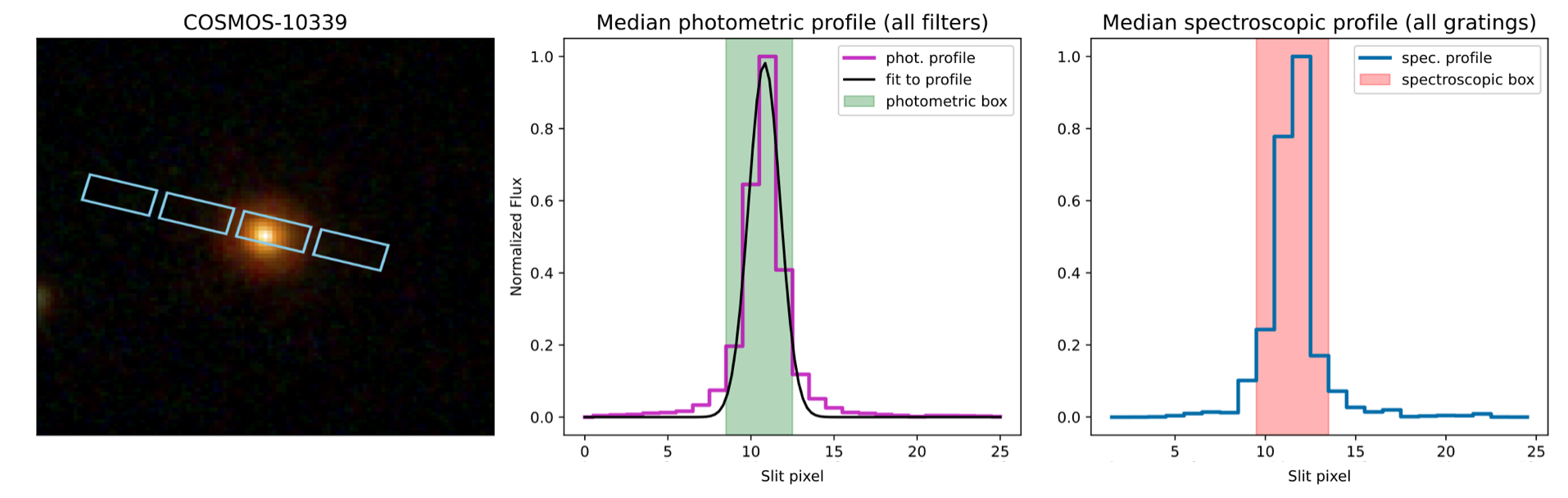}
\caption{Definition of the spectroscopic box for an example galaxy. \textit{Left}: JWST/NIRCam cutout with the footprint of the NIRSpec slitlet. \textit{Middle}: Photometric profile along the slitlet (purple line), obtained by calculating the median of all NIRCam filters resampled on the NIRSpec grid of $0\farcs1$ pixels. The black line is the best-fit Gaussian profile, and the box photometry is extracted from the green region. \textit{Right}: Spectroscopic profile along the slitlet (median of all three NIRSpec gratings), with the spectroscopic box shown in pink. There is a small offset between the photometric and spectroscopic box to account for the effect of bar shadows and misalignments.
\label{fig:phot_box}}
\end{figure*}

\subsection{Source photometry}
\label{sec:source_phot}

For each target, we used the NIRCam cutout in the F444W filter to calculate a segmentation map, from which we derived an elliptical aperture, based on the Kron radius, which encompasses more than 90\% of the emission belonging to the source. The segmentation map parameters were manually adjusted for complicated cases, for instance, when the galaxy is clumpy or is interacting with a companion.
Next, we PSF-matched all the JWST and HST images to the reddest filter available on the same telescope, that is, F444W on JWST and F160W on HST, which have nearly identical spatial resolution (full width at half maximum of 0.14~arcsec).
We then measured the photometry in the elliptical aperture (derived from F444W) for each PSF-matched image, and applied a theoretical aperture correction under the assumption of a point source.

The uncertainty on the source photometry was calculated from the flux error in each pixel for JWST/NIRCam data; while for HST bands we adopted the procedure developed by \citet{skelton14}, which is based on the photometry measured for empty apertures and thus accounts for correlated errors.
Finally, for each HST and JWST band we assessed the possible presence of nonzero background due to, for example, imperfect data reduction. We estimated the background flux by measuring the median flux in all pixels that are near the source, after aggressively masking all detected objects. We found that the background is systematically consistent with zero, and we did not subtract it from the measured photometry; however we included the uncertainty on the determination of the background flux in the final photometric uncertainty of each source.

\subsection{Box photometry}
\label{sec:box_photometry}

For each source, in addition to the elliptical aperture we also defined a smaller ``box'' aperture that is used for extracting both photometry and spectroscopy. We defined the box in the following way: we took the footprint of the NIRSpec slitlet and treated it as if it was a single, solid slit (i.e., we neglect the shutter bars), then we measured the flux profile along the slit in each JWST/NIRCam band. We resampled the flux profile on a grid matching the NIRSpec spatial pixels, which are $0\farcs10$ on the side, and then median-stacked the normalized profiles of all bands together; an example is shown in Fig.~\ref{fig:phot_box}. We fit the profile with a Gaussian profile and took the $\pm2\sigma$ limits, rounded to the closest pixels, as the box limits. This choice is made in order to optimize the S/N of the 1D spectrum, since a longer box would include more light but also more readout noise.
We measured the photometry and its uncertainty in the box aperture, for each of the HST and JWST filters that are available, using the PSF-matched cutouts.

Photometric measurements made through such a small aperture are extremely sensitive to the exact placement of the box on the cutout. We thus estimated the relative uncertainty on the box photometry due to a possible offset between the nominal and actual position of the NIRSpec slitlet on the galaxy image. There are two physical reasons for this offset:
\begin{enumerate}
    \item The nominal position is based on the reference DASH mosaic described in Sect.~\ref{sec:parent_catalog}, which has an uncertainty of 12~mas on the absolute astrometry as measured from the position of \textit{Gaia} stars. This means that each cutout used for the photometry measurement cannot be aligned to the box aperture with a precision higher than this.
    
    \item The NIRSpec ``MSATA'' acquisition procedure can position the slitlets with a limited accuracy, which depends on the astrometric precision of the reference mosaic used to prepare the observations\footnote{\url{https://jwst-docs.stsci.edu/jwst-near-infrared-spectrograph/nirspec-operations/nirspec-target-acquisition/nirspec-msa-target-acquisition}}. Given the astrometric precision of our parent catalog, the NIRSpec pointing should have an uncertainty of 20~mas. 
\end{enumerate}
As a result, the uncertainty on the placement of the NIRSpec shutter footprint on any given mosaic of HST or JWST imaging is 23~mas on each coordinate. This is almost ten times smaller than the shutter width, which is 200~mas. For a uniformly illuminated slit this uncertainty on the position does not introduce any additional uncertainty on the flux measured through the slit. However, for a point source placed near the edge of the shutter the uncertainty can be non-negligible. We estimated an approximate value for this uncertainty by assuming that the PSF is a uniformly illuminated disk with a 140~mas diameter, which is the full width at half maximum (FWHM) of the PSF-matched imaging (see Sect.~\ref{sec:source_phot}). If the PSF is placed near the edge of the shutter, and the edge position is changed by 23~mas, then the area of the PSF falling in the shutter changes by at most $23/140 \sim 16\%$, causing a similar change in the measured flux. Considering that our typical target is substantially more extended than the size of the PSF, we adopted this as a conservative estimate of the systematic uncertainty in the slit photometry, to be added in quadrature to the flux uncertainty.

\subsection{Extraction of 1D spectra}

We aimed to extract the NIRSpec 1D spectrum from the same box aperture used to extract the photometry. The first step was to obtain the median light profile along the slitlet, by collapsing the 2D spectrum on the spatial direction for all three gratings. When we compared this ``spectroscopic'' profile with the ``photometric'' profile we often found peaks at slightly different positions along the slitlet (see, e.g., Fig.~\ref{fig:phot_box}). This is due to different reasons, such as inconsistent definitions of the zero-th pixel in the NIRSpec versus NIRCam data, and the presence of bar shadows in the 2D spectrum. We measured a systematic offset in the profile peak of 0.7 pixels, which we applied to every spectroscopic profile. We then allowed for a further shift of the box by up to $\pm2$~pixels, by calculating the box position that maximizes the spectroscopic S/N. We repeated this procedure separately for each of the three gratings, and defined our final spectroscopic box as the one that maximizes the S/N for the majority of the gratings. This procedure ensured that both the photometry and the spectroscopy come from an equal-size box along the slitlet that encompasses the peak of the light emitted by the source.

Finally, 1D spectra were extracted using a simple boxcar extraction from the selected region along the slit. We decided not to use the optimal extraction algorithm \citep{horne86} because it makes it difficult to build a dataset of matching spectroscopic and photometric data extracted from the same region. The optimal extraction guarantees a higher S/N, but we verified that the increase in S/N compared to our boxcar extractions is marginal for the majority of the sample. We extracted separately the A and B nods for each object, and then combined the 1D spectra. This gave us the opportunity to detect and mask any remaining bad pixels, which appear as significant outliers in one nod but not the other.

\begin{figure}[tbp]
\includegraphics[width=0.5\textwidth]{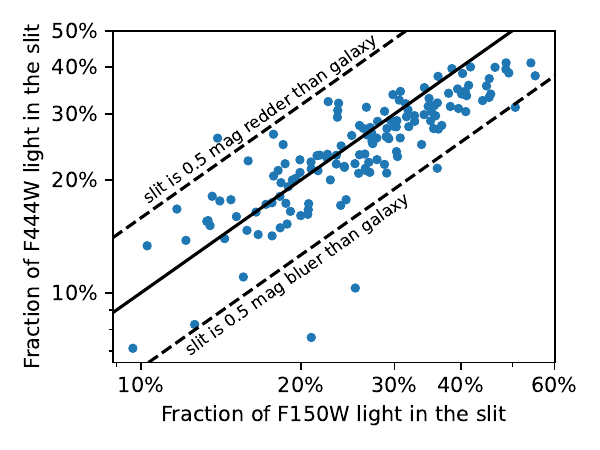}
\caption{Slit loss in F444W versus slit loss in F150W for the Blue Jay sample. The slit loss is defined here as the flux measured in the spectroscopic box from which the NIRSpec 1D spectrum is extracted, divided by the flux measured for the entire source. The solid line is the 1:1 relation, where the colors of the spectroscopic aperture are the same as the galaxy colors. Dashed lines indicate where the colors measured in the spectroscopic aperture deviate by 0.5 mag compared to the colors measured for the entire galaxy.
\label{fig:slitloss}}
\end{figure}

\subsection{Slit losses}
\label{sec:slit_losses}

Using the two sets of measurements (the source photometry and the box photometry), we were able to assess the fraction of light emitted from each source that is captured by the NIRSpec spectrum. We defined the ``slit loss'' in a given band as the ratio between the flux measured in the box aperture and the total source flux measured in the elliptical aperture. We note that this measurement does not account for the light lost to the shutter bars; moreover, slit loss is somewhat a misnomer because we did not consider the full extent of the slit, only the small box aperture that is used to extract the 1D spectrum.

Figure~\ref{fig:slitloss} shows the distribution of slit losses for the Blue Jay sample, calculated in the F444W and F150W NIRCam bands. Generally, less than 50\% of the light falls in the spectroscopic aperture, and this fraction can be as low as 7\% for the most extreme cases such as 18139 (one of the examples shown in Fig.~\ref{fig:cutouts}), where the slitlet goes through a small, HST-detected clump that belongs to a much larger, HST-dark galaxy.

By comparing the slit loss in different bands we could also assess whether the region targeted by NIRSpec is representative of the whole galaxy. The bulk of the sources in Fig.~\ref{fig:slitloss} lie near the 1:1 line, where the slit loss is the same in blue and red filters; however the scatter is large and can reach 0.5 magnitudes of difference in either direction. This means that the spectroscopic measurements of star formation rates, stellar ages, and so on are not those of the entire galaxy, but only of the region targeted by NIRSpec, which is often but not always the central region. Moreover, this introduces a systematic effect when fitting simultaneously the spectroscopy derived from the box aperture with the photometry derived from the large elliptical aperture: the two datasets are inconsistent as they belong to regions with different physical properties. This problem has generally been of limited impact for ground-based spectroscopic observations given the combination of large PSF and large slit width. However, the small size of the NIRSpec shutters compared to the typical size of galaxies at cosmic noon makes it particularly important to match the spectroscopic and photometric apertures. By employing the box photometry we were able to model both spectroscopic and photometric data in a self-consistent way; this analysis is presented in Bugiani et al. (in prep.).

\subsection{Data release}

We released the HST and JWST photometry for each Blue Jay target, measured from both the elliptical and the box apertures, in a public repository\footnote{\url{https://doi.org/10.5281/zenodo.17333505}}.
We also released the spectroscopic dataset, which includes the 2D spectra obtained both with the master background subtraction and with the $A-B$ subtraction, and the 1D spectra extracted from the box aperture using different methods, in a public repository\footnote{\url{https://doi.org/10.5281/zenodo.21813494}}. For each galaxy we defined a ``best'' version of the 1D spectrum; for the majority of the sample this is the master-background subtracted spectrum, but for a handful of special cases the presence of strong contamination made it necessary to adopt the A-B subtraction.
The spectroscopic data, together with the NIRCam cutouts showing the shutter footprint; catalogs for the target and parent samples including the spectroscopic weights; and the custom MSA metafiles used for the data reduction, were also included in the public repository.
The photometric and spectroscopic dataset released with the present work have been labeled \texttt{v2.3.1} to distinguish them from earlier versions adopted (and publicly released) by previous works \citep[e.g.,][]{belli24, park24}. The \texttt{v2.3.1} data reduction represents a substantial improvement over previous versions, which included fewer targets due to failures in the data reduction and did not employ the matched-aperture extraction of spectroscopy and photometry.

%%%%%%%%%%%%%%%%%%%%%%%%%%%%%%%%%%%%%%%%%%
\section{The Blue Jay sample} 
\label{sec:sample}
%%%%%%%%%%%%%%%%%%%%%%%%%%%%%%%%%%%%%%%%%%

Four of the 153 targets are high-redshift fillers, and are all confirmed as being at $z\sim 6-7$ via the detection of several emission lines; their spectra are shown in Appendix~\ref{sec:hiz}. We removed these targets from the Blue Jay sample due to their different selection method.
The cosmic noon sample thus consists of 149 galaxies, which we list in Table~\ref{tab:sample}. For 139 of these galaxies we were able to secure a spectroscopic redshift; we show a small portion for each of these spectra in Appendix~\ref{sec:spectra}. Clearly, the vast majority of the spectra contain a large number of emission and/or absorption lines, making it possible to measure redshifts without ambiguity.

\begin{table*}[t]
\caption{Blue Jay sample (extract).}
\label{tab:sample}
\centering
\begin{tabular}{ccccccccccc}
\hline \hline
\rule{0pt}{2ex}(1) & (2) & (3) & (4) & (5) & (6) & (7) & (8) & (9) & (10) & (11) \\
\rule{0pt}{2ex}ID & R.A. & Dec. & F160W & Weight & $\log M_\ast/M_\odot$ & $z_\mathrm{phot}$ & $z_\mathrm{spec}$ & S/N$_{140}$ & S/N$_{235}$ & S/N$_{395}$ \\
\hline
\rule{0pt}{3ex}
7102  &  150.089834  &  2.250410  &          25.2  &     5  &           9.4  &  3.31  &          3.264  	&  8.19 	 & 4.82 	& 4.39 \\ 
7136  &  150.106314  &  2.251556  &          23.2  &    19  &          11.1  &  2.00  &          1.850 	 &  12.87 	 & 25.08 	& 32.86 \\
7185  &  150.105835  &  2.251845  &          24.4  &     1  &           9.6  &  1.98  &          1.850  	&  15.1 	 & 9.89 	& 7.3 \\
7549  &  150.120110  &  2.255571  &          24.3  &    22  &          10.7  &  2.62  &          2.623	 &  8.78 	 & 16.11 	& 16.24 \\
7730  &  150.124647  &  2.257914  &          25.0  &     8  &           9.5  &  3.48  &          3.267  	&  6.09 	 & 3.91 	& 2.59 \\
\multicolumn{11}{c}{...} \\
\hline
\end{tabular}
\tablefoot{The full table with 149 rows is available at the CDS; the four high-redshift fillers are not included (see Appendix~\ref{sec:hiz}). (1) ID from the 3D-HST catalog for the COSMOS field (v4.1.5, \citealt{skelton14}); (2) and (3) coordinates measured from a mosaic of HST F160W data astrometrically registered with \textit{Gaia} stars (see Sect.~\ref{sec:parent_catalog}); (4) AB magnitude in F160W, from the 3D-HST catalog; (5) spectroscopic weight used for planning the MSA configuration (see Sect.~\ref{sec:weights}); (6) and (7) stellar mass and photometric redshift measured with a \texttt{Prospector} emulator (see Sect.~\ref{sec:parent_physical_prop}); (8) spectroscopic redshift measured from Blue Jay spectroscopy; (9), (10) and (11) median S/N per spectral pixel measured in each of the three gratings.}
\end{table*}

\begin{figure}[tbp]
\includegraphics[width=0.47\textwidth]{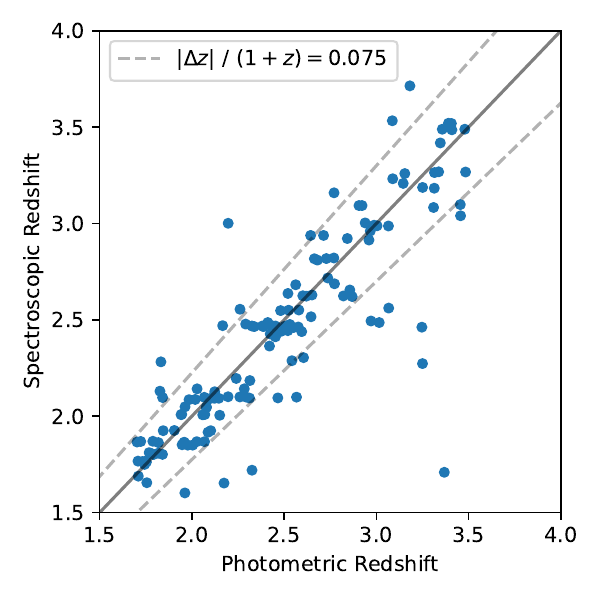}
\caption{Comparison of spectroscopic and photometric redshifts. The solid black line is the 1:1 relation, while the dashed lines mark the standard deviation of the discrepancy.
\label{fig:z_comparison}}
\end{figure}

\subsection{Spectroscopic redshifts}
\label{sec:specz}

Spectroscopic redshifts were measured for 139 out of 149 galaxies, corresponding to a success rate of 93\%. For the 10 failures we were unable to identify emission or absorption lines in the NIRSpec data, although a faint continuum is usually visible. Some of these spectra are affected by contamination due to stuck open shutters, which effectively reduces the wavelength range in which spectral features can be identified. In other cases, adding the NIRCam photometry (which was not available when the targets were selected) to the SED fit drastically changed the photometric redshift, revealing that four targets are likely at very low redshifts, $z\sim0.3$. In the end, only a handful of galaxies are probably in the correct redshift range and have emission lines that are simply too faint to be detected by NIRSpec.

Since the sample selection was based on the photometric redshifts, it is important to verify whether these redshifts are approximately correct for the majority of the sample. Figure~\ref{fig:z_comparison} shows that in most cases the photometric redshifts are in excellent agreement with the spectroscopic redshifts. The standard deviation of the discrepancy is $(z_\mathrm{phot} - z_\mathrm{spec})/(1+z_\mathrm{spec}) = 0.075$, and only four galaxies are outliers, that is, are discrepant beyond 3$\sigma$. Incidentally, the \texttt{Prospector} emulator performed substantially better than the photometric redshifts from the 3D-HST catalog, which have a standard deviation twice as large. The 3D-HST catalog includes four catastrophic outliers with $|\Delta z| \gtrsim 2$, which is likely similar to the number of catastrophic outliers for the \texttt{Prospector} emulator, if one includes the $z\sim0.3$ galaxies for which we could not secure a spectroscopic redshift.

\begin{figure}[tbp]
\includegraphics[width=0.47\textwidth]{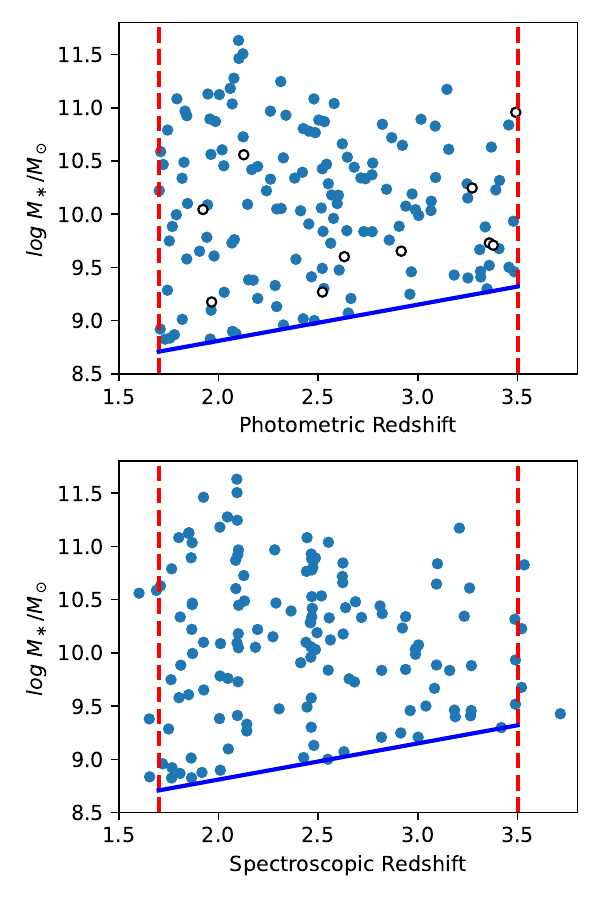}
\caption{Stellar mass versus redshift for the Blue Jay sample, using photometric redshifts (\textit{top}) and spectroscopic redshifts (\textit{bottom}). Empty circles mark galaxies without spectroscopic redshift. Lines showing the limits in mass and redshift for the sample selection are the same as those in Fig.~\ref{fig:weights}. The overdensities at $z\sim2.1$ and $z\sim2.45$ are visible in the spectroscopic redshift distribution.
\label{fig:sample}}
\end{figure}

\subsection{A representative sample}
\label{sec:representative_sample}

The top panel of Fig.~\ref{fig:sample} shows the Blue Jay sample in the stellar mass versus redshift plane used for the selection, and it confirms that our spectroscopic observations probe this parameter space in a roughly uniform manner. This was achieved by adopting appropriate spectroscopic weights during the NIRSpec MSA planning, otherwise the number of low-mass and low-redshift galaxies would have been overwhelmingly larger, following the distribution of the parent sample shown in Fig.~\ref{fig:weights}.

The 10 galaxies with no spectroscopic redshift measurement, shown as empty circles in the figure, tend to be at low masses, near the mass selection limit. Their faintness helps explain why these galaxies have undetected emission lines in the NIRSpec data, and also why in some cases their pre-JWST photometric redshifts were wrong.

The bottom panel of Fig.~\ref{fig:sample} shows the sample distribution on the same stellar mass versus redshift plane, but using spectroscopic instead of photometric redshifts. A few targets are now slightly outside the $1.7 < z < 3.5$ range, but this clearly does not substantially impact the overall redshift distribution of the sample. The most striking difference between the top and bottom panels, however, is the emergence of vertical patterns in the galaxy distribution when using the spectroscopic redshifts. These are overdense regions containing an excess of galaxies at the same redshift. Two overdensities are particularly noticeable: one at $z\sim2.1$ \citep{spitler12}, which was included in the North pointing by design in order to optimize the number of massive galaxies; and another at $z\sim2.45$, corresponding to the large-scale ``Hyperion'' protocluster \citep{cucciati18}.
While the presence of overdensities in Blue Jay makes it possible to study galaxy properties as a function of environment, it also represents a source of bias in the sample since galaxies living in dense environments are likely overrepresented. This bias must be taken into account when analyzing the number density of Blue Jay targets \citep[see, e.g.,][]{park24}. 
Nonetheless, this bias does not dominate the statistics of the sample, since the sum of the two overdensities accounts for only 30\% of the total number of galaxies in the Blue Jay sample and 40\% of the massive systems ($\log M_\ast / M_\odot > 10.5 $).
Ultimately, the number density of Blue Jay targets is not limited only by the selection bias due to the placement fo the pointing, but also by the substantial cosmic variance that affects small fields.

\begin{figure}[tbp]
\includegraphics[width=0.47\textwidth]{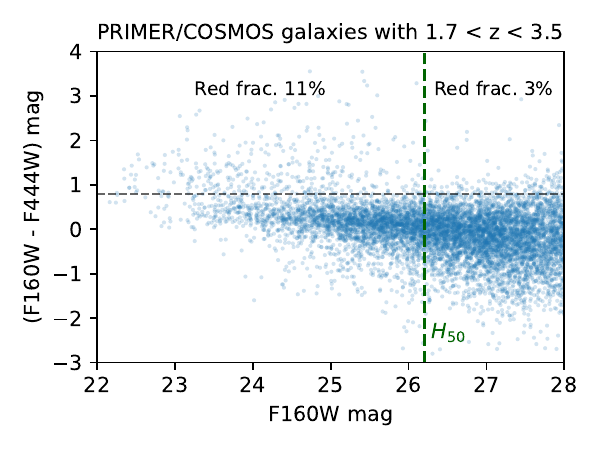}
\caption{Color-magnitude distribution of galaxies in the PRIMER/COSMOS catalog with $1.7 < z < 3.5$ from the \textsc{Astrodeep} catalog \citep{merlin24}. The dashed green line marks the completeness limit $H_{50}$ for the HST catalog. The fraction of red galaxies is reported for the two subsamples of sources that are brighter and fainter than $H_{50}$. Red galaxies are defined as having F160W-F444W$>0.8$, an empirical threshold that is shown as a horizontal dashed line.
\label{fig:color-mag}}
\end{figure}

Finally, we considered whether the selection of the parent sample based on the HST F160W filter, which is relatively blue, may introduce a bias against very red systems.
The launch of JWST enabled the discovery of a non-negligible population of sources that are bright in JWST/NIRCam imaging at 3-5~{\textmu}m, but were previously undetected in HST observations \citep[e.g.,][]{nelson23, rodighiero23, perezgonzalez23, barrufet23}. These so-called ``HST-dark galaxies'' are a mixed population, typically at $z>2$, which include dusty galaxies, often seen edge-on, massive quiescent systems, and emission-line dominated galaxies.
To investigate whether we are missing a large fraction of red galaxies from the parent sample, we made use of JWST/NIRCam observations in the COSMOS field by the PRIMER survey, which were not available when the Blue Jay sample selection was performed. Adopting the \textsc{Astrodeep} catalog by \citet{merlin24}, which includes all sources detected in F356W + F444W and is therefore not biased towards blue objects, we selected all galaxies with photometric redshift in the range $1.7 < z < 3.5$ and plot their F160W$-$F444W color as a function of their F160W magnitude in Fig.~\ref{fig:color-mag}. Some of the sources have colors that are much redder compared to those of the main population. There is a clear trend where the red sources (which we define here with an arbitrary and conservative threshold of F160W$-$F444W$>0.8$) represent a larger fraction of the population at the bright end, and nearly disappear at the low end. This is consistent with the physical nature of most of the red sources, which are either massive and quiescent or massive and dust-obscured galaxies. We draw two important conclusions from Fig.~\ref{fig:color-mag}:
\begin{enumerate}
    \item About $11\%$ of the galaxies in the parent sample (i.e., with F160W$ < H_{50}$) are red. These galaxies are, however, sufficiently bright to be detected by HST, and do not introduce a bias or incompleteness in our sample.
    \item On the other hand, red sources that are fainter than $H_{50}$ can potentially bias our sample, because they may be more massive than one would infer having access only to the HST data, and therefore might be above our mass limit $M_{min} \sim 10^9~M_\odot$ (see Sect.~\ref{sec:weights}), and yet not be part of the sample because of their faintness in HST imaging. However, at the faint end the fraction of red sources is low, only $3\%$ for galaxies with $H_{50}< \mathrm{F160W} < 28$, indicating that the vast majority of low-mass galaxies have blue colors.
\end{enumerate}
We conclude that the population of HST-dark galaxies does not represent a substantial source of bias for the Blue Jay sample, because at these redshifts deep HST observations are able to capture the bulk of the galaxy population irrespective of their color. This bias is likely much more prominent at higher redshifts.

Finally, we note that the Blue Jay sample naturally includes interesting objects such as active galactic nuclei (AGNs) and mergers, since no filtering was applied to the parent sample. Three galaxies in the sample are detected in the X-rays (ID 12020, 11337, 19572), two are broad-line AGNs (12020 and 18977, clearly identifiable in Fig.~\ref{fig:spectra3}), and several sources are involved in minor or major mergers. For a detailed study of the quiescent galaxy 12020 hosting a broad-line AGN detected in the X-rays, see \citet{ito25}.

\begin{figure*}[tbp]
\includegraphics[width=\textwidth]{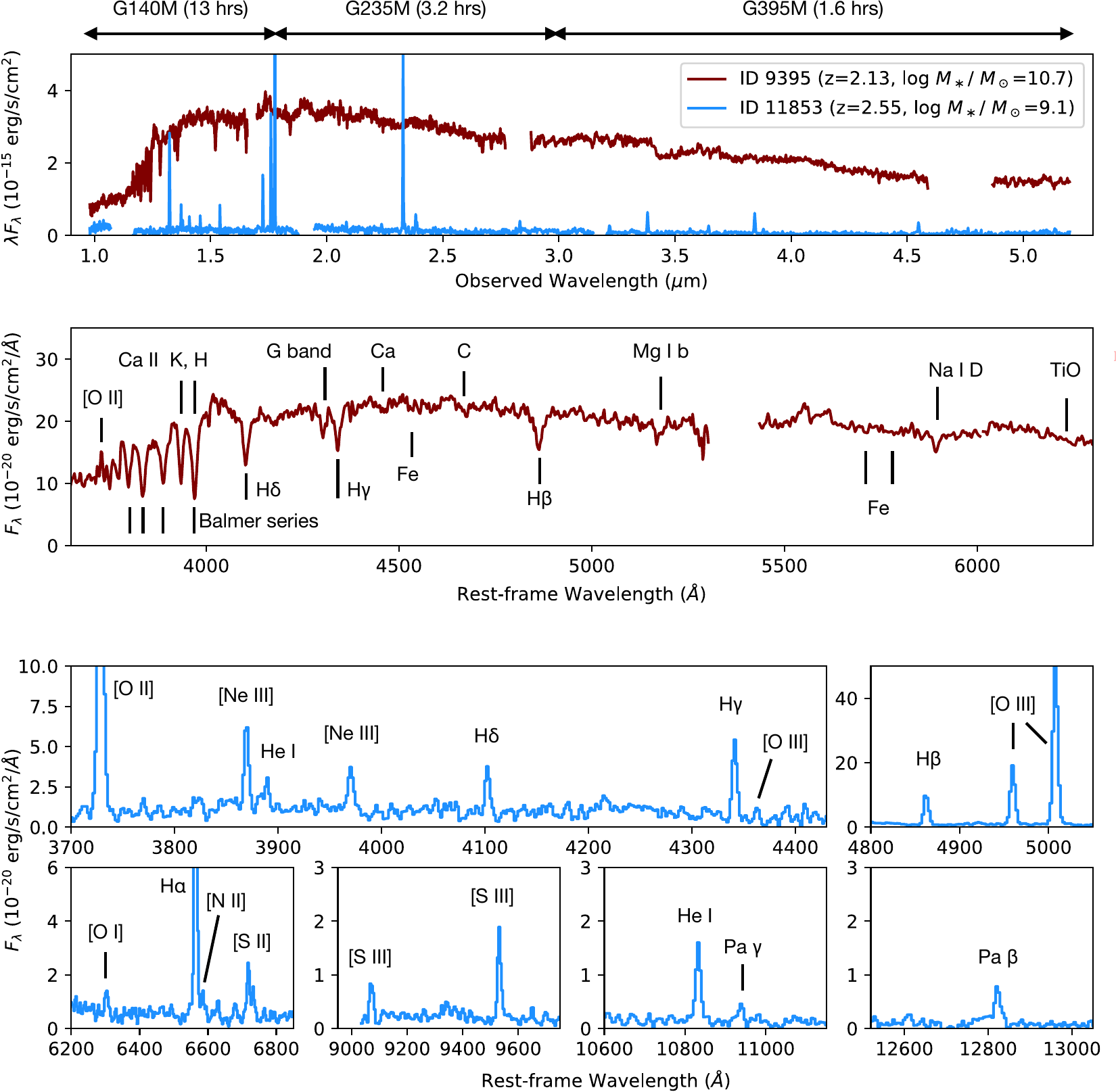}
\caption{Example spectra for a massive quiescent galaxy (ID 9395, in red) and a low-mass star-forming galaxy (ID 11853, in blue). \emph{Top}: Full spectrum (plotted as $\lambda F_\lambda$) for both galaxies covering the full range 1-5~{\textmu}m. The wavelength range and exposure time for each of the three medium-resolution gratings are shown above the panel. 
\emph{Middle}: Zoom-in of the quiescent galaxy spectrum in the rest-frame optical region observed with the G140M grating, showing numerous absorption lines.
\emph{Bottom panels}: Zoom-ins on several emission lines detected in the star-forming galaxy, ranging from the rest-frame UV to the rest-frame near-infrared.
\label{fig:spectra}}
\end{figure*}

\subsection{Overview of the spectroscopic data}
\label{sec:science}

The Blue Jay survey obtained deep rest-optical spectroscopy for a representative sample of galaxies at cosmic noon. The key strengths of this program are (1) the wavelength coverage and depth of the spectroscopic observations, which enable the detection of a large number of spectral features in both absorption and emission; and (2) the careful selection of the targets, which allows one to study several types of galaxies free from strong selection biases.

We show two examples of Blue Jay spectra in Fig.~\ref{fig:spectra}, one for a massive quiescent galaxy (in red) and one for a low-mass star-forming galaxy (in blue). The top panel shows the full wavelength range, obtained by combining the three medium-resolution gratings. The bluer grating, G140M, has a much deeper exposure and enables the study of numerous absorption lines for massive galaxies. These mostly originate from either hydrogen (Balmer series) or metals (Ca, Mg, Fe, and others) in the stellar photospheres, and can be used to measure the distribution of stellar ages and metal abundances. Moreover, the Na~I~D and Ca~II~H,K absorption lines are resonant (i.e., are transitions involving the ground level), and can therefore originate in cold gas in addition to the stellar photospheres. For this reason these lines can be used to probe cold neutral gas in or around galaxies.

The typical spectrum of star-forming galaxies, on the other hand, contains dozens of emission lines that can be used to study the physical properties of the interstellar medium. In the example spectrum shown in Fig.~\ref{fig:spectra} we detect many emission lines, from [O~II]$\lambda 3727$ in the rest-frame UV to Pa~$\beta$ in the rest-frame near-infrared, including the full rest-frame optical spectrum. The Pa~$\beta$ line is outside the NIRSpec range for galaxies at the high-redshift end of the Blue Jay sample, but coverage up to Pa~$\gamma$ is observed for every target in the $1.7 < z < 3.5$ range. Hydrogen lines in the Paschen and Balmer series can be used to derive the extinction curve and the star formation rate; while sets of lines corresponding to different ionization stages of the same element, such as [O~II] and [O~III] or [S~II] and [S~III], offer a direct probe of the ionization state of the gas. By combining metal and hydrogen lines it is also possible to measure metal abundances and detect the presence of shocks and/or photoionization by AGNs.

%%%%%%%%%%%%%%%%%%%%%%%%%%%%%%%%%%%%%%%%%%
\section{Science objectives} 
\label{sec:objectives}
%%%%%%%%%%%%%%%%%%%%%%%%%%%%%%%%%%%%%%%%%%

The Blue Jay sample is, by construction, ideal for the study of a wide range of galaxy properties at cosmic noon. In this section we list some of the key results that have already been obtained with this dataset, or that we are currently pursuing. We split them into three broad topics: the study of massive quiescent galaxies; the role of AGN-driven outflows in galaxy quenching; and the interplay between star formation and the interstellar medium.

\subsection{Physical properties of massive quiescent galaxies}

The science goal driving the design of Blue Jay is the detailed study of massive quiescent galaxies, with a focus on their stellar properties. This requires having access to several absorption lines, each detected with a high S/N, which was nearly impossible before the advent of JWST.

\subsubsection{Star formation histories}

Reconstructing the past history of massive quiescent galaxies is crucial for our understanding of their formation and evolution. Measuring the timescale over which star formation declines provides an important constraint on physical models of galaxy quenching, which are typically split into rapid processes, such as ejective feedback or merger-driven starburst, and slow processes, such as radio-mode AGN feedback or virial shock heating \citep[][and references therein]{man18}. At low redshifts ($z<1$) there is evidence for both slow and fast quenching \citep[e.g.][]{schawinski14, wu18, tacchella22}; however the population of galaxies that are observed after a rapid shutdown, dubbed ``post-starburst galaxies,'' is rare. Moving to higher redshifts, the fraction of post-starburst galaxies increases sharply \citep[e.g.,][]{wild16, belli19, park23}, suggesting that rapid quenching plays a prominent role in the formation of the first generation of quiescent galaxies. Analysis of the Blue Jay spectra for 14 massive quiescent galaxies by \citet{park24} has led to one of the most comprehensive studies of star formation histories at cosmic noon. The measured quenching timescales are mostly rapid, and in many cases a pre-quenching starburst phase can also be detected.
Moreover, the reconstructed star formation histories can be used to link the population of massive quiescent galaxies with their progenitors at higher redshifts. \citet{park24} find that two galaxies in the Blue Jay sample are likely descendants of the earliest quiescent galaxies that formed at $z>6$, which are one of the most puzzling discoveries made by JWST \citep{glazebrook24,carnall24,weibel25}.

\subsubsection{Stellar abundances}

Measuring the stellar chemical abundances of massive galaxies can shed light on their formation in a way that is highly complementary to the reconstruction of star formation histories. The metallicity locked up in stars can be used as a natural clock to time the formation epoch of a galaxy, since at earlier times, the interstellar medium (ISM) out of which stars were formed was less enriched. Moreover, different chemical elements trace different timescales: $\alpha$ elements (e.g., O, Mg, and Si) are mostly produced by core-collapse supernovae on short timescales, while Fe can also be produced on much longer timescales by type Ia supernovae \citep{nomoto13}. Thus, the [$\alpha$/Fe] abundance ratio is a key probe of how rapid the star formation process was \citep{thomas05}. So far, only a handful of studies have been able to detect metal absorption lines with sufficient S/N to measure stellar abundances beyond $z\sim2$ \citep[e.g.,][]{kriek16, jafa20, jafa25, beverage24}. These observations are generally showing lower metallicities and higher $\alpha$ enhancement for massive quiescent galaxies at $z>2$ compared to the local Universe, but with a large diversity within the high-redshift population. 
JWST is now offering the ability to obtain multi-element abundance measurements for increasingly larger samples \citep{beverage25}. Despite the availability of high-quality JWST data, these measurements are still challenging, partly because of systematic uncertainties in the models, such as the assumption of solar-scaled chemical abundance patterns. The development of new models, in which the $\alpha$ enhancement is treated self-consistently in both the spectral templates and the isochrone library \citep{park25}, will greatly improve our ability to derive robust stellar abundances.
Using these new models, we are planning to derive Fe and $\alpha$ abundances from the Blue Jay spectra, and explore how [$\alpha$/Fe] and star formation histories are related (Park et al., in prep.).

\subsubsection{Morphology and kinematics}

Before quenching, massive galaxies at cosmic noon typically grow inside out, thus forming a mostly quiescent central bulge \citep{tacchella15,nelson16}. However, it is not clear what happens during quenching, and whether the effective size must have decreased in order to match the remarkably small sizes measured for quiescent systems \citep[e.g.,][]{vanderwel14}. One possibility is that of a ``compaction'' phase involving gas-rich processes such as instabilities or interactions \citep{dekel14}. However, the kinematics of high-redshift quiescent galaxies do not necessarily require a violent process, since these systems appear to be mostly rotationally supported, similarly to the star-forming population \citep[e.g.,][]{newman18}. This may be consistent with alternative scenarios in which compaction is due to gentler processes, such as transient spiral arms in galaxy disks \citep{vanderwel25}.
The inability to match quiescent galaxies to their direct progenitors (the so-called progenitor bias) limits our understanding of how morphology and kinematics evolve through the quenching process.
The availability of high-quality spectra from the Blue Jay survey will allow us to explore these questions by studying galaxy morphology as a function of the star formation histories. We will study the compactness and central mass concentration, derived by JWST/NIRCam imaging, for quiescent galaxies with fast- versus~slow-quenching timescales (Brown et al. in prep.). Moreover, we will measure the stellar velocity dispersion from the absorption lines in order to investigate the galaxy kinematics. The NIRSpec shutters can only probe small regions of each galaxy, but by forward-modeling the mass distribution we will be able to obtain statistical constraints on the amount of rotational support and on the dynamical masses (Mendel et al., in prep.). These are particularly important to test the stellar masses of high-redshift quiescent galaxies \citep[e.g.,][]{slob25}, which in some studies appear to be too high compared to expectations from the Lambda cold dark matter model \citep{glazebrook24}.

\subsubsection{Environment}

It is likely that the environment plays a key role in the formation and quenching of massive galaxies, but the physical processes involved are still poorly understood. In the local Universe, galaxy quenching appears to independently correlate with stellar mass and local overdensity, thus suggesting two distinct quenching channels \citep{peng10}. However, the situation at high redshifts is less clear. Theoretical models make different predictions for the environment of early quenched galaxies \citep[e.g.,][]{kimmig25, weller25, delucia26}, which can be observationally tested. 
One advantage of the Blue Jay sample is that it is located in the popular COSMOS field, offering a wealth of ancillary data that can be used to estimate the overdensity using different methods, from galaxy counts \citep{fossati17, hatamnia25} to Lyman-$\alpha$ tomography, which probes the neutral intergalactic medium \citep{newman25}. 
We plan to use these public datasets to explore how the star formation histories measured for the Blue Jay galaxies vary with halo mass (Gao et al. in prep.), and also to investigate how the physical properties of quiescent galaxies depend on the local density as measured from both galaxy counts and Lyman-$\alpha$ tomography (Khoram et al. in prep.).

\subsection{AGN activity and outflows}

Studying the properties of quiescent galaxies can give us important insights on the physical processes responsible for quenching; however, a more direct approach consists in observing those processes in action. For the past two decades, theoretical models have favored AGN feedback as the main driver for galaxy quenching, and this has motivated a large observational effort directed toward the detection of AGN activity in galaxies that are becoming quiescent. Unfortunately, these observations have not been successful in establishing a clear link between AGN activity and quenching \citep{harrison17}, and the measured mass outflow rates have systematically been more than one order of magnitude smaller than what predicted by models of ejective feedback \citep[e.g.,][]{concas22}.

A key limitation of existing observations has been the inability to trace the different gas phases that make up galaxy outflows. High-redshift measurements are typically based on strong emission lines, probing the ionized phase, which is warmer, less dense and, thus, less important in the outflow mass budget. Colder phases are expected to carry the majority of the mass, but are substantially more difficult to observe. Beyond the local Universe, very deep integrations with the Atacama Large Millimeter/submillimeter Array (ALMA) are able to detect outflows in the molecular phase only for the most extreme objects such as quasars \citep{veilleux20}.

The high quality of the Blue Jay spectra allowed us to probe the neutral phase via the detection of the Na~D absorption line, for the first time in a representative sample of high-redshift galaxies. \citet{davies24} show that about half of the massive galaxies in Blue Jay have strong Na~D absorption due to neutral gas, and in many cases this line is blueshifted, indicating the widespread presence of neutral outflows. The wide wavelength range of our spectra also enables a test of this result using an independent tracer of neutral gas, the Ca~II H and K absorption lines, which broadly confirm the Na~D results, as shown by \citet{liboni25}. These neutral outflows are only present in massive galaxies and are likely caused by AGN activity, because the observed star formation rates are too low to explain the outflow velocities, energetics, and masses \citep{davies24}. It is noteworthy that neutral outflows are common also among quiescent galaxies, suggesting a link with quenching. Moreover, the analysis of the emission lines performed by \citet{bugiani25} reveals that quiescent galaxies present a remarkably high incidence of low-level luminosity AGNs.

When considering these results together with the star formation history measurements, a consistent picture is emerging: massive galaxies likely experience a rapid shutdown of star formation following rapid evacuation of their cold gas reservoir via powerful AGN-driven outflows. In at least one case, the massive galaxy 11142 at $z=2.45$ analyzed by \citet{belli24}, the Blue Jay spectrum clearly reveals all the elements of this physical picture: a rapid and recent quenching, a low-luminosity AGN, a weak ionized outflow probed by emission lines, and a powerful neutral outflow probed by resonant absorption lines. The measurement of these physical properties in the same galaxy, which happens to be observed immediately after a rapid shutdown of its star formation, represents a rare case in which the link between AGN feedback, outflows, and quenching can be directly established.
The relevance of neutral outflows for galaxy quenching has been independently confirmed by other JWST observations of recently quenched galaxies at cosmic noon and even higher redshifts \citep[e.g.][]{deugenio24, wu25, valentino25, sun25}.

\subsection{The interplay of gas, dust, and stars}

Star formation is tightly connected to the gas and dust cycle, and a comprehensive view of these processes requires measuring a variety of physical properties in the same galaxy sample. The combination of sample selection, wide spectral coverage, and deep exposures makes Blue Jay uniquely suited for linking the properties of stars, gas and dust in the galaxy population at cosmic noon.

\subsubsection{Star formation}

Star-forming galaxies lie on a scaling relation between stellar mass and star formation rate, the so-called main sequence \citep{noeske07}. This is one of the fundamental scaling relations, as it ties the growth rate of a galaxy with the total amount of mass formed in the past. Measuring the scatter of this relation can give us important information on the variability of individual galaxies and can constrain the burstiness, which is thought to dominate the star formation process at high redshifts \citep{tacchella16, simmonds25}.
Blue Jay offers the opportunity to measure the main sequence for a representative sample using several different indicators of star formation rate, which are sensitive to different timescales. We plan to compare the  star formation rates derived from Balmer emission lines (corrected for dust attenuation using the H$\alpha$/H$\beta$ ratio, as explained in Sect.~\ref{sec:dust}), which probe timescales of $\sim10$~Myr, with those derived from the star formation history, which can probe $\gtrsim100$~Myr, thus constraining the burstiness of galaxies on different timescales (Bugiani et al. in prep.). Moreover, using the star formation histories we can evolve each galaxy back in time and study its path on the main sequence relation; this offers a novel approach to the study of mass growth and quenching.

\subsubsection{ISM properties}

The wealth of emission lines detected in each Blue Jay spectrum can be used to constrain the physical properties of gas, such as the metallicity and the ionization source \citep{kewley19}.
The traditional approach to derive these measurements has been to use line ratios \citep[e.g.,][]{sanders24}; however, as the number of available emission lines increases, more advanced methods are required to make full use of the observations. Moreover, line ratios are typically calibrated on local H~II regions, which may not reflect the ionization and chemical environments found in star-forming galaxies at cosmic noon. We plan to use \texttt{Cue}, a fast and flexible model that can be used to fit the observed line ratios to derive gas properties \citep{li25}. \texttt{Cue} is based on neural networks and does not make strong assumptions on the ionizing source, which is approximated as a piecewise power law. By combining the emission line fluxes with the information on the stellar populations, we will be able to constrain the sources of the ionizing radiation and their relation to the galaxy physical properties (Li et al. in prep.)

\subsubsection{Dust attenuation}
\label{sec:dust}

The study of dust attenuation has important practical implications, and can also be used to derive physical insights on the physics, chemistry, and geometry of galaxies. One of the main limitations in this field is the poorly constrained dust attenuation curve, particularly beyond the local Universe \citep{salim20}. Using \texttt{Prospector} fits to the Blue Jay data, \citet{maheson25} study how the dust attenuation curve varies with galaxy properties, finding systematic trends with specific star formation rate, amount of dust attenuation, and geometry. Using the H$\alpha$/H$\beta$ flux ratio, which has a known and fixed intrinsic value, they also measure the nebular attenuation, finding that it correlates with the stellar attenuation. We plan to extend this method \citep[see, e.g.,][]{reddy20} to the full series of Balmer and Paschen lines detected in the Blue Jay spectra so that the dust attenuation curve can be derived for the nebular emission (Maheson et al., in prep.). This is a crucial ingredient when analyzing observed emission line fluxes with the goal of deriving physical properties such as star formation rate and gas metallicity.

%%%%%%%%%%%%%%%%%%%%%%%%%%%%%%%%%%%%%%%%%%
\section{Summary and discussion} 
\label{sec:summary}
%%%%%%%%%%%%%%%%%%%%%%%%%%%%%%%%%%%%%%%%%%

We presented the Blue Jay survey, a cycle 1 JWST program that obtained deep NIRSpec spectroscopy for 153 sources in the COSMOS field, with the goal of studying the physical properties of galaxies at cosmic noon. We briefly summarize the most important aspects of this survey below.
\begin{enumerate}
    \item The sample was carefully selected in order to obtain a roughly uniform coverage of the mass-redshift plane within the range $1.7 < z < 3.5$ and $M_\ast > M_\mathrm{min}(z)$, where the stellar mass limit $M_\mathrm{min}$ varies between $10^{8.7}~M_\odot$ and $10^{9.3}~M_\odot$, and this ensures that all types of galaxies at a given mass were included. The Blue Jay sample is representative of the cosmic noon population, without strong biases in any physical properties, and it includes both star-forming and quiescent galaxies. The sample also spans a range of environments, with about 30\% of the targets belonging to two known overdensities at $z=2.1$ and $z=2.45$.  
    \item We employed a nonstandard observing strategy with the NIRSpec MSA, where each target was placed on a two-shutter slitlet, and additional empty shutters were opened until all the available space on the detector was filled. We observed with an $A-B$ nodding pattern, which was used as a dither in order to avoid self-subtraction of large targets, however, background subtraction was performed via a master background spectrum, constructed from the empty shutters. We validated the master background spectrum via comparison to the JBT model and to the small-scale features of the zodiacal light.
    \item In order to perform a self-consistent analysis of the spectroscopic data together with HST and JWST broadband photometry, we extracted the photometry from the same box aperture that we used to extract the 1D spectrum from the NIRSpec observations. We publicly release the matched spectrophotometric dataset, together with photometry measured in a standard elliptical aperture.
    \item Each target was observed with the G140M, G235M, and G395M gratings for 13 hours, 3.2 hours, and 1.6 hours, respectively. The observations cover the full NIRSpec wavelength range, 1 to 5~{\textmu}m, corresponding to the entire rest-frame optical wavelength range. The $R\sim1000$ resolution allowed us to detect a wealth of emission lines, including the full range between [O~II]~$\lambda$3727 and Pa$\gamma$ for every observed galaxy. The lines outside this range were observed in many galaxies, according to their redshift. The deeper observations in G140M enabled us to detect Balmer and metal absorption lines in massive galaxies. We unambiguously identified the spectroscopic redshift for 93\% of the sample.
\end{enumerate}

We have reviewed several scientific results that have already been obtained with the Blue Jay spectra, with an emphasis on constraining the physical processes that cause galaxy quenching. The results include the discovery of widespread neutral outflows in massive galaxies, the measurement of rapid quenching timescales, and the detection of low-luminosity AGN activity in a large fraction of quiescent systems \citep{davies24, park24, bugiani25}. The emerging picture indicates a key role of AGN ejective feedback in galaxy quenching, with the cold-gas phase dominating the mass outflow rate \citep{belli24}. Further analysis of the Blue Jay sample is ongoing, including studies of the stellar abundances and kinematics, morphology, environment, dust attenuation, and properties of the interstellar medium. 

The Blue Jay sample is also ideal for building follow-up campaigns with JWST and other facilities to expand the set of physical properties that can be measured for the same galaxies. Scheduled ALMA observations will soon enable the detection of molecular gas reservoirs in some of the Blue Jay massive galaxies, while ground-based optical spectroscopy is being obtained to extend the study of absorption lines of interstellar and stellar origin into the rest-frame UV. Nearly the full sample has been reobserved with JWST/NIRSpec in cycle 3 using the high-resolution G235H grating, with the goal of spectrally resolving the Na~I~D and H$\alpha$ lines and obtaining a detailed measurement of the gas kinematics in the neutral and ionized phases. In cycle 5, three of the Blue Jay galaxies with the most powerful neutral outflows will be observed with integral field spectroscopy by NIRSpec in order to spatially map the multiphase outflows and constrain the geometry, which will enable us to measure substantially more robust mass-outflow rates.

%%%%%%%%%%%%%%%%%%%%%%%%%%%%%%%%%%%%%%%%%%

\begin{acknowledgements}
We acknowledge the anonymous referee for their useful comments, which made this article clearer and more complete.
SB acknowledges helpful discussions about the NIRSpec data reduction with Ryan Trainor, Gwen Rudie, and the CECILIA team, and also with Yang Sun.
This research was supported by the International Space Science Institute (ISSI) in Bern, through ISSI International Team project 24-602 "Multiphase Outflows in Galaxies at Cosmic Noon".
The Blue Jay Survey is funded in part by STScI Grant JWST-GO-01810.
SB, LB, AHK and CL are supported by ERC grant 101076080 ``Red Cardinal''. 
MP acknowledges support from the Kavli Institute for Cosmology, Cambridge and the Kavli Foundation.
RLD is supported by the Australian Research Council through the Discovery Early Career Researcher Award (DECRA) Fellowship DE240100136 funded by the Australian Government. 
RW acknowledges funding of a Leibniz Junior Research
Group (project number J131/2022).   
This work is based on observations made with the NASA/ESA/CSA \textit{James Webb} Space Telescope. The data were obtained from the Mikulski Archive for Space Telescopes at the Space Telescope Science Institute, which is operated by the Association of Universities for Research in Astronomy, Inc., under NASA contract NAS 5-03127 for JWST. These observations are associated with program GO 1810 and GO 1837. This work also makes use of observations taken by the 3D-HST Treasury Program (GO 12177 and 12328) with the NASA/ESA HST, which is operated by the Association of Universities for Research in Astronomy, Inc., under NASA contract NAS 5-26555.
\end{acknowledgements}

%%%%%%%%%%%%%%%%%%%%%%%%%%%%%%%%%%%%%%%%%%

\bibliographystyle{aa}
\bibliography{aa57679-25}

@software{jwst24,
  author       = {Bushouse, Howard and
                  Eisenhamer, Jonathan and
                  Dencheva, Nadia and
                  Davies, James and
                  Greenfield, Perry and
                  Morrison, Jane and
                  Hodge, Phil and
                  Simon, Bernie and
                  Grumm, David and
                  Droettboom, Michael and
                  Slavich, Edward and
                  Sosey, Megan and
                  Pauly, Tyler and
                  Miller, Todd and
                  Jedrzejewski, Robert and
                  Hack, Warren and
                  Davis, David and
                  Crawford, Steven and
                  Law, David and
                  Gordon, Karl and
                  Regan, Michael and
                  Cara, Mihai and
                  MacDonald, Ken and
                  Bradley, Larry and
                  Shanahan, Clare and
                  Jamieson, William and
                  Teodoro, Mairan and
                  Williams, Thomas and
                  Pena-Guerrero, Maria},
  title        = {JWST Calibration Pipeline},
  month        = jul,
  year         = 2024,
  publisher    = {Zenodo},
  version      = {1.15.1},
  doi          = {10.5281/zenodo.12692459},
  url          = {https://doi.org/10.5281/zenodo.12692459}
}

@ARTICLE{conroy24,
       author = {{Conroy}, Charlie and {Johnson}, Benjamin D. and {van Dokkum}, Pieter and {Deason}, Alis and {Tacchella}, Sandro and {Belli}, Sirio and {Bowman}, William P. and {Naidu}, Rohan P. and {Park}, Minjung and {Abraham}, Roberto and {Emami}, Razieh},
        title = "{Detection of Accretion Shelves Out to the Virial Radius of a Low-mass Galaxy with JWST}",
      journal = {\apj},
         year = 2024,
        month = jun,
       volume = {968},
       number = {2},
          eid = {129},
        pages = {129},
          doi = {10.3847/1538-4357/ad414a},
archivePrefix = {arXiv},
       eprint = {2310.13048},
 primaryClass = {astro-ph.GA},
       adsurl = {https://ui.adsabs.harvard.edu/abs/2024ApJ...968..129C}
}

@ARTICLE{brammer12,
       author = {{Brammer}, Gabriel B. and {van Dokkum}, Pieter G. and {Franx}, Marijn and {Fumagalli}, Mattia and {Patel}, Shannon and {Rix}, Hans-Walter and {Skelton}, Rosalind E. and {Kriek}, Mariska and {Nelson}, Erica and {Schmidt}, Kasper B. and {Bezanson}, Rachel and {da Cunha}, Elisabete and {Erb}, Dawn K. and {Fan}, Xiaohui and {F{\"o}rster Schreiber}, Natascha and {Illingworth}, Garth D. and {Labb{\'e}}, Ivo and {Leja}, Joel and {Lundgren}, Britt and {Magee}, Dan and {Marchesini}, Danilo and {McCarthy}, Patrick and {Momcheva}, Ivelina and {Muzzin}, Adam and {Quadri}, Ryan and {Steidel}, Charles C. and {Tal}, Tomer and {Wake}, David and {Whitaker}, Katherine E. and {Williams}, Anna},
        title = "{3D-HST: A Wide-field Grism Spectroscopic Survey with the Hubble Space Telescope}",
      journal = {\apjs},
         year = 2012,
        month = jun,
       volume = {200},
       number = {2},
          eid = {13},
        pages = {13},
          doi = {10.1088/0067-0049/200/2/13},
archivePrefix = {arXiv},
       eprint = {1204.2829},
 primaryClass = {astro-ph.CO},
       adsurl = {https://ui.adsabs.harvard.edu/abs/2012ApJS..200...13B}
}

@ARTICLE{momcheva16,
       author = {{Momcheva}, Ivelina G. and {Brammer}, Gabriel B. and {van Dokkum}, Pieter G. and {Skelton}, Rosalind E. and {Whitaker}, Katherine E. and {Nelson}, Erica J. and {Fumagalli}, Mattia and {Maseda}, Michael V. and {Leja}, Joel and {Franx}, Marijn and {Rix}, Hans-Walter and {Bezanson}, Rachel and {Da Cunha}, Elisabete and {Dickey}, Claire and {F{\"o}rster Schreiber}, Natascha M. and {Illingworth}, Garth and {Kriek}, Mariska and {Labb{\'e}}, Ivo and {Ulf Lange}, Johannes and {Lundgren}, Britt F. and {Magee}, Daniel and {Marchesini}, Danilo and {Oesch}, Pascal and {Pacifici}, Camilla and {Patel}, Shannon G. and {Price}, Sedona and {Tal}, Tomer and {Wake}, David A. and {van der Wel}, Arjen and {Wuyts}, Stijn},
        title = "{The 3D-HST Survey: Hubble Space Telescope WFC3/G141 Grism Spectra, Redshifts, and Emission Line Measurements for \raisebox{-0.5ex}\textasciitilde 100,000 Galaxies}",
      journal = {\apjs},
         year = 2016,
        month = aug,
       volume = {225},
       number = {2},
          eid = {27},
        pages = {27},
          doi = {10.3847/0067-0049/225/2/27},
archivePrefix = {arXiv},
       eprint = {1510.02106},
 primaryClass = {astro-ph.GA},
       adsurl = {https://ui.adsabs.harvard.edu/abs/2016ApJS..225...27M}
}

@ARTICLE{skelton14,
       author = {{Skelton}, Rosalind E. and {Whitaker}, Katherine E. and {Momcheva}, Ivelina G. and {Brammer}, Gabriel B. and {van Dokkum}, Pieter G. and {Labb{\'e}}, Ivo and {Franx}, Marijn and {van der Wel}, Arjen and {Bezanson}, Rachel and {Da Cunha}, Elisabete and {Fumagalli}, Mattia and {F{\"o}rster Schreiber}, Natascha and {Kriek}, Mariska and {Leja}, Joel and {Lundgren}, Britt F. and {Magee}, Daniel and {Marchesini}, Danilo and {Maseda}, Michael V. and {Nelson}, Erica J. and {Oesch}, Pascal and {Pacifici}, Camilla and {Patel}, Shannon G. and {Price}, Sedona and {Rix}, Hans-Walter and {Tal}, Tomer and {Wake}, David A. and {Wuyts}, Stijn},
        title = "{3D-HST WFC3-selected Photometric Catalogs in the Five CANDELS/3D-HST Fields: Photometry, Photometric Redshifts, and Stellar Masses}",
      journal = {\apjs},
         year = 2014,
        month = oct,
       volume = {214},
       number = {2},
          eid = {24},
        pages = {24},
          doi = {10.1088/0067-0049/214/2/24},
archivePrefix = {arXiv},
       eprint = {1403.3689},
 primaryClass = {astro-ph.GA},
       adsurl = {https://ui.adsabs.harvard.edu/abs/2014ApJS..214...24S}
}

@ARTICLE{belli19,
       author = {{Belli}, Sirio and {Newman}, Andrew B. and {Ellis}, Richard S.},
        title = "{MOSFIRE Spectroscopy of Quiescent Galaxies at 1.5 < z < 2.5. II. Star Formation Histories and Galaxy Quenching}",
      journal = {\apj},
         year = 2019,
        month = mar,
       volume = {874},
       number = {1},
          eid = {17},
        pages = {17},
          doi = {10.3847/1538-4357/ab07af},
archivePrefix = {arXiv},
       eprint = {1810.00008},
 primaryClass = {astro-ph.GA},
       adsurl = {https://ui.adsabs.harvard.edu/abs/2019ApJ...874...17B}
}

@ARTICLE{grogin11,
       author = {{Grogin}, Norman A. and {Kocevski}, Dale D. and {Faber}, S.~M. and {Ferguson}, Henry C. and {Koekemoer}, Anton M. and {Riess}, Adam G. and {Acquaviva}, Viviana and {Alexander}, David M. and {Almaini}, Omar and {Ashby}, Matthew L.~N. and {Barden}, Marco and {Bell}, Eric F. and {Bournaud}, Fr{\'e}d{\'e}ric and {Brown}, Thomas M. and {Caputi}, Karina I. and {Casertano}, Stefano and {Cassata}, Paolo and {Castellano}, Marco and {Challis}, Peter and {Chary}, Ranga-Ram and {Cheung}, Edmond and {Cirasuolo}, Michele and {Conselice}, Christopher J. and {Roshan Cooray}, Asantha and {Croton}, Darren J. and {Daddi}, Emanuele and {Dahlen}, Tomas and {Dav{\'e}}, Romeel and {de Mello}, Du{\'\i}lia F. and {Dekel}, Avishai and {Dickinson}, Mark and {Dolch}, Timothy and {Donley}, Jennifer L. and {Dunlop}, James S. and {Dutton}, Aaron A. and {Elbaz}, David and {Fazio}, Giovanni G. and {Filippenko}, Alexei V. and {Finkelstein}, Steven L. and {Fontana}, Adriano and {Gardner}, Jonathan P. and {Garnavich}, Peter M. and {Gawiser}, Eric and {Giavalisco}, Mauro and {Grazian}, Andrea and {Guo}, Yicheng and {Hathi}, Nimish P. and {H{\"a}ussler}, Boris and {Hopkins}, Philip F. and {Huang}, Jia-Sheng and {Huang}, Kuang-Han and {Jha}, Saurabh W. and {Kartaltepe}, Jeyhan S. and {Kirshner}, Robert P. and {Koo}, David C. and {Lai}, Kamson and {Lee}, Kyoung-Soo and {Li}, Weidong and {Lotz}, Jennifer M. and {Lucas}, Ray A. and {Madau}, Piero and {McCarthy}, Patrick J. and {McGrath}, Elizabeth J. and {McIntosh}, Daniel H. and {McLure}, Ross J. and {Mobasher}, Bahram and {Moustakas}, Leonidas A. and {Mozena}, Mark and {Nandra}, Kirpal and {Newman}, Jeffrey A. and {Niemi}, Sami-Matias and {Noeske}, Kai G. and {Papovich}, Casey J. and {Pentericci}, Laura and {Pope}, Alexandra and {Primack}, Joel R. and {Rajan}, Abhijith and {Ravindranath}, Swara and {Reddy}, Naveen A. and {Renzini}, Alvio and {Rix}, Hans-Walter and {Robaina}, Aday R. and {Rodney}, Steven A. and {Rosario}, David J. and {Rosati}, Piero and {Salimbeni}, Sara and {Scarlata}, Claudia and {Siana}, Brian and {Simard}, Luc and {Smidt}, Joseph and {Somerville}, Rachel S. and {Spinrad}, Hyron and {Straughn}, Amber N. and {Strolger}, Louis-Gregory and {Telford}, Olivia and {Teplitz}, Harry I. and {Trump}, Jonathan R. and {van der Wel}, Arjen and {Villforth}, Carolin and {Wechsler}, Risa H. and {Weiner}, Benjamin J. and {Wiklind}, Tommy and {Wild}, Vivienne and {Wilson}, Grant and {Wuyts}, Stijn and {Yan}, Hao-Jing and {Yun}, Min S.},
        title = "{CANDELS: The Cosmic Assembly Near-infrared Deep Extragalactic Legacy Survey}",
      journal = {\apjs},
         year = 2011,
        month = dec,
       volume = {197},
       number = {2},
          eid = {35},
        pages = {35},
          doi = {10.1088/0067-0049/197/2/35},
archivePrefix = {arXiv},
       eprint = {1105.3753},
 primaryClass = {astro-ph.CO},
       adsurl = {https://ui.adsabs.harvard.edu/abs/2011ApJS..197...35G}
}

@ARTICLE{koekemoer11,
       author = {{Koekemoer}, Anton M. and {Faber}, S.~M. and {Ferguson}, Henry C. and {Grogin}, Norman A. and {Kocevski}, Dale D. and {Koo}, David C. and {Lai}, Kamson and {Lotz}, Jennifer M. and {Lucas}, Ray A. and {McGrath}, Elizabeth J. and {Ogaz}, Sara and {Rajan}, Abhijith and {Riess}, Adam G. and {Rodney}, Steve A. and {Strolger}, Louis and {Casertano}, Stefano and {Castellano}, Marco and {Dahlen}, Tomas and {Dickinson}, Mark and {Dolch}, Timothy and {Fontana}, Adriano and {Giavalisco}, Mauro and {Grazian}, Andrea and {Guo}, Yicheng and {Hathi}, Nimish P. and {Huang}, Kuang-Han and {van der Wel}, Arjen and {Yan}, Hao-Jing and {Acquaviva}, Viviana and {Alexander}, David M. and {Almaini}, Omar and {Ashby}, Matthew L.~N. and {Barden}, Marco and {Bell}, Eric F. and {Bournaud}, Fr{\'e}d{\'e}ric and {Brown}, Thomas M. and {Caputi}, Karina I. and {Cassata}, Paolo and {Challis}, Peter J. and {Chary}, Ranga-Ram and {Cheung}, Edmond and {Cirasuolo}, Michele and {Conselice}, Christopher J. and {Roshan Cooray}, Asantha and {Croton}, Darren J. and {Daddi}, Emanuele and {Dav{\'e}}, Romeel and {de Mello}, Duilia F. and {de Ravel}, Loic and {Dekel}, Avishai and {Donley}, Jennifer L. and {Dunlop}, James S. and {Dutton}, Aaron A. and {Elbaz}, David and {Fazio}, Giovanni G. and {Filippenko}, Alexei V. and {Finkelstein}, Steven L. and {Frazer}, Chris and {Gardner}, Jonathan P. and {Garnavich}, Peter M. and {Gawiser}, Eric and {Gruetzbauch}, Ruth and {Hartley}, Will G. and {H{\"a}ussler}, Boris and {Herrington}, Jessica and {Hopkins}, Philip F. and {Huang}, Jia-Sheng and {Jha}, Saurabh W. and {Johnson}, Andrew and {Kartaltepe}, Jeyhan S. and {Khostovan}, Ali A. and {Kirshner}, Robert P. and {Lani}, Caterina and {Lee}, Kyoung-Soo and {Li}, Weidong and {Madau}, Piero and {McCarthy}, Patrick J. and {McIntosh}, Daniel H. and {McLure}, Ross J. and {McPartland}, Conor and {Mobasher}, Bahram and {Moreira}, Heidi and {Mortlock}, Alice and {Moustakas}, Leonidas A. and {Mozena}, Mark and {Nandra}, Kirpal and {Newman}, Jeffrey A. and {Nielsen}, Jennifer L. and {Niemi}, Sami and {Noeske}, Kai G. and {Papovich}, Casey J. and {Pentericci}, Laura and {Pope}, Alexandra and {Primack}, Joel R. and {Ravindranath}, Swara and {Reddy}, Naveen A. and {Renzini}, Alvio and {Rix}, Hans-Walter and {Robaina}, Aday R. and {Rosario}, David J. and {Rosati}, Piero and {Salimbeni}, Sara and {Scarlata}, Claudia and {Siana}, Brian and {Simard}, Luc and {Smidt}, Joseph and {Snyder}, Diana and {Somerville}, Rachel S. and {Spinrad}, Hyron and {Straughn}, Amber N. and {Telford}, Olivia and {Teplitz}, Harry I. and {Trump}, Jonathan R. and {Vargas}, Carlos and {Villforth}, Carolin and {Wagner}, Cory R. and {Wandro}, Pat and {Wechsler}, Risa H. and {Weiner}, Benjamin J. and {Wiklind}, Tommy and {Wild}, Vivienne and {Wilson}, Grant and {Wuyts}, Stijn and {Yun}, Min S.},
        title = "{CANDELS: The Cosmic Assembly Near-infrared Deep Extragalactic Legacy Survey{\textemdash}The Hubble Space Telescope Observations, Imaging Data Products, and Mosaics}",
      journal = {\apjs},
         year = 2011,
        month = dec,
       volume = {197},
       number = {2},
          eid = {36},
        pages = {36},
          doi = {10.1088/0067-0049/197/2/36},
archivePrefix = {arXiv},
       eprint = {1105.3754},
 primaryClass = {astro-ph.CO},
       adsurl = {https://ui.adsabs.harvard.edu/abs/2011ApJS..197...36K}
}

@ARTICLE{tal14,
       author = {{Tal}, Tomer and {Dekel}, Avishai and {Oesch}, Pascal and {Muzzin}, Adam and {Brammer}, Gabriel B. and {van Dokkum}, Pieter G. and {Franx}, Marijn and {Illingworth}, Garth D. and {Leja}, Joel and {Magee}, Daniel and {Marchesini}, Danilo and {Momcheva}, Ivelina and {Nelson}, Erica J. and {Patel}, Shannon G. and {Quadri}, Ryan F. and {Rix}, Hans-Walter and {Skelton}, Rosalind E. and {Wake}, David A. and {Whitaker}, Katherine E.},
        title = "{Observations of Environmental Quenching in Groups in the 11 GYR since z = 2.5: Different Quenching for Central and Satellite Galaxies}",
      journal = {\apj},
         year = 2014,
        month = jul,
       volume = {789},
       number = {2},
          eid = {164},
        pages = {164},
          doi = {10.1088/0004-637X/789/2/164},
archivePrefix = {arXiv},
       eprint = {1401.2984},
 primaryClass = {astro-ph.CO},
       adsurl = {https://ui.adsabs.harvard.edu/abs/2014ApJ...789..164T}
}

@ARTICLE{leja19sed,
       author = {{Leja}, Joel and {Johnson}, Benjamin D. and {Conroy}, Charlie and {van Dokkum}, Pieter and {Speagle}, Joshua S. and {Brammer}, Gabriel and {Momcheva}, Ivelina and {Skelton}, Rosalind and {Whitaker}, Katherine E. and {Franx}, Marijn and {Nelson}, Erica J.},
        title = "{An Older, More Quiescent Universe from Panchromatic SED Fitting of the 3D-HST Survey}",
      journal = {\apj},
         year = 2019,
        month = jun,
       volume = {877},
       number = {2},
          eid = {140},
        pages = {140},
          doi = {10.3847/1538-4357/ab1d5a},
archivePrefix = {arXiv},
       eprint = {1812.05608},
 primaryClass = {astro-ph.GA},
       adsurl = {https://ui.adsabs.harvard.edu/abs/2019ApJ...877..140L}
}

@ARTICLE{pozzetti10,
       author = {{Pozzetti}, L. and {Bolzonella}, M. and {Zucca}, E. and {Zamorani}, G. and {Lilly}, S. and {Renzini}, A. and {Moresco}, M. and {Mignoli}, M. and {Cassata}, P. and {Tasca}, L. and {Lamareille}, F. and {Maier}, C. and {Meneux}, B. and {Halliday}, C. and {Oesch}, P. and {Vergani}, D. and {Caputi}, K. and {Kova{\v{c}}}, K. and {Cimatti}, A. and {Cucciati}, O. and {Iovino}, A. and {Peng}, Y. and {Carollo}, M. and {Contini}, T. and {Kneib}, J. -P. and {Le F{\'e}vre}, O. and {Mainieri}, V. and {Scodeggio}, M. and {Bardelli}, S. and {Bongiorno}, A. and {Coppa}, G. and {de la Torre}, S. and {de Ravel}, L. and {Franzetti}, P. and {Garilli}, B. and {Kampczyk}, P. and {Knobel}, C. and {Le Borgne}, J. -F. and {Le Brun}, V. and {Pell{\`o}}, R. and {Perez Montero}, E. and {Ricciardelli}, E. and {Silverman}, J.~D. and {Tanaka}, M. and {Tresse}, L. and {Abbas}, U. and {Bottini}, D. and {Cappi}, A. and {Guzzo}, L. and {Koekemoer}, A.~M. and {Leauthaud}, A. and {Maccagni}, D. and {Marinoni}, C. and {McCracken}, H.~J. and {Memeo}, P. and {Porciani}, C. and {Scaramella}, R. and {Scarlata}, C. and {Scoville}, N.},
        title = "{zCOSMOS - 10k-bright spectroscopic sample. The bimodality in the galaxy stellar mass function: exploring its evolution with redshift}",
      journal = {\aap},
         year = 2010,
        month = nov,
       volume = {523},
          eid = {A13},
        pages = {A13},
          doi = {10.1051/0004-6361/200913020},
archivePrefix = {arXiv},
       eprint = {0907.5416},
 primaryClass = {astro-ph.CO},
       adsurl = {https://ui.adsabs.harvard.edu/abs/2010A&A...523A..13P}
}

@ARTICLE{spitler12,
       author = {{Spitler}, Lee R. and {Labb{\'e}}, Ivo and {Glazebrook}, Karl and {Persson}, S. Eric and {Monson}, Andy and {Papovich}, Casey and {Tran}, Kim-Vy H. and {Poole}, Gregory B. and {Quadri}, Ryan and {van Dokkum}, Pieter and {Kelson}, Daniel D. and {Kacprzak}, Glenn G. and {McCarthy}, Patrick J. and {Murphy}, David and {Straatman}, Caroline M.~S. and {Tilvi}, Vithal},
        title = "{First Results from Z -FOURGE: Discovery of a Candidate Cluster at z = 2.2 in COSMOS}",
      journal = {\apjl},
         year = 2012,
        month = apr,
       volume = {748},
       number = {2},
          eid = {L21},
        pages = {L21},
          doi = {10.1088/2041-8205/748/2/L21},
archivePrefix = {arXiv},
       eprint = {1112.2691},
 primaryClass = {astro-ph.CO},
       adsurl = {https://ui.adsabs.harvard.edu/abs/2012ApJ...748L..21S}
}

@ARTICLE{mowla22,
       author = {{Mowla}, Lamiya A. and {Cutler}, Sam E. and {Brammer}, Gabriel B. and {Momcheva}, Ivelina G. and {Whitaker}, Katherine E. and {van Dokkum}, Pieter G. and {Bezanson}, Rachel S. and {F{\"o}rster Schreiber}, Natascha M. and {Franx}, Marijn and {Iyer}, Kartheik G. and {Marchesini}, Danilo and {Muzzin}, Adam and {Nelson}, Erica J. and {Skelton}, Rosalind E. and {Snyder}, Gregory F. and {Wake}, David A. and {Wuyts}, Stijn and {van der Wel}, Arjen},
        title = "{3D-DASH: The Widest Near-infrared Hubble Space Telescope Survey}",
      journal = {\apj},
         year = 2022,
        month = jul,
       volume = {933},
       number = {2},
          eid = {129},
        pages = {129},
          doi = {10.3847/1538-4357/ac71af},
archivePrefix = {arXiv},
       eprint = {2206.01156},
 primaryClass = {astro-ph.GA},
       adsurl = {https://ui.adsabs.harvard.edu/abs/2022ApJ...933..129M}
}

@ARTICLE{gaia23,
       author = {{Gaia Collaboration} and {Vallenari}, A. and {Brown}, A.~G.~A. and {Prusti}, T. and {de Bruijne}, J.~H.~J. and {Arenou}, F. and {Babusiaux}, C. and {Biermann}, M. and {Creevey}, O.~L. and {Ducourant}, C. and {Evans}, D.~W. and {Eyer}, L. and {Guerra}, R. and {Hutton}, A. and {Jordi}, C. and {Klioner}, S.~A. and {Lammers}, U.~L. and {Lindegren}, L. and {Luri}, X. and {Mignard}, F. and {Panem}, C. and {Pourbaix}, D. and {Randich}, S. and {Sartoretti}, P. and {Soubiran}, C. and {Tanga}, P. and {Walton}, N.~A. and {Bailer-Jones}, C.~A.~L. and {Bastian}, U. and {Drimmel}, R. and {Jansen}, F. and {Katz}, D. and {Lattanzi}, M.~G. and {van Leeuwen}, F. and {Bakker}, J. and {Cacciari}, C. and {Casta{\~n}eda}, J. and {De Angeli}, F. and {Fabricius}, C. and {Fouesneau}, M. and {Fr{\'e}mat}, Y. and {Galluccio}, L. and {Guerrier}, A. and {Heiter}, U. and {Masana}, E. and {Messineo}, R. and {Mowlavi}, N. and {Nicolas}, C. and {Nienartowicz}, K. and {Pailler}, F. and {Panuzzo}, P. and {Riclet}, F. and {Roux}, W. and {Seabroke}, G.~M. and {Sordo}, R. and {Th{\'e}venin}, F. and {Gracia-Abril}, G. and {Portell}, J. and {Teyssier}, D. and {Altmann}, M. and {Andrae}, R. and {Audard}, M. and {Bellas-Velidis}, I. and {Benson}, K. and {Berthier}, J. and {Blomme}, R. and {Burgess}, P.~W. and {Busonero}, D. and {Busso}, G. and {C{\'a}novas}, H. and {Carry}, B. and {Cellino}, A. and {Cheek}, N. and {Clementini}, G. and {Damerdji}, Y. and {Davidson}, M. and {de Teodoro}, P. and {Nu{\~n}ez Campos}, M. and {Delchambre}, L. and {Dell'Oro}, A. and {Esquej}, P. and {Fern{\'a}ndez-Hern{\'a}ndez}, J. and {Fraile}, E. and {Garabato}, D. and {Garc{\'\i}a-Lario}, P. and {Gosset}, E. and {Haigron}, R. and {Halbwachs}, J. -L. and {Hambly}, N.~C. and {Harrison}, D.~L. and {Hern{\'a}ndez}, J. and {Hestroffer}, D. and {Hodgkin}, S.~T. and {Holl}, B. and {Jan{\ss}en}, K. and {Jevardat de Fombelle}, G. and {Jordan}, S. and {Krone-Martins}, A. and {Lanzafame}, A.~C. and {L{\"o}ffler}, W. and {Marchal}, O. and {Marrese}, P.~M. and {Moitinho}, A. and {Muinonen}, K. and {Osborne}, P. and {Pancino}, E. and {Pauwels}, T. and {Recio-Blanco}, A. and {Reyl{\'e}}, C. and {Riello}, M. and {Rimoldini}, L. and {Roegiers}, T. and {Rybizki}, J. and {Sarro}, L.~M. and {Siopis}, C. and {Smith}, M. and {Sozzetti}, A. and {Utrilla}, E. and {van Leeuwen}, M. and {Abbas}, U. and {{\'A}brah{\'a}m}, P. and {Abreu Aramburu}, A. and {Aerts}, C. and {Aguado}, J.~J. and {Ajaj}, M. and {Aldea-Montero}, F. and {Altavilla}, G. and {{\'A}lvarez}, M.~A. and {Alves}, J. and {Anders}, F. and {Anderson}, R.~I. and {Anglada Varela}, E. and {Antoja}, T. and {Baines}, D. and {Baker}, S.~G. and {Balaguer-N{\'u}{\~n}ez}, L. and {Balbinot}, E. and {Balog}, Z. and {Barache}, C. and {Barbato}, D. and {Barros}, M. and {Barstow}, M.~A. and {Bartolom{\'e}}, S. and {Bassilana}, J. -L. and {Bauchet}, N. and {Becciani}, U. and {Bellazzini}, M. and {Berihuete}, A. and {Bernet}, M. and {Bertone}, S. and {Bianchi}, L. and {Binnenfeld}, A. and {Blanco-Cuaresma}, S. and {Blazere}, A. and {Boch}, T. and {Bombrun}, A. and {Bossini}, D. and {Bouquillon}, S. and {Bragaglia}, A. and {Bramante}, L. and {Breedt}, E. and {Bressan}, A. and {Brouillet}, N. and {Brugaletta}, E. and {Bucciarelli}, B. and {Burlacu}, A. and {Butkevich}, A.~G. and {Buzzi}, R. and {Caffau}, E. and {Cancelliere}, R. and {Cantat-Gaudin}, T. and {Carballo}, R. and {Carlucci}, T. and {Carnerero}, M.~I. and {Carrasco}, J.~M. and {Casamiquela}, L. and {Castellani}, M. and {Castro-Ginard}, A. and {Chaoul}, L. and {Charlot}, P. and {Chemin}, L. and {Chiaramida}, V. and {Chiavassa}, A. and {Chornay}, N. and {Comoretto}, G. and {Contursi}, G. and {Cooper}, W.~J. and {Cornez}, T. and {Cowell}, S. and {Crifo}, F. and {Cropper}, M. and {Crosta}, M. and {Crowley}, C. and {Dafonte}, C. and {Dapergolas}, A. and {David}, M. and {David}, P. and {de Laverny}, P. and {De Luise}, F. and {De March}, R.},
        title = "{Gaia Data Release 3. Summary of the content and survey properties}",
      journal = {\aap},
         year = 2023,
        month = jun,
       volume = {674},
          eid = {A1},
        pages = {A1},
          doi = {10.1051/0004-6361/202243940},
archivePrefix = {arXiv},
       eprint = {2208.00211},
 primaryClass = {astro-ph.GA},
       adsurl = {https://ui.adsabs.harvard.edu/abs/2023A&A...674A...1G}
}

@ARTICLE{brammer08,
       author = {{Brammer}, Gabriel B. and {van Dokkum}, Pieter G. and {Coppi}, Paolo},
        title = "{EAZY: A Fast, Public Photometric Redshift Code}",
      journal = {\apj},
         year = 2008,
        month = oct,
       volume = {686},
       number = {2},
        pages = {1503-1513},
          doi = {10.1086/591786},
archivePrefix = {arXiv},
       eprint = {0807.1533},
 primaryClass = {astro-ph},
       adsurl = {https://ui.adsabs.harvard.edu/abs/2008ApJ...686.1503B}
}

@ARTICLE{kriek09,
       author = {{Kriek}, Mariska and {van Dokkum}, Pieter G. and {Labb{\'e}}, Ivo and {Franx}, Marijn and {Illingworth}, Garth D. and {Marchesini}, Danilo and {Quadri}, Ryan F.},
        title = "{An Ultra-Deep Near-Infrared Spectrum of a Compact Quiescent Galaxy at z = 2.2}",
      journal = {\apj},
         year = 2009,
        month = jul,
       volume = {700},
       number = {1},
        pages = {221-231},
          doi = {10.1088/0004-637X/700/1/221},
archivePrefix = {arXiv},
       eprint = {0905.1692},
 primaryClass = {astro-ph.CO},
       adsurl = {https://ui.adsabs.harvard.edu/abs/2009ApJ...700..221K}
}

@ARTICLE{mathews23,
       author = {{Mathews}, Elijah P. and {Leja}, Joel and {Speagle}, Joshua S. and {Johnson}, Benjamin D. and {Gibson}, Justus and {Nelson}, Erica J. and {Suess}, Katherine A. and {Tacchella}, Sandro and {Whitaker}, Katherine E. and {Wang}, Bingjie},
        title = "{As Simple as Possible but No Simpler: Optimizing the Performance of Neural Net Emulators for Galaxy SED Fitting}",
      journal = {\apj},
         year = 2023,
        month = sep,
       volume = {954},
       number = {2},
          eid = {132},
        pages = {132},
          doi = {10.3847/1538-4357/ace720},
archivePrefix = {arXiv},
       eprint = {2306.16442},
 primaryClass = {astro-ph.IM},
       adsurl = {https://ui.adsabs.harvard.edu/abs/2023ApJ...954..132M}
}

@article{johnson21,
   title={Stellar Population Inference with Prospector},
   volume={254},
   ISSN={1538-4365},
   url={http://dx.doi.org/10.3847/1538-4365/abef67},
   DOI={10.3847/1538-4365/abef67},
   number={2},
   journal={The Astrophysical Journal Supplement Series},
   publisher={American Astronomical Society},
   author={Johnson, Benjamin D. and Leja, Joel and Conroy, Charlie and Speagle, Joshua S.},
   year={2021},
   month=may, pages={22} }

@ARTICLE{conroy10,
       author = {{Conroy}, Charlie and {Gunn}, James E.},
        title = "{The Propagation of Uncertainties in Stellar Population Synthesis Modeling. III. Model Calibration, Comparison, and Evaluation}",
      journal = {\apj},
         year = 2010,
        month = apr,
       volume = {712},
       number = {2},
        pages = {833-857},
          doi = {10.1088/0004-637X/712/2/833},
archivePrefix = {arXiv},
       eprint = {0911.3151},
 primaryClass = {astro-ph.CO},
       adsurl = {https://ui.adsabs.harvard.edu/abs/2010ApJ...712..833C}
}

@ARTICLE{conroy09,
       author = {{Conroy}, Charlie and {Gunn}, James E. and {White}, Martin},
        title = "{The Propagation of Uncertainties in Stellar Population Synthesis Modeling. I. The Relevance of Uncertain Aspects of Stellar Evolution and the Initial Mass Function to the Derived Physical Properties of Galaxies}",
      journal = {\apj},
         year = 2009,
        month = jul,
       volume = {699},
       number = {1},
        pages = {486-506},
          doi = {10.1088/0004-637X/699/1/486},
archivePrefix = {arXiv},
       eprint = {0809.4261},
 primaryClass = {astro-ph},
       adsurl = {https://ui.adsabs.harvard.edu/abs/2009ApJ...699..486C}
}

@ARTICLE{leja17,
       author = {{Leja}, Joel and {Johnson}, Benjamin D. and {Conroy}, Charlie and {van Dokkum}, Pieter G. and {Byler}, Nell},
        title = "{Deriving Physical Properties from Broadband Photometry with Prospector: Description of the Model and a Demonstration of its Accuracy Using 129 Galaxies in the Local Universe}",
      journal = {\apj},
         year = 2017,
        month = mar,
       volume = {837},
       number = {2},
          eid = {170},
        pages = {170},
          doi = {10.3847/1538-4357/aa5ffe},
archivePrefix = {arXiv},
       eprint = {1609.09073},
 primaryClass = {astro-ph.GA},
       adsurl = {https://ui.adsabs.harvard.edu/abs/2017ApJ...837..170L}
}

@ARTICLE{maheson25,
       author = {{Maheson}, Gabriel and {Tacchella}, Sandro and {Belli}, Sirio and {Park}, Minjung and {Danhaive}, A. Lola and {Bugiani}, Letizia and {Davies}, Rebecca and {Emami}, Razieh and {Khoram}, Amir H. and {Lam}, Laurence and {Leja}, Joel and {Mendel}, Trevor and {Nelson}, Erica June},
        title = "{Big, Dusty Galaxies in Blue Jay: Insights into the Relationship Between Morphology and Dust Attenuation at Cosmic Noon}",
      journal = {MNRAS, submitted},
         year = 2025,
        month = apr,
          eid = {arXiv:2504.15346},
        pages = {arXiv:2504.15346},
          doi = {10.48550/arXiv.2504.15346},
archivePrefix = {arXiv},
       eprint = {2504.15346},
 primaryClass = {astro-ph.GA},
       adsurl = {https://ui.adsabs.harvard.edu/abs/2025arXiv250415346M}
}

@ARTICLE{liboni25,
       author = {{Liboni}, Caterina and {Belli}, Sirio and {Bugiani}, Letizia and {Davies}, Rebecca and {Park}, Minjung and {Conroy}, Charlie and {Emami}, Razieh and {Johnson}, Benjamin D. and {Khoram}, Amir H. and {Leja}, Joel and {Maheson}, Gabriel and {Sapori}, Matteo and {Mendel}, Trevor and {Tacchella}, Sandro and {Weinberger}, Rainer},
        title = "{Probing neutral outflows in z {\ensuremath{\sim}} 2 galaxies using JWST observations of Ca II H and K absorption lines}",
      journal = {\aap},
         year = 2026,
        month = jan,
       volume = {705},
          eid = {A125},
        pages = {A125},
          doi = {10.1051/0004-6361/202555845},
archivePrefix = {arXiv},
       eprint = {2506.05470},
 primaryClass = {astro-ph.GA},
       adsurl = {https://ui.adsabs.harvard.edu/abs/2026A&A...705A.125L}
}

@ARTICLE{bugiani25,
       author = {{Bugiani}, Letizia and {Belli}, Sirio and {Park}, Minjung and {Davies}, Rebecca L. and {Mendel}, J. Trevor and {Johnson}, Benjamin D. and {Khoram}, Amir H. and {Benton}, Chlo{\"e} and {Cimatti}, Andrea and {Conroy}, Charlie and {Emami}, Razieh and {Leja}, Joel and {Li}, Yijia and {Maheson}, Gabriel and {Mathews}, Elijah P. and {Naidu}, Rohan P. and {Nelson}, Erica J. and {Tacchella}, Sandro and {Terrazas}, Bryan A. and {Weinberger}, Rainer},
        title = "{Active Galactic Nucleus Feedback in Quiescent Galaxies at Cosmic Noon Traced by Ionized Gas Emission}",
      journal = {\apj},
         year = 2025,
        month = mar,
       volume = {981},
       number = {1},
          eid = {25},
        pages = {25},
          doi = {10.3847/1538-4357/adaeaf},
archivePrefix = {arXiv},
       eprint = {2406.08547},
 primaryClass = {astro-ph.GA},
       adsurl = {https://ui.adsabs.harvard.edu/abs/2025ApJ...981...25B}
}

@ARTICLE{park24,
       author = {{Park}, Minjung and {Belli}, Sirio and {Conroy}, Charlie and {Johnson}, Benjamin D. and {Davies}, Rebecca L. and {Leja}, Joel and {Tacchella}, Sandro and {Mendel}, J. Trevor and {Benton}, Chlo{\"e} and {Bugiani}, Letizia and {Emami}, Razieh and {Khoram}, Amir H. and {Li}, Yijia and {Maheson}, Gabriel and {Mathews}, Elijah P. and {Naidu}, Rohan P. and {Nelson}, Erica J. and {Terrazas}, Bryan A. and {Weinberger}, Rainer},
        title = "{Widespread Rapid Quenching at Cosmic Noon Revealed by JWST Deep Spectroscopy}",
      journal = {\apj},
         year = 2024,
        month = nov,
       volume = {976},
       number = {1},
          eid = {72},
        pages = {72},
          doi = {10.3847/1538-4357/ad7e15},
archivePrefix = {arXiv},
       eprint = {2404.17945},
 primaryClass = {astro-ph.GA},
       adsurl = {https://ui.adsabs.harvard.edu/abs/2024ApJ...976...72P}
}

@ARTICLE{belli24,
       author = {{Belli}, Sirio and {Park}, Minjung and {Davies}, Rebecca L. and {Mendel}, J. Trevor and {Johnson}, Benjamin D. and {Conroy}, Charlie and {Benton}, Chlo{\"e} and {Bugiani}, Letizia and {Emami}, Razieh and {Leja}, Joel and {Li}, Yijia and {Maheson}, Gabriel and {Mathews}, Elijah P. and {Naidu}, Rohan P. and {Nelson}, Erica J. and {Tacchella}, Sandro and {Terrazas}, Bryan A. and {Weinberger}, Rainer},
        title = "{Star formation shut down by multiphase gas outflow in a galaxy at a redshift of 2.45}",
      journal = {\nat},
         year = 2024,
        month = jun,
       volume = {630},
       number = {8015},
        pages = {54-58},
          doi = {10.1038/s41586-024-07412-1},
archivePrefix = {arXiv},
       eprint = {2308.05795},
 primaryClass = {astro-ph.GA},
       adsurl = {https://ui.adsabs.harvard.edu/abs/2024Natur.630...54B}
}

@ARTICLE{davies24,
       author = {{Davies}, Rebecca L. and {Belli}, Sirio and {Park}, Minjung and {Mendel}, J. Trevor and {Johnson}, Benjamin D. and {Conroy}, Charlie and {Benton}, Chlo{\"e} and {Bugiani}, Letizia and {Emami}, Razieh and {Leja}, Joel and {Li}, Yijia and {Maheson}, Gabriel and {Mathews}, Elijah P. and {Naidu}, Rohan P. and {Nelson}, Erica J. and {Tacchella}, Sandro and {Terrazas}, Bryan A. and {Weinberger}, Rainer},
        title = "{JWST reveals widespread AGN-driven neutral gas outflows in massive z   2 galaxies}",
      journal = {\mnras},
         year = 2024,
        month = mar,
       volume = {528},
       number = {3},
        pages = {4976-4992},
          doi = {10.1093/mnras/stae327},
archivePrefix = {arXiv},
       eprint = {2310.17939},
 primaryClass = {astro-ph.GA},
       adsurl = {https://ui.adsabs.harvard.edu/abs/2024MNRAS.528.4976D}
}

@ARTICLE{forster-schreiber20,
       author = {{F{\"o}rster Schreiber}, Natascha M. and {Wuyts}, Stijn},
        title = "{Star-Forming Galaxies at Cosmic Noon}",
      journal = {\araa},
         year = 2020,
        month = aug,
       volume = {58},
        pages = {661-725},
          doi = {10.1146/annurev-astro-032620-021910},
archivePrefix = {arXiv},
       eprint = {2010.10171},
 primaryClass = {astro-ph.GA},
       adsurl = {https://ui.adsabs.harvard.edu/abs/2020ARA&A..58..661F}
}

@ARTICLE{strom23,
       author = {{Strom}, Allison L. and {Rudie}, Gwen C. and {Trainor}, Ryan F. and {Brammer}, Gabriel B. and {Maseda}, Michael V. and {Raptis}, Menelaos and {Rogers}, Noah S.~J. and {Steidel}, Charles C. and {Chen}, Yuguang and {Law}, David R.},
        title = "{CECILIA: The Faint Emission Line Spectrum of z   2-3 Star-forming Galaxies}",
      journal = {\apjl},
         year = 2023,
        month = nov,
       volume = {958},
       number = {1},
          eid = {L11},
        pages = {L11},
          doi = {10.3847/2041-8213/ad07dc},
archivePrefix = {arXiv},
       eprint = {2308.13508},
 primaryClass = {astro-ph.GA},
       adsurl = {https://ui.adsabs.harvard.edu/abs/2023ApJ...958L..11S}
}

@ARTICLE{shapley25,
       author = {{Shapley}, Alice E. and {Sanders}, Ryan L. and {Topping}, Michael W. and {Reddy}, Naveen A. and {Berg}, Danielle A. and {Bouwens}, Rychard J. and {Brammer}, Gabriel and {Carnall}, Adam C. and {Cullen}, Fergus and {Dav{\'e}}, Romeel and {Dunlop}, James S. and {Ellis}, Richard S. and {F{\"o}rster Schreiber}, N.~M. and {Furlanetto}, Steven R. and {Glazebrook}, Karl and {Illingworth}, Garth D. and {Jones}, Tucker and {Kriek}, Mariska and {McLeod}, Derek J. and {McLure}, Ross J. and {Narayanan}, Desika and {Oesch}, Pascal and {Pahl}, Anthony J. and {Pettini}, Max and {Schaerer}, Daniel and {Stark}, Daniel P. and {Steidel}, Charles C. and {Tang}, Mengtao and {Clarke}, Leonardo and {Donnan}, Callum T. and {Kehoe}, Emily},
        title = "{The AURORA Survey: A New Era of Emission-line Diagrams with JWST/NIRSpec}",
      journal = {\apj},
         year = 2025,
        month = feb,
       volume = {980},
       number = {2},
          eid = {242},
        pages = {242},
          doi = {10.3847/1538-4357/adad68},
archivePrefix = {arXiv},
       eprint = {2407.00157},
 primaryClass = {astro-ph.GA},
       adsurl = {https://ui.adsabs.harvard.edu/abs/2025ApJ...980..242S}
}

@ARTICLE{cataldi25,
       author = {{Cataldi}, E. and {Belfiore}, F. and {Curti}, M. and {Moreschini}, B. and {Mannucci}, F. and {D'Amato}, Q. and {Cresci}, G. and {Feltre}, A. and {Ginolfi}, M. and {Marconi}, A. and {Amiri}, A. and {Arnaboldi}, M. and {Bertola}, E. and {Bracci}, C. and {Carniani}, S. and {Ceci}, M. and {Chakraborty}, A. and {Cirasuolo}, M. and {Cullen}, F. and {Kobayashi}, C. and {Kumari}, N. and {Maiolino}, R. and {Marconcini}, C. and {Scialpi}, M. and {Ulivi}, L.},
        title = "{MARTA: Temperature-temperature relationships and strong-line metallicity calibrations from multiple auroral-line detections at cosmic noon}",
      journal = {\aap},
         year = 2025,
        month = nov,
       volume = {703},
          eid = {A208},
        pages = {A208},
          doi = {10.1051/0004-6361/202554843},
archivePrefix = {arXiv},
       eprint = {2504.03839},
 primaryClass = {astro-ph.GA},
       adsurl = {https://ui.adsabs.harvard.edu/abs/2025A&A...703A.208C}
}

@ARTICLE{slob24,
       author = {{Slob}, Martje and {Kriek}, Mariska and {Beverage}, Aliza G. and {Suess}, Katherine A. and {Barro}, Guillermo and {Bezanson}, Rachel and {Brammer}, Gabriel and {Cheng}, Chloe M. and {Conroy}, Charlie and {de Graaff}, Anna and {F{\"o}rster Schreiber}, Natascha M. and {Franx}, Marijn and {Lorenz}, Brian and {Mancera Pi{\~n}a}, Pavel E. and {Marchesini}, Danilo and {Muzzin}, Adam and {Newman}, Andrew B. and {Price}, Sedona H. and {Shapley}, Alice E. and {Stefanon}, Mauro and {van Dokkum}, Pieter and {Weisz}, Daniel R.},
        title = "{The JWST-SUSPENSE Ultradeep Spectroscopic Program: Survey Overview and Star Formation Histories of Quiescent Galaxies at 1 < z < 3}",
      journal = {\apj},
         year = 2024,
        month = oct,
       volume = {973},
       number = {2},
          eid = {131},
        pages = {131},
          doi = {10.3847/1538-4357/ad65ff},
archivePrefix = {arXiv},
       eprint = {2404.12432},
 primaryClass = {astro-ph.GA},
       adsurl = {https://ui.adsabs.harvard.edu/abs/2024ApJ...973..131S}
}

@ARTICLE{madau14,
       author = {{Madau}, Piero and {Dickinson}, Mark},
        title = "{Cosmic Star-Formation History}",
      journal = {\araa},
         year = 2014,
        month = aug,
       volume = {52},
        pages = {415-486},
          doi = {10.1146/annurev-astro-081811-125615},
archivePrefix = {arXiv},
       eprint = {1403.0007},
 primaryClass = {astro-ph.CO},
       adsurl = {https://ui.adsabs.harvard.edu/abs/2014ARA&A..52..415M}
}

@ARTICLE{delvecchio14,
       author = {{Delvecchio}, I. and {Gruppioni}, C. and {Pozzi}, F. and {Berta}, S. and {Zamorani}, G. and {Cimatti}, A. and {Lutz}, D. and {Scott}, D. and {Vignali}, C. and {Cresci}, G. and {Feltre}, A. and {Cooray}, A. and {Vaccari}, M. and {Fritz}, J. and {Le Floc'h}, E. and {Magnelli}, B. and {Popesso}, P. and {Oliver}, S. and {Bock}, J. and {Carollo}, M. and {Contini}, T. and {Le F{\'e}vre}, O. and {Lilly}, S. and {Mainieri}, V. and {Renzini}, A. and {Scodeggio}, M.},
        title = "{Tracing the cosmic growth of supermassive black holes to z {\ensuremath{\sim}} 3 with Herschel}",
      journal = {\mnras},
         year = 2014,
        month = apr,
       volume = {439},
       number = {3},
        pages = {2736-2754},
          doi = {10.1093/mnras/stu130},
archivePrefix = {arXiv},
       eprint = {1401.4503},
 primaryClass = {astro-ph.GA},
       adsurl = {https://ui.adsabs.harvard.edu/abs/2014MNRAS.439.2736D}
}

@ARTICLE{muzzin13,
       author = {{Muzzin}, Adam and {Marchesini}, Danilo and {Stefanon}, Mauro and {Franx}, Marijn and {McCracken}, Henry J. and {Milvang-Jensen}, Bo and {Dunlop}, James S. and {Fynbo}, J.~P.~U. and {Brammer}, Gabriel and {Labb{\'e}}, Ivo and {van Dokkum}, Pieter G.},
        title = "{The Evolution of the Stellar Mass Functions of Star-forming and Quiescent Galaxies to z = 4 from the COSMOS/UltraVISTA Survey}",
      journal = {\apj},
         year = 2013,
        month = nov,
       volume = {777},
       number = {1},
          eid = {18},
        pages = {18},
          doi = {10.1088/0004-637X/777/1/18},
archivePrefix = {arXiv},
       eprint = {1303.4409},
 primaryClass = {astro-ph.CO},
       adsurl = {https://ui.adsabs.harvard.edu/abs/2013ApJ...777...18M}
}

@ARTICLE{steidel04,
       author = {{Steidel}, Charles C. and {Shapley}, Alice E. and {Pettini}, Max and {Adelberger}, Kurt L. and {Erb}, Dawn K. and {Reddy}, Naveen A. and {Hunt}, Matthew P.},
        title = "{A Survey of Star-forming Galaxies in the 1.4<\raisebox{-0.5ex}\textasciitildeZ<\raisebox{-0.5ex}\textasciitilde 2.5 Redshift Desert: Overview}",
      journal = {\apj},
         year = 2004,
        month = apr,
       volume = {604},
       number = {2},
        pages = {534-550},
          doi = {10.1086/381960},
archivePrefix = {arXiv},
       eprint = {astro-ph/0401439},
 primaryClass = {astro-ph},
       adsurl = {https://ui.adsabs.harvard.edu/abs/2004ApJ...604..534S}
}

@ARTICLE{stern99,
       author = {{Stern}, Daniel and {Spinrad}, Hyron},
        title = "{Search Techniques for Distant Galaxies}",
      journal = {\pasp},
         year = 1999,
        month = dec,
       volume = {111},
       number = {766},
        pages = {1475-1502},
          doi = {10.1086/316471},
archivePrefix = {arXiv},
       eprint = {astro-ph/9912082},
 primaryClass = {astro-ph},
       adsurl = {https://ui.adsabs.harvard.edu/abs/1999PASP..111.1475S}
}

@ARTICLE{kriek15,
       author = {{Kriek}, Mariska and {Shapley}, Alice E. and {Reddy}, Naveen A. and {Siana}, Brian and {Coil}, Alison L. and {Mobasher}, Bahram and {Freeman}, William R. and {de Groot}, Laura and {Price}, Sedona H. and {Sanders}, Ryan and {Shivaei}, Irene and {Brammer}, Gabriel B. and {Momcheva}, Ivelina G. and {Skelton}, Rosalind E. and {van Dokkum}, Pieter G. and {Whitaker}, Katherine E. and {Aird}, James and {Azadi}, Mojegan and {Kassis}, Marc and {Bullock}, James S. and {Conroy}, Charlie and {Dav{\'e}}, Romeel and {Kere{\v{s}}}, Du{\v{s}}an and {Krumholz}, Mark},
        title = "{The MOSFIRE Deep Evolution Field (MOSDEF) Survey: Rest-frame Optical Spectroscopy for \raisebox{-0.5ex}\textasciitilde1500 H-selected Galaxies at 1.37 < z < 3.8}",
      journal = {\apjs},
         year = 2015,
        month = jun,
       volume = {218},
       number = {2},
          eid = {15},
        pages = {15},
          doi = {10.1088/0067-0049/218/2/15},
archivePrefix = {arXiv},
       eprint = {1412.1835},
 primaryClass = {astro-ph.GA},
       adsurl = {https://ui.adsabs.harvard.edu/abs/2015ApJS..218...15K}
}

@ARTICLE{steidel14,
       author = {{Steidel}, Charles C. and {Rudie}, Gwen C. and {Strom}, Allison L. and {Pettini}, Max and {Reddy}, Naveen A. and {Shapley}, Alice E. and {Trainor}, Ryan F. and {Erb}, Dawn K. and {Turner}, Monica L. and {Konidaris}, Nicholas P. and {Kulas}, Kristin R. and {Mace}, Gregory and {Matthews}, Keith and {McLean}, Ian S.},
        title = "{Strong Nebular Line Ratios in the Spectra of z \raisebox{-0.5ex}\textasciitilde 2-3 Star Forming Galaxies: First Results from KBSS-MOSFIRE}",
      journal = {\apj},
         year = 2014,
        month = nov,
       volume = {795},
       number = {2},
          eid = {165},
        pages = {165},
          doi = {10.1088/0004-637X/795/2/165},
archivePrefix = {arXiv},
       eprint = {1405.5473},
 primaryClass = {astro-ph.GA},
       adsurl = {https://ui.adsabs.harvard.edu/abs/2014ApJ...795..165S}
}

@ARTICLE{wisnioski15,
       author = {{Wisnioski}, E. and {F{\"o}rster Schreiber}, N.~M. and {Wuyts}, S. and {Wuyts}, E. and {Bandara}, K. and {Wilman}, D. and {Genzel}, R. and {Bender}, R. and {Davies}, R. and {Fossati}, M. and {Lang}, P. and {Mendel}, J.~T. and {Beifiori}, A. and {Brammer}, G. and {Chan}, J. and {Fabricius}, M. and {Fudamoto}, Y. and {Kulkarni}, S. and {Kurk}, J. and {Lutz}, D. and {Nelson}, E.~J. and {Momcheva}, I. and {Rosario}, D. and {Saglia}, R. and {Seitz}, S. and {Tacconi}, L.~J. and {van Dokkum}, P.~G.},
        title = "{The KMOS$^{3D}$ Survey: Design, First Results, and the Evolution of Galaxy Kinematics from 0.7 <= z <= 2.7}",
      journal = {\apj},
         year = 2015,
        month = feb,
       volume = {799},
       number = {2},
          eid = {209},
        pages = {209},
          doi = {10.1088/0004-637X/799/2/209},
archivePrefix = {arXiv},
       eprint = {1409.6791},
 primaryClass = {astro-ph.GA},
       adsurl = {https://ui.adsabs.harvard.edu/abs/2015ApJ...799..209W}
}

@ARTICLE{curti20,
       author = {{Curti}, Mirko and {Maiolino}, Roberto and {Cirasuolo}, Michele and {Mannucci}, Filippo and {Williams}, Rebecca J. and {Auger}, Matt and {Mercurio}, Amata and {Hayden-Pawson}, Connor and {Cresci}, Giovanni and {Marconi}, Alessandro and {Belfiore}, Francesco and {Cappellari}, Michele and {Cicone}, Claudia and {Cullen}, Fergus and {Meneghetti}, Massimo and {Ota}, Kazuaki and {Peng}, Yingjie and {Pettini}, Max and {Swinbank}, Mark and {Troncoso}, Paulina},
        title = "{The KLEVER Survey: spatially resolved metallicity maps and gradients in a sample of 1.2 < z < 2.5 lensed galaxies}",
      journal = {\mnras},
         year = 2020,
        month = feb,
       volume = {492},
       number = {1},
        pages = {821-842},
          doi = {10.1093/mnras/stz3379},
archivePrefix = {arXiv},
       eprint = {1910.13451},
 primaryClass = {astro-ph.GA},
       adsurl = {https://ui.adsabs.harvard.edu/abs/2020MNRAS.492..821C}
}

@ARTICLE{forster-schreiber09,
       author = {{F{\"o}rster Schreiber}, N.~M. and {Genzel}, R. and {Bouch{\'e}}, N. and {Cresci}, G. and {Davies}, R. and {Buschkamp}, P. and {Shapiro}, K. and {Tacconi}, L.~J. and {Hicks}, E.~K.~S. and {Genel}, S. and {Shapley}, A.~E. and {Erb}, D.~K. and {Steidel}, C.~C. and {Lutz}, D. and {Eisenhauer}, F. and {Gillessen}, S. and {Sternberg}, A. and {Renzini}, A. and {Cimatti}, A. and {Daddi}, E. and {Kurk}, J. and {Lilly}, S. and {Kong}, X. and {Lehnert}, M.~D. and {Nesvadba}, N. and {Verma}, A. and {McCracken}, H. and {Arimoto}, N. and {Mignoli}, M. and {Onodera}, M.},
        title = "{The SINS Survey: SINFONI Integral Field Spectroscopy of z \raisebox{-0.5ex}\textasciitilde 2 Star-forming Galaxies}",
      journal = {\apj},
         year = 2009,
        month = dec,
       volume = {706},
       number = {2},
        pages = {1364-1428},
          doi = {10.1088/0004-637X/706/2/1364},
archivePrefix = {arXiv},
       eprint = {0903.1872},
 primaryClass = {astro-ph.CO},
       adsurl = {https://ui.adsabs.harvard.edu/abs/2009ApJ...706.1364F}
}

@ARTICLE{kriek24,
       author = {{Kriek}, Mariska and {Beverage}, Aliza G. and {Price}, Sedona H. and {Suess}, Katherine A. and {Barro}, Guillermo and {Bezanson}, Rachel S. and {Conroy}, Charlie and {Cutler}, Sam E. and {Franx}, Marijn and {Lin}, Jamie and {Lorenz}, Brian and {Ma}, Yilun and {Momcheva}, Ivelina G. and {Mowla}, Lamiya A. and {Pasha}, Imad and {van Dokkum}, Pieter and {Whitaker}, Katherine E.},
        title = "{The Heavy Metal Survey: Star Formation Constraints and Dynamical Masses of 21 Massive Quiescent Galaxies at z = 1.3{\textendash}2.3}",
      journal = {\apj},
         year = 2024,
        month = may,
       volume = {966},
       number = {1},
          eid = {36},
        pages = {36},
          doi = {10.3847/1538-4357/ad2df9},
archivePrefix = {arXiv},
       eprint = {2311.16232},
 primaryClass = {astro-ph.GA},
       adsurl = {https://ui.adsabs.harvard.edu/abs/2024ApJ...966...36K}
}

@ARTICLE{belli17mosfire,
       author = {{Belli}, Sirio and {Newman}, Andrew B. and {Ellis}, Richard S.},
        title = "{MOSFIRE Spectroscopy of Quiescent Galaxies at 1.5 < z < 2.5. I. Evolution of Structural and Dynamical Properties}",
      journal = {\apj},
         year = 2017,
        month = jan,
       volume = {834},
       number = {1},
          eid = {18},
        pages = {18},
          doi = {10.3847/1538-4357/834/1/18},
archivePrefix = {arXiv},
       eprint = {1608.00608},
 primaryClass = {astro-ph.GA},
       adsurl = {https://ui.adsabs.harvard.edu/abs/2017ApJ...834...18B}
}

@ARTICLE{carnall19,
       author = {{Carnall}, A.~C. and {McLure}, R.~J. and {Dunlop}, J.~S. and {Cullen}, F. and {McLeod}, D.~J. and {Wild}, V. and {Johnson}, B.~D. and {Appleby}, S. and {Dav{\'e}}, R. and {Amorin}, R. and {Bolzonella}, M. and {Castellano}, M. and {Cimatti}, A. and {Cucciati}, O. and {Gargiulo}, A. and {Garilli}, B. and {Marchi}, F. and {Pentericci}, L. and {Pozzetti}, L. and {Schreiber}, C. and {Talia}, M. and {Zamorani}, G.},
        title = "{The VANDELS survey: the star-formation histories of massive quiescent galaxies at 1.0 < z < 1.3}",
      journal = {\mnras},
         year = 2019,
        month = nov,
       volume = {490},
       number = {1},
        pages = {417-439},
          doi = {10.1093/mnras/stz2544},
archivePrefix = {arXiv},
       eprint = {1903.11082},
 primaryClass = {astro-ph.GA},
       adsurl = {https://ui.adsabs.harvard.edu/abs/2019MNRAS.490..417C}
}

@ARTICLE{mendel20,
       author = {{Mendel}, J. Trevor and {Beifiori}, Alessandra and {Saglia}, Roberto P. and {Bender}, Ralf and {Brammer}, Gabriel B. and {Chan}, Jeffrey and {F{\"o}rster Schreiber}, Natascha M. and {Fossati}, Matteo and {Galametz}, Audrey and {Momcheva}, Ivelina G. and {Nelson}, Erica J. and {Wilman}, David J. and {Wuyts}, Stijn},
        title = "{The Kinematics of Massive Quiescent Galaxies at 1.4 < z < 2.1: Dark Matter Fractions, IMF Variation, and the Relation to Local Early-type Galaxies}",
      journal = {\apj},
         year = 2020,
        month = aug,
       volume = {899},
       number = {1},
          eid = {87},
        pages = {87},
          doi = {10.3847/1538-4357/ab9ffc},
archivePrefix = {arXiv},
       eprint = {2006.13949},
 primaryClass = {astro-ph.GA},
       adsurl = {https://ui.adsabs.harvard.edu/abs/2020ApJ...899...87M}
}

@ARTICLE{belli14,
       author = {{Belli}, Sirio and {Newman}, Andrew B. and {Ellis}, Richard S.},
        title = "{Velocity Dispersions and Dynamical Masses for a Large Sample of Quiescent Galaxies at z > 1: Improved Measures of the Growth in Mass and Size}",
      journal = {\apj},
         year = 2014,
        month = mar,
       volume = {783},
       number = {2},
          eid = {117},
        pages = {117},
          doi = {10.1088/0004-637X/783/2/117},
archivePrefix = {arXiv},
       eprint = {1311.3317},
 primaryClass = {astro-ph.CO},
       adsurl = {https://ui.adsabs.harvard.edu/abs/2014ApJ...783..117B}
}

@ARTICLE{vandesande13,
       author = {{van de Sande}, Jesse and {Kriek}, Mariska and {Franx}, Marijn and {van Dokkum}, Pieter G. and {Bezanson}, Rachel and {Bouwens}, Rychard J. and {Quadri}, Ryan F. and {Rix}, Hans-Walter and {Skelton}, Rosalind E.},
        title = "{Stellar Kinematics of z \raisebox{-0.5ex}\textasciitilde 2 Galaxies and the Inside-out Growth of Quiescent Galaxies}",
      journal = {\apj},
         year = 2013,
        month = jul,
       volume = {771},
       number = {2},
          eid = {85},
        pages = {85},
          doi = {10.1088/0004-637X/771/2/85},
archivePrefix = {arXiv},
       eprint = {1211.3424},
 primaryClass = {astro-ph.CO},
       adsurl = {https://ui.adsabs.harvard.edu/abs/2013ApJ...771...85V}
}

@ARTICLE{bezanson13,
       author = {{Bezanson}, Rachel and {van Dokkum}, Pieter and {van de Sande}, Jesse and {Franx}, Marijn and {Kriek}, Mariska},
        title = "{Massive and Newly Dead: Discovery of a Significant Population of Galaxies with High-velocity Dispersions and Strong Balmer Lines at z \raisebox{-0.5ex}\textasciitilde 1.5 from Deep Keck Spectra and HST/WFC3 Imaging}",
      journal = {\apjl},
         year = 2013,
        month = feb,
       volume = {764},
       number = {1},
          eid = {L8},
        pages = {L8},
          doi = {10.1088/2041-8205/764/1/L8},
archivePrefix = {arXiv},
       eprint = {1210.7236},
 primaryClass = {astro-ph.CO},
       adsurl = {https://ui.adsabs.harvard.edu/abs/2013ApJ...764L...8B}
}

@ARTICLE{onodera12,
       author = {{Onodera}, M. and {Renzini}, A. and {Carollo}, M. and {Cappellari}, M. and {Mancini}, C. and {Strazzullo}, V. and {Daddi}, E. and {Arimoto}, N. and {Gobat}, R. and {Yamada}, Y. and {McCracken}, H.~J. and {Ilbert}, O. and {Capak}, P. and {Cimatti}, A. and {Giavalisco}, M. and {Koekemoer}, A.~M. and {Kong}, X. and {Lilly}, S. and {Motohara}, K. and {Ohta}, K. and {Sanders}, D.~B. and {Scoville}, N. and {Tamura}, N. and {Taniguchi}, Y.},
        title = "{Deep Near-infrared Spectroscopy of Passively Evolving Galaxies at z >\raisebox{-0.5ex}\textasciitilde 1.4}",
      journal = {\apj},
         year = 2012,
        month = aug,
       volume = {755},
       number = {1},
          eid = {26},
        pages = {26},
          doi = {10.1088/0004-637X/755/1/26},
archivePrefix = {arXiv},
       eprint = {1206.1540},
 primaryClass = {astro-ph.CO},
       adsurl = {https://ui.adsabs.harvard.edu/abs/2012ApJ...755...26O}
}

@ARTICLE{degraaff25,
       author = {{de Graaff}, Anna and {Brammer}, Gabriel and {Weibel}, Andrea and {Lewis}, Zach and {Maseda}, Michael V. and {Oesch}, Pascal A. and {Bezanson}, Rachel and {Boogaard}, Leindert A. and {Cleri}, Nikko J. and {Cooper}, Olivia R. and {Gottumukkala}, Rashmi and {Greene}, Jenny E. and {Hirschmann}, Michaela and {Hviding}, Raphael E. and {Katz}, Harley and {Labb{\'e}}, Ivo and {Leja}, Joel and {Matthee}, Jorryt and {McConachie}, Ian and {Miller}, Tim B. and {Naidu}, Rohan P. and {Price}, Sedona H. and {Rix}, Hans-Walter and {Setton}, David J. and {Suess}, Katherine A. and {Wang}, Bingjie and {Whitaker}, Katherine E. and {Williams}, Christina C.},
        title = "{RUBIES: A complete census of the bright and red distant Universe with JWST/NIRSpec}",
      journal = {\aap},
         year = 2025,
        month = may,
       volume = {697},
          eid = {A189},
        pages = {A189},
          doi = {10.1051/0004-6361/202452186},
archivePrefix = {arXiv},
       eprint = {2409.05948},
 primaryClass = {astro-ph.GA},
       adsurl = {https://ui.adsabs.harvard.edu/abs/2025A&A...697A.189D}
}

@ARTICLE{vanderwel14,
       author = {{van der Wel}, A. and {Franx}, M. and {van Dokkum}, P.~G. and {Skelton}, R.~E. and {Momcheva}, I.~G. and {Whitaker}, K.~E. and {Brammer}, G.~B. and {Bell}, E.~F. and {Rix}, H. -W. and {Wuyts}, S. and {Ferguson}, H.~C. and {Holden}, B.~P. and {Barro}, G. and {Koekemoer}, A.~M. and {Chang}, Yu-Yen and {McGrath}, E.~J. and {H{\"a}ussler}, B. and {Dekel}, A. and {Behroozi}, P. and {Fumagalli}, M. and {Leja}, J. and {Lundgren}, B.~F. and {Maseda}, M.~V. and {Nelson}, E.~J. and {Wake}, D.~A. and {Patel}, S.~G. and {Labb{\'e}}, I. and {Faber}, S.~M. and {Grogin}, N.~A. and {Kocevski}, D.~D.},
        title = "{3D-HST+CANDELS: The Evolution of the Galaxy Size-Mass Distribution since z = 3}",
      journal = {\apj},
         year = 2014,
        month = jun,
       volume = {788},
       number = {1},
          eid = {28},
        pages = {28},
          doi = {10.1088/0004-637X/788/1/28},
archivePrefix = {arXiv},
       eprint = {1404.2844},
 primaryClass = {astro-ph.GA},
       adsurl = {https://ui.adsabs.harvard.edu/abs/2014ApJ...788...28V}
}

@ARTICLE{cucciati18,
       author = {{Cucciati}, O. and {Lemaux}, B.~C. and {Zamorani}, G. and {Le F{\`e}vre}, O. and {Tasca}, L.~A.~M. and {Hathi}, N.~P. and {Lee}, K. -G. and {Bardelli}, S. and {Cassata}, P. and {Garilli}, B. and {Le Brun}, V. and {Maccagni}, D. and {Pentericci}, L. and {Thomas}, R. and {Vanzella}, E. and {Zucca}, E. and {Lubin}, L.~M. and {Amorin}, R. and {Cassar{\`a}}, L.~P. and {Cimatti}, A. and {Talia}, M. and {Vergani}, D. and {Koekemoer}, A. and {Pforr}, J. and {Salvato}, M.},
        title = "{The progeny of a cosmic titan: a massive multi-component proto-supercluster in formation at z = 2.45 in VUDS}",
      journal = {\aap},
         year = 2018,
        month = nov,
       volume = {619},
          eid = {A49},
        pages = {A49},
          doi = {10.1051/0004-6361/201833655},
archivePrefix = {arXiv},
       eprint = {1806.06073},
 primaryClass = {astro-ph.CO},
       adsurl = {https://ui.adsabs.harvard.edu/abs/2018A&A...619A..49C}
}

@ARTICLE{merlin24,
       author = {{Merlin}, E. and {Santini}, P. and {Paris}, D. and {Castellano}, M. and {Fontana}, A. and {Treu}, T. and {Finkelstein}, S.~L. and {Dunlop}, J.~S. and {Arrabal Haro}, P. and {Bagley}, M. and {Boyett}, K. and {Calabr{\`o}}, A. and {Correnti}, M. and {Davis}, K. and {Dickinson}, M. and {Donnan}, C.~T. and {Ferguson}, H.~C. and {Fortuni}, F. and {Giavalisco}, M. and {Glazebrook}, K. and {Grazian}, A. and {Grogin}, N.~A. and {Hathi}, N. and {Hirschmann}, M. and {Kartaltepe}, J.~S. and {Kewley}, L.~J. and {Kirkpatrick}, A. and {Kocevski}, D.~D. and {Koekemoer}, A.~M. and {Leung}, G. and {Lotz}, J.~M. and {Lucas}, R.~A. and {Magee}, D.~K. and {Marchesini}, D. and {Mascia}, S. and {McLeod}, D.~J. and {McLure}, R.~J. and {Nanayakkara}, T. and {Napolitano}, L. and {Nonino}, M. and {Papovich}, C. and {Pentericci}, L. and {P{\'e}rez-Gonz{\'a}lez}, P.~G. and {Pirzkal}, N. and {Ravindranath}, S. and {Roberts-Borsani}, G. and {Somerville}, R.~S. and {Trenti}, M. and {Trump}, J.~R. and {Vulcani}, B. and {Wang}, X. and {Watson}, P.~J. and {Wilkins}, S.~M. and {Yang}, G. and {Yung}, L.~Y.~A.},
        title = "{ASTRODEEP-JWST: NIRCam-HST multi-band photometry and redshifts for half a million sources in six extragalactic deep fields}",
      journal = {\aap},
         year = 2024,
        month = nov,
       volume = {691},
          eid = {A240},
        pages = {A240},
          doi = {10.1051/0004-6361/202451409},
archivePrefix = {arXiv},
       eprint = {2409.00169},
 primaryClass = {astro-ph.GA},
       adsurl = {https://ui.adsabs.harvard.edu/abs/2024A&A...691A.240M}
}

@ARTICLE{rodighiero23,
       author = {{Rodighiero}, Giulia and {Bisigello}, Laura and {Iani}, Edoardo and {Marasco}, Antonino and {Grazian}, Andrea and {Sinigaglia}, Francesco and {Cassata}, Paolo and {Gruppioni}, Carlotta},
        title = "{JWST unveils heavily obscured (active and passive) sources up to z   13}",
      journal = {\mnras},
         year = 2023,
        month = jan,
       volume = {518},
       number = {1},
        pages = {L19-L24},
          doi = {10.1093/mnrasl/slac115},
archivePrefix = {arXiv},
       eprint = {2208.02825},
 primaryClass = {astro-ph.GA},
       adsurl = {https://ui.adsabs.harvard.edu/abs/2023MNRAS.518L..19R}
}

@ARTICLE{nelson23,
       author = {{Nelson}, Erica J. and {Suess}, Katherine A. and {Bezanson}, Rachel and {Price}, Sedona H. and {van Dokkum}, Pieter and {Leja}, Joel and {Wang}, Bingjie and {Whitaker}, Katherine E. and {Labb{\'e}}, Ivo and {Barrufet}, Laia and {Brammer}, Gabriel and {Eisenstein}, Daniel J. and {Gibson}, Justus and {Hartley}, Abigail I. and {Johnson}, Benjamin D. and {Heintz}, Kasper E. and {Mathews}, Elijah and {Miller}, Tim B. and {Oesch}, Pascal A. and {Sandles}, Lester and {Setton}, David J. and {Speagle}, Joshua S. and {Tacchella}, Sandro and {Tadaki}, Ken-ichi and {{\"U}bler}, Hannah and {Weaver}, John. R.},
        title = "{JWST Reveals a Population of Ultrared, Flattened Galaxies at 2 {\ensuremath{\lesssim}} z {\ensuremath{\lesssim}} 6 Previously Missed by HST}",
      journal = {\apjl},
         year = 2023,
        month = may,
       volume = {948},
       number = {2},
          eid = {L18},
        pages = {L18},
          doi = {10.3847/2041-8213/acc1e1},
archivePrefix = {arXiv},
       eprint = {2208.01630},
 primaryClass = {astro-ph.GA},
       adsurl = {https://ui.adsabs.harvard.edu/abs/2023ApJ...948L..18N}
}

@ARTICLE{barrufet23,
       author = {{Barrufet}, L. and {Oesch}, P.~A. and {Weibel}, A. and {Brammer}, G. and {Bezanson}, R. and {Bouwens}, R. and {Fudamoto}, Y. and {Gonzalez}, V. and {Gottumukkala}, R. and {Illingworth}, G. and {Heintz}, K.~E. and {Holden}, B. and {Labbe}, I. and {Magee}, D. and {Naidu}, R.~P. and {Nelson}, E. and {Stefanon}, M. and {Smit}, R. and {van Dokkum}, P. and {Weaver}, J.~R. and {Williams}, C.~C.},
        title = "{Unveiling the nature of infrared bright, optically dark galaxies with early JWST data}",
      journal = {\mnras},
         year = 2023,
        month = jun,
       volume = {522},
       number = {1},
        pages = {449-456},
          doi = {10.1093/mnras/stad947},
archivePrefix = {arXiv},
       eprint = {2207.14733},
 primaryClass = {astro-ph.GA},
       adsurl = {https://ui.adsabs.harvard.edu/abs/2023MNRAS.522..449B}
}

@ARTICLE{perezgonzalez23,
       author = {{P{\'e}rez-Gonz{\'a}lez}, Pablo G. and {Barro}, Guillermo and {Annunziatella}, Marianna and {Costantin}, Luca and {Garc{\'\i}a-Argum{\'a}nez}, {\'A}ngela and {McGrath}, Elizabeth J. and {M{\'e}rida}, Rosa M. and {Zavala}, Jorge A. and {Arrabal Haro}, Pablo and {Bagley}, Micaela B. and {Backhaus}, Bren E. and {Behroozi}, Peter and {Bell}, Eric F. and {Bisigello}, Laura and {Buat}, V{\'e}ronique and {Calabr{\`o}}, Antonello and {Casey}, Caitlin M. and {Cleri}, Nikko J. and {Coogan}, Rosemary T. and {Cooper}, M.~C. and {Cooray}, Asantha R. and {Dekel}, Avishai and {Dickinson}, Mark and {Elbaz}, David and {Ferguson}, Henry C. and {Finkelstein}, Steven L. and {Fontana}, Adriano and {Franco}, Maximilien and {Gardner}, Jonathan P. and {Giavalisco}, Mauro and {G{\'o}mez-Guijarro}, Carlos and {Grazian}, Andrea and {Grogin}, Norman A. and {Guo}, Yuchen and {Huertas-Company}, Marc and {Jogee}, Shardha and {Kartaltepe}, Jeyhan S. and {Kewley}, Lisa J. and {Kirkpatrick}, Allison and {Kocevski}, Dale D. and {Koekemoer}, Anton M. and {Long}, Arianna S. and {Lotz}, Jennifer M. and {Lucas}, Ray A. and {Papovich}, Casey and {Pirzkal}, Nor and {Ravindranath}, Swara and {Somerville}, Rachel S. and {Tacchella}, Sandro and {Trump}, Jonathan R. and {Wang}, Weichen and {Wilkins}, Stephen M. and {Wuyts}, Stijn and {Yang}, Guang and {Yung}, L.~Y. Aaron},
        title = "{CEERS Key Paper. IV. A Triality in the Nature of HST-dark Galaxies}",
      journal = {\apjl},
         year = 2023,
        month = mar,
       volume = {946},
       number = {1},
          eid = {L16},
        pages = {L16},
          doi = {10.3847/2041-8213/acb3a5},
archivePrefix = {arXiv},
       eprint = {2211.00045},
 primaryClass = {astro-ph.GA},
       adsurl = {https://ui.adsabs.harvard.edu/abs/2023ApJ...946L..16P}
}

@ARTICLE{horne86,
       author = {{Horne}, K.},
        title = "{An optimal extraction algorithm for CCD spectroscopy.}",
      journal = {\pasp},
         year = 1986,
        month = jun,
       volume = {98},
        pages = {609-617},
          doi = {10.1086/131801},
       adsurl = {https://ui.adsabs.harvard.edu/abs/1986PASP...98..609H}
}

@ARTICLE{ito25,
       author = {{Ito}, Kei and {Tanaka}, Takumi S. and {Shimasaku}, Kazuhiro and {Ando}, Makoto and {Onoue}, Masafusa and {Tanaka}, Masayuki and {Matsui}, Suin and {Kakimoto}, Takumi and {Valentino}, Francesco},
        title = "{Black hole mass of a quiescent galaxy hosting a Type 1 AGN at z = 2.09: implications for black hole-galaxy coevolution and AGN quenching at high redshift}",
      journal = {\mnras},
         year = 2025,
        month = apr,
       volume = {538},
       number = {3},
        pages = {1501-1516},
          doi = {10.1093/mnras/staf372},
archivePrefix = {arXiv},
       eprint = {2408.08492},
 primaryClass = {astro-ph.GA},
       adsurl = {https://ui.adsabs.harvard.edu/abs/2025MNRAS.538.1501I}
}

@ARTICLE{deugenio24,
       author = {{D'Eugenio}, Francesco and {P{\'e}rez-Gonz{\'a}lez}, Pablo G. and {Maiolino}, Roberto and {Scholtz}, Jan and {Perna}, Michele and {Circosta}, Chiara and {{\"U}bler}, Hannah and {Arribas}, Santiago and {B{\"o}ker}, Torsten and {Bunker}, Andrew J. and {Carniani}, Stefano and {Charlot}, Stephane and {Chevallard}, Jacopo and {Cresci}, Giovanni and {Curtis-Lake}, Emma and {Jones}, Gareth C. and {Kumari}, Nimisha and {Lamperti}, Isabella and {Looser}, Tobias J. and {Parlanti}, Eleonora and {Rix}, Hans-Walter and {Robertson}, Brant and {Rodr{\'\i}guez Del Pino}, Bruno and {Tacchella}, Sandro and {Venturi}, Giacomo and {Willott}, Chris J.},
        title = "{A fast-rotator post-starburst galaxy quenched by supermassive black-hole feedback at z = 3}",
      journal = {Nature Astronomy},
         year = 2024,
        month = nov,
       volume = {8},
        pages = {1443-1456},
          doi = {10.1038/s41550-024-02345-1},
archivePrefix = {arXiv},
       eprint = {2308.06317},
 primaryClass = {astro-ph.GA},
       adsurl = {https://ui.adsabs.harvard.edu/abs/2024NatAs...8.1443D}
}

@ARTICLE{wu25,
       author = {{Wu}, Po-Feng},
        title = "{Ejective Feedback as a Quenching Mechanism in the First 1.5 Billion Years of the Universe: Detection of Neutral Gas Outflow in a z = 4 Recently Quenched Galaxy}",
      journal = {\apj},
         year = 2025,
        month = jan,
       volume = {978},
       number = {2},
          eid = {131},
        pages = {131},
          doi = {10.3847/1538-4357/ad98ef},
archivePrefix = {arXiv},
       eprint = {2409.00471},
 primaryClass = {astro-ph.GA},
       adsurl = {https://ui.adsabs.harvard.edu/abs/2025ApJ...978..131W}
}

@ARTICLE{valentino25,
       author = {{Valentino}, F. and {Heintz}, K.~E. and {Brammer}, G. and {Ito}, K. and {Kokorev}, V. and {Whitaker}, K.~E. and {Gallazzi}, A. and {de Graaff}, A. and {Weibel}, A. and {Frye}, B.~L. and {Kamieneski}, P.~S. and {Jin}, S. and {Ceverino}, D. and {Faisst}, A. and {Farcy}, M. and {Fujimoto}, S. and {Gillman}, S. and {Gottumukkala}, R. and {Hamadouche}, M. and {Harrington}, K.~C. and {Hirschmann}, M. and {Jespersen}, C.~K. and {Kakimoto}, T. and {Kubo}, M. and {Lagos}, C. d. P. and {Lee}, M. and {Magdis}, G.~E. and {Man}, A.~W.~S. and {Onodera}, M. and {Rizzo}, F. and {Shimakawa}, R. and {Setton}, D.~J. and {Tanaka}, M. and {Toft}, S. and {Wu}, P. -F. and {Zhu}, P.},
        title = "{Gas outflows in two recently quenched galaxies at z = 4 and 7}",
      journal = {\aap},
         year = 2025,
        month = jul,
       volume = {699},
          eid = {A358},
        pages = {A358},
          doi = {10.1051/0004-6361/202553908},
archivePrefix = {arXiv},
       eprint = {2503.01990},
 primaryClass = {astro-ph.GA},
       adsurl = {https://ui.adsabs.harvard.edu/abs/2025A&A...699A.358V}
}

@ARTICLE{sun25,
       author = {{Sun}, Yang and {Ji}, Zhiyuan and {Rieke}, George H. and {D'Eugenio}, Francesco and {Zhu}, Yongda and {Sun}, Fengwu and {Lin}, Xiaojing and {Bunker}, Andrew J. and {Lyu}, Jianwei and {Rinaldi}, Pierluigi and {Willmer}, Christopher N.~A.},
        title = "{Extreme Neutral Outflow in a Non-active Galactic Nucleus Quiescent Galaxy at z {\ensuremath{\sim}} 1.3}",
      journal = {\apj},
         year = 2026,
        month = feb,
       volume = {997},
       number = {2},
          eid = {140},
        pages = {140},
          doi = {10.3847/1538-4357/ae2c83},
archivePrefix = {arXiv},
       eprint = {2504.14682},
 primaryClass = {astro-ph.GA},
       adsurl = {https://ui.adsabs.harvard.edu/abs/2026ApJ...997..140S}
}

@BOOK{kurucz93,
       author = {{Kurucz}, Robert L.},
        title = "{SYNTHE spectrum synthesis programs and line data}",
         year = 1993,
       adsurl = {https://ui.adsabs.harvard.edu/abs/1993sssp.book.....K}
}

@ARTICLE{maseda23,
       author = {{Maseda}, Michael V. and {Lewis}, Zach and {Matthee}, Jorryt and {Hennawi}, Joseph F. and {Boogaard}, Leindert and {Feltre}, Anna and {Nanayakkara}, Themiya and {Bacon}, Roland and {Barger}, Amy and {Brinchmann}, Jarle and {Franx}, Marijn and {Hashimoto}, Takuya and {Inami}, Hanae and {Kusakabe}, Haruka and {Leclercq}, Floriane and {Rowland}, Lucie and {Taylor}, Anthony J. and {Tremonti}, Christy and {Urrutia}, Tanya and {Schaye}, Joop and {Simmonds}, Charlotte and {Vitte}, Elo{\"\i}se},
        title = "{JWST/NIRSpec Measurements of Extremely Low Metallicities in High Equivalent Width Ly{\ensuremath{\alpha}} Emitters}",
      journal = {\apj},
         year = 2023,
        month = oct,
       volume = {956},
       number = {1},
          eid = {11},
        pages = {11},
          doi = {10.3847/1538-4357/acf12b},
archivePrefix = {arXiv},
       eprint = {2304.08511},
 primaryClass = {astro-ph.GA},
       adsurl = {https://ui.adsabs.harvard.edu/abs/2023ApJ...956...11M}
}

@ARTICLE{rogers25,
       author = {{Rogers}, Noah S.~J. and {Strom}, Allison L. and {Rudie}, Gwen C. and {Trainor}, Ryan F. and {von Raesfeld}, Caroline and {Raptis}, Menelaos and {Korhonen Cuestas}, Nathalie A. and {Miller}, Tim B. and {Steidel}, Charles C. and {Maseda}, Michael V. and {Chen}, Yuguang and {Law}, David R.},
        title = "{CECILIA: Gas-phase Physical Conditions and Multielement Chemistry at Cosmic Noon}",
      journal = {\apjl},
         year = 2026,
        month = feb,
       volume = {997},
       number = {2},
          eid = {L44},
        pages = {L44},
          doi = {10.3847/2041-8213/ae31f3},
archivePrefix = {arXiv},
       eprint = {2509.18257},
 primaryClass = {astro-ph.GA},
       adsurl = {https://ui.adsabs.harvard.edu/abs/2026ApJ...997L..44R}
}

@ARTICLE{rigby23,
       author = {{Rigby}, Jane R. and {Lightsey}, Paul A. and {Garc{\'\i}a Mar{\'\i}n}, Macarena and {Bowers}, Charles W. and {Smith}, Erin C. and {Glasse}, Alistair and {McElwain}, Michael W. and {Rieke}, George H. and {Chary}, Ranga-Ram and {Liu}, Xiang (Cate) and {Clampin}, Mark and {Kimble}, Randy A. and {Kinzel}, Wayne and {Laidler}, Vicki and {Mehalick}, Kimberly I. and {Noriega-Crespo}, Alberto and {Shivaei}, Irene and {Skelton}, Dennis and {Stark}, Christopher and {Temim}, Tea and {Wei}, Zongying and {Willott}, Chris J.},
        title = "{How Dark the Sky: The JWST Backgrounds}",
      journal = {\pasp},
         year = 2023,
        month = apr,
       volume = {135},
       number = {1046},
          eid = {048002},
        pages = {048002},
          doi = {10.1088/1538-3873/acbcf4},
archivePrefix = {arXiv},
       eprint = {2211.09890},
 primaryClass = {astro-ph.IM},
       adsurl = {https://ui.adsabs.harvard.edu/abs/2023PASP..135d8002R}
}

@ARTICLE{ilbert13,
       author = {{Ilbert}, O. and {McCracken}, H.~J. and {Le F{\`e}vre}, O. and {Capak}, P. and {Dunlop}, J. and {Karim}, A. and {Renzini}, M.~A. and {Caputi}, K. and {Boissier}, S. and {Arnouts}, S. and {Aussel}, H. and {Comparat}, J. and {Guo}, Q. and {Hudelot}, P. and {Kartaltepe}, J. and {Kneib}, J.~P. and {Krogager}, J.~K. and {Le Floc'h}, E. and {Lilly}, S. and {Mellier}, Y. and {Milvang-Jensen}, B. and {Moutard}, T. and {Onodera}, M. and {Richard}, J. and {Salvato}, M. and {Sanders}, D.~B. and {Scoville}, N. and {Silverman}, J.~D. and {Taniguchi}, Y. and {Tasca}, L. and {Thomas}, R. and {Toft}, S. and {Tresse}, L. and {Vergani}, D. and {Wolk}, M. and {Zirm}, A.},
        title = "{Mass assembly in quiescent and star-forming galaxies since z ≃ 4 from UltraVISTA}",
      journal = {\aap},
         year = 2013,
        month = aug,
       volume = {556},
          eid = {A55},
        pages = {A55},
          doi = {10.1051/0004-6361/201321100},
archivePrefix = {arXiv},
       eprint = {1301.3157},
 primaryClass = {astro-ph.CO},
       adsurl = {https://ui.adsabs.harvard.edu/abs/2013A&A...556A..55I}
}

@ARTICLE{wang24,
       author = {{Wang}, Bingjie and {Leja}, Joel and {Labb{\'e}}, Ivo and {Bezanson}, Rachel and {Whitaker}, Katherine E. and {Brammer}, Gabriel and {Furtak}, Lukas J. and {Weaver}, John R. and {Price}, Sedona H. and {Zitrin}, Adi and {Atek}, Hakim and {Coe}, Dan and {Cutler}, Sam E. and {Dayal}, Pratika and {van Dokkum}, Pieter and {Feldmann}, Robert and {Marchesini}, Danilo and {Franx}, Marijn and {F{\"o}rster Schreiber}, Natascha and {Fujimoto}, Seiji and {Geha}, Marla and {Glazebrook}, Karl and {de Graaff}, Anna and {Greene}, Jenny E. and {Juneau}, St{\'e}phanie and {Kassin}, Susan and {Kriek}, Mariska and {Khullar}, Gourav and {Maseda}, Michael and {Mowla}, Lamiya A. and {Muzzin}, Adam and {Nanayakkara}, Themiya and {Nelson}, Erica J. and {Oesch}, Pascal A. and {Pacifici}, Camilla and {Pan}, Richard and {Papovich}, Casey and {Setton}, David J. and {Shapley}, Alice E. and {Smit}, Renske and {Stefanon}, Mauro and {Suess}, Katherine A. and {Taylor}, Edward N. and {Williams}, Christina C.},
        title = "{The UNCOVER Survey: A First-look HST+JWST Catalog of Galaxy Redshifts and Stellar Population Properties Spanning 0.2 {\ensuremath{\lesssim}} z {\ensuremath{\lesssim}} 15}",
      journal = {\apjs},
         year = 2024,
        month = jan,
       volume = {270},
       number = {1},
          eid = {12},
        pages = {12},
          doi = {10.3847/1538-4365/ad0846},
archivePrefix = {arXiv},
       eprint = {2310.01276},
 primaryClass = {astro-ph.GA},
       adsurl = {https://ui.adsabs.harvard.edu/abs/2024ApJS..270...12W}
}

@ARTICLE{wild16,
       author = {{Wild}, Vivienne and {Almaini}, Omar and {Dunlop}, Jim and {Simpson}, Chris and {Rowlands}, Kate and {Bowler}, Rebecca and {Maltby}, David and {McLure}, Ross},
        title = "{The evolution of post-starburst galaxies from z=2 to 0.5}",
      journal = {\mnras},
         year = 2016,
        month = nov,
       volume = {463},
       number = {1},
        pages = {832-844},
          doi = {10.1093/mnras/stw1996},
archivePrefix = {arXiv},
       eprint = {1608.00588},
 primaryClass = {astro-ph.GA},
       adsurl = {https://ui.adsabs.harvard.edu/abs/2016MNRAS.463..832W}
}

@ARTICLE{park25,
       author = {{Park}, Minjung and {Conroy}, Charlie and {Johnson}, Benjamin D. and {Leja}, Joel and {Dotter}, Aaron and {Cargile}, Phillip A.},
        title = "{{\ensuremath{\alpha}}-MC: Self-consistent {\ensuremath{\alpha}}-enhanced Stellar Population Models Covering a Wide Range of Age, Metallicity, and Wavelength}",
      journal = {\apj},
         year = 2025,
        month = dec,
       volume = {994},
       number = {2},
          eid = {165},
        pages = {165},
          doi = {10.3847/1538-4357/ae0cba},
archivePrefix = {arXiv},
       eprint = {2410.21375},
 primaryClass = {astro-ph.GA},
       adsurl = {https://ui.adsabs.harvard.edu/abs/2025ApJ...994..165P}
}

@ARTICLE{beverage24,
       author = {{Beverage}, Aliza G. and {Kriek}, Mariska and {Suess}, Katherine A. and {Conroy}, Charlie and {Price}, Sedona H. and {Barro}, Guillermo and {Bezanson}, Rachel and {Franx}, Marijn and {Lorenz}, Brian and {Ma}, Yilun and {Mowla}, Lamiya A. and {Pasha}, Imad and {van Dokkum}, Pieter and {Weisz}, Daniel R.},
        title = "{The Heavy Metal Survey: The Evolution of Stellar Metallicities, Abundance Ratios, and Ages of Massive Quiescent Galaxies since z {\ensuremath{\sim}} 2}",
      journal = {\apj},
         year = 2024,
        month = may,
       volume = {966},
       number = {2},
          eid = {234},
        pages = {234},
          doi = {10.3847/1538-4357/ad372d},
archivePrefix = {arXiv},
       eprint = {2312.05307},
 primaryClass = {astro-ph.GA},
       adsurl = {https://ui.adsabs.harvard.edu/abs/2024ApJ...966..234B}
}

@ARTICLE{beverage25,
       author = {{Beverage}, Aliza G. and {Slob}, Martje and {Kriek}, Mariska and {Conroy}, Charlie and {Barro}, Guillermo and {Bezanson}, Rachel and {Brammer}, Gabriel and {Cheng}, Chloe M. and {de Graaff}, Anna and {F{\"o}rster Schreiber}, Natascha M. and {Franx}, Marijn and {Lorenz}, Brian and {Mancera Pi{\~n}a}, Pavel E. and {Marchesini}, Danilo and {Muzzin}, Adam and {Newman}, Andrew B. and {Price}, Sedona H. and {Shapley}, Alice E. and {Stefanon}, Mauro and {Suess}, Katherine A. and {van Dokkum}, Pieter and {Weinberg}, David and {Weisz}, Daniel R.},
        title = "{Carbon and Iron Deficiencies in Quiescent Galaxies at z = 1─3 from JWST-SUSPENSE: Implications for the Formation Histories of Massive Galaxies}",
      journal = {\apj},
         year = 2025,
        month = feb,
       volume = {979},
       number = {2},
          eid = {249},
        pages = {249},
          doi = {10.3847/1538-4357/ad96b6},
archivePrefix = {arXiv},
       eprint = {2407.02556},
 primaryClass = {astro-ph.GA},
       adsurl = {https://ui.adsabs.harvard.edu/abs/2025ApJ...979..249B}
}

@ARTICLE{jafa25,
       author = {{Jafariyazani}, Marziye and {Newman}, Andrew B. and {Mobasher}, Bahram and {Belli}, Sirio and {Ellis}, Richard S. and {Faisst}, Andreas L.},
        title = "{Chemical Abundances of Early Quiescent Galaxies: New Observations and Modeling Impacts}",
      journal = {\apj},
         year = 2025,
        month = jun,
       volume = {986},
       number = {2},
          eid = {148},
        pages = {148},
          doi = {10.3847/1538-4357/addbdc},
archivePrefix = {arXiv},
       eprint = {2406.03549},
 primaryClass = {astro-ph.GA},
       adsurl = {https://ui.adsabs.harvard.edu/abs/2025ApJ...986..148J}
}

@ARTICLE{jafa20,
       author = {{Jafariyazani}, Marziye and {Newman}, Andrew B. and {Mobasher}, Bahram and {Belli}, Sirio and {Ellis}, Richard S. and {Patel}, Shannon G.},
        title = "{Resolved Multi-element Stellar Chemical Abundances in the Brightest Quiescent Galaxy at z {\ensuremath{\sim}} 2}",
      journal = {\apjl},
         year = 2020,
        month = jul,
       volume = {897},
       number = {2},
          eid = {L42},
        pages = {L42},
          doi = {10.3847/2041-8213/aba11c},
archivePrefix = {arXiv},
       eprint = {2007.00205},
 primaryClass = {astro-ph.GA},
       adsurl = {https://ui.adsabs.harvard.edu/abs/2020ApJ...897L..42J}
}

@ARTICLE{kriek16,
       author = {{Kriek}, Mariska and {Conroy}, Charlie and {van Dokkum}, Pieter G. and {Shapley}, Alice E. and {Choi}, Jieun and {Reddy}, Naveen A. and {Siana}, Brian and {van de Voort}, Freeke and {Coil}, Alison L. and {Mobasher}, Bahram},
        title = "{A massive, quiescent, population II galaxy at a redshift of 2.1}",
      journal = {\nat},
         year = 2016,
        month = dec,
       volume = {540},
       number = {7632},
        pages = {248-251},
          doi = {10.1038/nature20570},
archivePrefix = {arXiv},
       eprint = {1612.02001},
 primaryClass = {astro-ph.GA},
       adsurl = {https://ui.adsabs.harvard.edu/abs/2016Natur.540..248K}
}

@ARTICLE{schawinski14,
       author = {{Schawinski}, Kevin and {Urry}, C. Megan and {Simmons}, Brooke D. and {Fortson}, Lucy and {Kaviraj}, Sugata and {Keel}, William C. and {Lintott}, Chris J. and {Masters}, Karen L. and {Nichol}, Robert C. and {Sarzi}, Marc and {Skibba}, Ramin and {Treister}, Ezequiel and {Willett}, Kyle W. and {Wong}, O. Ivy and {Yi}, Sukyoung K.},
        title = "{The green valley is a red herring: Galaxy Zoo reveals two evolutionary pathways towards quenching of star formation in early- and late-type galaxies}",
      journal = {\mnras},
         year = 2014,
        month = may,
       volume = {440},
       number = {1},
        pages = {889-907},
          doi = {10.1093/mnras/stu327},
archivePrefix = {arXiv},
       eprint = {1402.4814},
 primaryClass = {astro-ph.GA},
       adsurl = {https://ui.adsabs.harvard.edu/abs/2014MNRAS.440..889S}
}

@ARTICLE{tacchella22,
       author = {{Tacchella}, Sandro and {Conroy}, Charlie and {Faber}, S.~M. and {Johnson}, Benjamin D. and {Leja}, Joel and {Barro}, Guillermo and {Cunningham}, Emily C. and {Deason}, Alis J. and {Guhathakurta}, Puragra and {Guo}, Yicheng and {Hernquist}, Lars and {Koo}, David C. and {McKinnon}, Kevin and {Rockosi}, Constance M. and {Speagle}, Joshua S. and {van Dokkum}, Pieter and {Yesuf}, Hassen M.},
        title = "{Fast, Slow, Early, Late: Quenching Massive Galaxies at z {\ensuremath{\sim}} 0.8}",
      journal = {\apj},
         year = 2022,
        month = feb,
       volume = {926},
       number = {2},
          eid = {134},
        pages = {134},
          doi = {10.3847/1538-4357/ac449b},
archivePrefix = {arXiv},
       eprint = {2102.12494},
 primaryClass = {astro-ph.GA},
       adsurl = {https://ui.adsabs.harvard.edu/abs/2022ApJ...926..134T}
}

@ARTICLE{man18,
       author = {{Man}, Allison and {Belli}, Sirio},
        title = "{Star formation quenching in massive galaxies}",
      journal = {Nature Astronomy},
         year = 2018,
        month = sep,
       volume = {2},
        pages = {695-697},
          doi = {10.1038/s41550-018-0558-1},
archivePrefix = {arXiv},
       eprint = {1809.00722},
 primaryClass = {astro-ph.GA},
       adsurl = {https://ui.adsabs.harvard.edu/abs/2018NatAs...2..695M}
}

@ARTICLE{park23,
       author = {{Park}, Minjung and {Belli}, Sirio and {Conroy}, Charlie and {Tacchella}, Sandro and {Leja}, Joel and {Cutler}, Sam E. and {Johnson}, Benjamin D. and {Nelson}, Erica J. and {Emami}, Razieh},
        title = "{Rapid Quenching of Galaxies at Cosmic Noon}",
      journal = {\apj},
         year = 2023,
        month = aug,
       volume = {953},
       number = {1},
          eid = {119},
        pages = {119},
          doi = {10.3847/1538-4357/acd54a},
archivePrefix = {arXiv},
       eprint = {2210.03747},
 primaryClass = {astro-ph.GA},
       adsurl = {https://ui.adsabs.harvard.edu/abs/2023ApJ...953..119P}
}

@ARTICLE{glazebrook24,
       author = {{Glazebrook}, Karl and {Nanayakkara}, Themiya and {Schreiber}, Corentin and {Lagos}, Claudia and {Kawinwanichakij}, Lalitwadee and {Jacobs}, Colin and {Chittenden}, Harry and {Brammer}, Gabriel and {Kacprzak}, Glenn G. and {Labbe}, Ivo and {Marchesini}, Danilo and {Marsan}, Z. Cemile and {Oesch}, Pascal A. and {Papovich}, Casey and {Remus}, Rhea-Silvia and {Tran}, Kim-Vy H. and {Esdaile}, James and {Chandro-Gomez}, Angel},
        title = "{A massive galaxy that formed its stars at z {\ensuremath{\approx}} 11}",
      journal = {\nat},
         year = 2024,
        month = apr,
       volume = {628},
       number = {8007},
        pages = {277-281},
          doi = {10.1038/s41586-024-07191-9},
archivePrefix = {arXiv},
       eprint = {2308.05606},
 primaryClass = {astro-ph.GA},
       adsurl = {https://ui.adsabs.harvard.edu/abs/2024Natur.628..277G}
}

@ARTICLE{carnall24,
       author = {{Carnall}, A.~C. and {Cullen}, F. and {McLure}, R.~J. and {McLeod}, D.~J. and {Begley}, R. and {Donnan}, C.~T. and {Dunlop}, J.~S. and {Shapley}, A.~E. and {Rowlands}, K. and {Almaini}, O. and {Arellano-C{\'o}rdova}, K.~Z. and {Barrufet}, L. and {Cimatti}, A. and {Ellis}, R.~S. and {Grogin}, N.~A. and {Hamadouche}, M.~L. and {Illingworth}, G.~D. and {Koekemoer}, A.~M. and {Leung}, H.-H. and {Lovell}, C.~C. and {P{\'e}rez-Gonz{\'a}lez}, P.~G. and {Santini}, P. and {Stanton}, T.~M. and {Wild}, V.},
        title = "{The JWST EXCELS survey: too much, too young, too fast? Ultra-massive quiescent galaxies at 3 < z < 5}",
      journal = {\mnras},
         year = 2024,
        month = oct,
       volume = {534},
       number = {1},
        pages = {325-348},
          doi = {10.1093/mnras/stae2092},
archivePrefix = {arXiv},
       eprint = {2405.02242},
 primaryClass = {astro-ph.GA},
       adsurl = {https://ui.adsabs.harvard.edu/abs/2024MNRAS.534..325C}
}

@ARTICLE{weibel25,
       author = {{Weibel}, Andrea and {de Graaff}, Anna and {Setton}, David J. and {Miller}, Tim B. and {Oesch}, Pascal A. and {Brammer}, Gabriel and {Lagos}, Claudia D.~P. and {Whitaker}, Katherine E. and {Williams}, Christina C. and {Baggen}, Josephine F.~W. and {Bezanson}, Rachel and {Boogaard}, Leindert A. and {Cleri}, Nikko J. and {Greene}, Jenny E. and {Hirschmann}, Michaela and {Hviding}, Raphael E. and {Kuruvanthodi}, Adarsh and {Labb{\'e}}, Ivo and {Leja}, Joel and {Maseda}, Michael V. and {Matthee}, Jorryt and {McConachie}, Ian and {Naidu}, Rohan P. and {Roberts-Borsani}, Guido and {Schaerer}, Daniel and {Suess}, Katherine A. and {Valentino}, Francesco and {van Dokkum}, Pieter and {Wang}, Bingjie},
        title = "{RUBIES Reveals a Massive Quiescent Galaxy at z = 7.3}",
      journal = {\apj},
         year = 2025,
        month = apr,
       volume = {983},
       number = {1},
          eid = {11},
        pages = {11},
          doi = {10.3847/1538-4357/adab7a},
archivePrefix = {arXiv},
       eprint = {2409.03829},
 primaryClass = {astro-ph.GA},
       adsurl = {https://ui.adsabs.harvard.edu/abs/2025ApJ...983...11W}
}

@ARTICLE{wu18,
       author = {{Wu}, Po-Feng and {van der Wel}, Arjen and {Bezanson}, Rachel and {Gallazzi}, Anna and {Pacifici}, Camilla and {Straatman}, Caroline M.~S. and {Bari{\v{s}}i{\'c}}, Ivana and {Bell}, Eric F. and {Chauke}, Priscilla and {van Houdt}, Josha and {Franx}, Marijn and {Muzzin}, Adam and {Sobral}, David and {Wild}, Vivienne},
        title = "{Fast and Slow Paths to Quiescence: Ages and Sizes of 400 Quiescent Galaxies from the LEGA-C Survey}",
      journal = {\apj},
         year = 2018,
        month = nov,
       volume = {868},
       number = {1},
          eid = {37},
        pages = {37},
          doi = {10.3847/1538-4357/aae822},
archivePrefix = {arXiv},
       eprint = {1809.01211},
 primaryClass = {astro-ph.GA},
       adsurl = {https://ui.adsabs.harvard.edu/abs/2018ApJ...868...37W}
}

@ARTICLE{tacchella15,
       author = {{Tacchella}, S. and {Carollo}, C.~M. and {Renzini}, A. and {F{\"o}rster Schreiber}, N.~M. and {Lang}, P. and {Wuyts}, S. and {Cresci}, G. and {Dekel}, A. and {Genzel}, R. and {Lilly}, S.~J. and {Mancini}, C. and {Newman}, S. and {Onodera}, M. and {Shapley}, A. and {Tacconi}, L. and {Woo}, J. and {Zamorani}, G.},
        title = "{Evidence for mature bulges and an inside-out quenching phase 3 billion years after the Big Bang}",
      journal = {Science},
         year = 2015,
        month = apr,
       volume = {348},
       number = {6232},
        pages = {314-317},
          doi = {10.1126/science.1261094},
archivePrefix = {arXiv},
       eprint = {1504.04021},
 primaryClass = {astro-ph.GA},
       adsurl = {https://ui.adsabs.harvard.edu/abs/2015Sci...348..314T}
}

@ARTICLE{nelson16,
       author = {{Nelson}, Erica June and {van Dokkum}, Pieter G. and {F{\"o}rster Schreiber}, Natascha M. and {Franx}, Marijn and {Brammer}, Gabriel B. and {Momcheva}, Ivelina G. and {Wuyts}, Stijn and {Whitaker}, Katherine E. and {Skelton}, Rosalind E. and {Fumagalli}, Mattia and {Hayward}, Christopher C. and {Kriek}, Mariska and {Labb{\'e}}, Ivo and {Leja}, Joel and {Rix}, Hans-Walter and {Tacconi}, Linda J. and {van der Wel}, Arjen and {van den Bosch}, Frank C. and {Oesch}, Pascal A. and {Dickey}, Claire and {Ulf Lange}, Johannes},
        title = "{Where Stars Form: Inside-out Growth and Coherent Star Formation from HST H{\ensuremath{\alpha}} Maps of 3200 Galaxies across the Main Sequence at 0.7 < z < 1.5}",
      journal = {\apj},
         year = 2016,
        month = sep,
       volume = {828},
       number = {1},
          eid = {27},
        pages = {27},
          doi = {10.3847/0004-637X/828/1/27},
archivePrefix = {arXiv},
       eprint = {1507.03999},
 primaryClass = {astro-ph.GA},
       adsurl = {https://ui.adsabs.harvard.edu/abs/2016ApJ...828...27N}
}

@ARTICLE{dekel14,
       author = {{Dekel}, A. and {Burkert}, A.},
        title = "{Wet disc contraction to galactic blue nuggets and quenching to red nuggets}",
      journal = {\mnras},
         year = 2014,
        month = feb,
       volume = {438},
       number = {2},
        pages = {1870-1879},
          doi = {10.1093/mnras/stt2331},
archivePrefix = {arXiv},
       eprint = {1310.1074},
 primaryClass = {astro-ph.CO},
       adsurl = {https://ui.adsabs.harvard.edu/abs/2014MNRAS.438.1870D}
}

@ARTICLE{slob25,
       author = {{Slob}, Martje and {Kriek}, Mariska and {de Graaff}, Anna and {Cheng}, Chloe M. and {Beverage}, Aliza G. and {Bezanson}, Rachel and {F{\"o}rster Schreiber}, Natascha M. and {Lorenz}, Brian and {Mancera Pi{\~n}a}, Pavel E. and {Marchesini}, Danilo and {Muzzin}, Adam and {Newman}, Andrew B. and {Price}, Sedona H. and {Suess}, Katherine A. and {van de Sande}, Jesse and {van Dokkum}, Pieter and {Weisz}, Daniel R.},
        title = "{Fast rotators at cosmic noon: Stellar kinematics for 15 quiescent galaxies from JWST-SUSPENSE}",
      journal = {\aap},
         year = 2025,
        month = oct,
       volume = {702},
          eid = {A110},
        pages = {A110},
          doi = {10.1051/0004-6361/202555812},
archivePrefix = {arXiv},
       eprint = {2506.04310},
 primaryClass = {astro-ph.GA},
       adsurl = {https://ui.adsabs.harvard.edu/abs/2025A&A...702A.110S}
}

@ARTICLE{newman18,
       author = {{Newman}, Andrew B. and {Belli}, Sirio and {Ellis}, Richard S. and {Patel}, Shannon G.},
        title = "{Resolving Quiescent Galaxies at z {\ensuremath{\gtrsim}} 2. II. Direct Measures of Rotational Support}",
      journal = {\apj},
         year = 2018,
        month = aug,
       volume = {862},
       number = {2},
          eid = {126},
        pages = {126},
          doi = {10.3847/1538-4357/aacd4f},
archivePrefix = {arXiv},
       eprint = {1806.06815},
 primaryClass = {astro-ph.GA},
       adsurl = {https://ui.adsabs.harvard.edu/abs/2018ApJ...862..126N}
}

@ARTICLE{peng10,
       author = {{Peng}, Ying-jie and {Lilly}, Simon J. and {Kova{\v{c}}}, Katarina and {Bolzonella}, Micol and {Pozzetti}, Lucia and {Renzini}, Alvio and {Zamorani}, Gianni and {Ilbert}, Olivier and {Knobel}, Christian and {Iovino}, Angela and {Maier}, Christian and {Cucciati}, Olga and {Tasca}, Lidia and {Carollo}, C. Marcella and {Silverman}, John and {Kampczyk}, Pawel and {de Ravel}, Loic and {Sanders}, David and {Scoville}, Nicholas and {Contini}, Thierry and {Mainieri}, Vincenzo and {Scodeggio}, Marco and {Kneib}, Jean-Paul and {Le F{\`e}vre}, Olivier and {Bardelli}, Sandro and {Bongiorno}, Angela and {Caputi}, Karina and {Coppa}, Graziano and {de la Torre}, Sylvain and {Franzetti}, Paolo and {Garilli}, Bianca and {Lamareille}, Fabrice and {Le Borgne}, Jean-Francois and {Le Brun}, Vincent and {Mignoli}, Marco and {Perez Montero}, Enrique and {Pello}, Roser and {Ricciardelli}, Elena and {Tanaka}, Masayuki and {Tresse}, Laurence and {Vergani}, Daniela and {Welikala}, Niraj and {Zucca}, Elena and {Oesch}, Pascal and {Abbas}, Ummi and {Barnes}, Luke and {Bordoloi}, Rongmon and {Bottini}, Dario and {Cappi}, Alberto and {Cassata}, Paolo and {Cimatti}, Andrea and {Fumana}, Marco and {Hasinger}, Gunther and {Koekemoer}, Anton and {Leauthaud}, Alexei and {Maccagni}, Dario and {Marinoni}, Christian and {McCracken}, Henry and {Memeo}, Pierdomenico and {Meneux}, Baptiste and {Nair}, Preethi and {Porciani}, Cristiano and {Presotto}, Valentina and {Scaramella}, Roberto},
        title = "{Mass and Environment as Drivers of Galaxy Evolution in SDSS and zCOSMOS and the Origin of the Schechter Function}",
      journal = {\apj},
         year = 2010,
        month = sep,
       volume = {721},
       number = {1},
        pages = {193-221},
          doi = {10.1088/0004-637X/721/1/193},
archivePrefix = {arXiv},
       eprint = {1003.4747},
 primaryClass = {astro-ph.CO},
       adsurl = {https://ui.adsabs.harvard.edu/abs/2010ApJ...721..193P}
}

@ARTICLE{kimmig25,
       author = {{Kimmig}, Lucas C. and {Remus}, Rhea-Silvia and {Seidel}, Benjamin and {Valenzuela}, Lucas M. and {Dolag}, Klaus and {Burkert}, Andreas},
        title = "{Blowing Out the Candle: How to Quench Galaxies at High Redshift{\textemdash}An Ensemble of Rapid Starbursts, AGN Feedback, and Environment}",
      journal = {\apj},
         year = 2025,
        month = jan,
       volume = {979},
       number = {1},
          eid = {15},
        pages = {15},
          doi = {10.3847/1538-4357/ad9472},
archivePrefix = {arXiv},
       eprint = {2310.16085},
 primaryClass = {astro-ph.GA},
       adsurl = {https://ui.adsabs.harvard.edu/abs/2025ApJ...979...15K}
}

@ARTICLE{delucia26,
       author = {{De Lucia}, Gabriella and {Xie}, Lizhi and {Hirschmann}, Michaela and {Fontanot}, Fabio},
        title = "{Where giants dwell: Probing the environments of early massive quiescent galaxies}",
      journal = {\aap},
         year = 2026,
        month = mar,
       volume = {708},
          eid = {A54},
        pages = {A54},
          doi = {10.1051/0004-6361/202557912},
archivePrefix = {arXiv},
       eprint = {2511.08674},
 primaryClass = {astro-ph.GA},
       adsurl = {https://ui.adsabs.harvard.edu/abs/2026A&A...708A..54D}
}

@ARTICLE{weller25,
       author = {{Weller}, Emma Jane and {Pacucci}, Fabio and {Ni}, Yueying and {Hernquist}, Lars and {Park}, Minjung},
        title = "{Discrepancies between JWST Observations and Simulations of Quenched Massive Galaxies at z > 3: A Comparative Study with IllustrisTNG and ASTRID}",
      journal = {\apj},
         year = 2025,
        month = feb,
       volume = {979},
       number = {2},
          eid = {181},
        pages = {181},
          doi = {10.3847/1538-4357/ada360},
archivePrefix = {arXiv},
       eprint = {2406.02664},
 primaryClass = {astro-ph.GA},
       adsurl = {https://ui.adsabs.harvard.edu/abs/2025ApJ...979..181W}
}

@ARTICLE{fossati17,
       author = {{Fossati}, M. and {Wilman}, D.~J. and {Mendel}, J.~T. and {Saglia}, R.~P. and {Galametz}, A. and {Beifiori}, A. and {Bender}, R. and {Chan}, J.~C.~C. and {Fabricius}, M. and {Bandara}, K. and {Brammer}, G.~B. and {Davies}, R. and {F{\"o}rster Schreiber}, N.~M. and {Genzel}, R. and {Hartley}, W. and {Kulkarni}, S.~K. and {Lang}, P. and {Momcheva}, I.~G. and {Nelson}, E.~J. and {Skelton}, R. and {Tacconi}, L.~J. and {Tadaki}, K. and {{\"U}bler}, H. and {van Dokkum}, P.~G. and {Wisnioski}, E. and {Whitaker}, K.~E. and {Wuyts}, E. and {Wuyts}, S.},
        title = "{Galaxy Environment in the 3D-HST Fields: Witnessing the Onset of Satellite Quenching at z {\ensuremath{\sim}} 1-2}",
      journal = {\apj},
         year = 2017,
        month = feb,
       volume = {835},
       number = {2},
          eid = {153},
        pages = {153},
          doi = {10.3847/1538-4357/835/2/153},
archivePrefix = {arXiv},
       eprint = {1611.07524},
 primaryClass = {astro-ph.GA},
       adsurl = {https://ui.adsabs.harvard.edu/abs/2017ApJ...835..153F}
}

@ARTICLE{hatamnia25,
       author = {{Hatamnia}, Hossein and {Mobasher}, Bahram and {Taamoli}, Sina and {Kartaltepe}, Jeyhan S. and {Casey}, Caitlin M. and {Akins}, Hollis B. and {Brinch}, Malte and {Chartab}, Nima and {Drakos}, Nicole E. and {Faisst}, Andreas L. and {Finkelstein}, Steven L. and {Franco}, Maximilien and {Giddings}, Finn and {Gozaliasl}, Ghassem and {Hadi}, Ali and {Haghjoo}, Aryana and {Harish}, Santosh and {Ilbert}, Olivier and {Jablonka}, Pascale L. and {Jin}, Shuowen and {Khostovan}, Ali Ahmad and {Koekemoer}, Anton M. and {Laishram}, Ronaldo and {Liu}, Daizhong and {Maturi}, Matteo and {McCracken}, Henry Joy and {Martin}, Crystal L. and {Moscardini}, Lauro and {Scognamiglio}, Diana and {Shuntov}, Marko and {Toni}, Greta and {de la Vega}, Alexander and {Weaver}, John R. and {Yang}, Lilan},
        title = "{Large-scale Structure in COSMOS-Web: Tracing Galaxy Evolution in the Cosmic Web up to z {\ensuremath{\sim}} 7 with the Largest JWST Survey}",
      journal = {\apj},
         year = 2026,
        month = may,
       volume = {1002},
       number = {2},
          eid = {192},
        pages = {192},
          doi = {10.3847/1538-4357/ae5bac},
archivePrefix = {arXiv},
       eprint = {2511.10727},
 primaryClass = {astro-ph.GA},
       adsurl = {https://ui.adsabs.harvard.edu/abs/2026ApJ..1002..192H}
}

@ARTICLE{newman25,
       author = {{Newman}, Andrew B. and {Rudie}, Gwen C. and {Blanc}, Guillermo A. and {Kelson}, Daniel D. and {Chartab}, Nima and {Congiu}, Enrico and {P{\'e}rez}, Victoria and {Qezlou}, Mahdi and {Bird}, Simeon and {Lemaux}, Brian C. and {Cucciati}, Olga},
        title = "{LATIS Data Release: {\ensuremath{\sim}}4200 Spectra of z {\ensuremath{\sim}} 2─3 Galaxies, Redshifts, and Intergalactic Medium Tomography Maps}",
      journal = {\apjs},
         year = 2025,
        month = dec,
       volume = {281},
       number = {2},
          eid = {32},
        pages = {32},
          doi = {10.3847/1538-4365/ae0cc5},
archivePrefix = {arXiv},
       eprint = {2510.08815},
 primaryClass = {astro-ph.GA},
       adsurl = {https://ui.adsabs.harvard.edu/abs/2025ApJS..281...32N}
}

@ARTICLE{harrison17,
       author = {{Harrison}, C.~M.},
        title = "{Impact of supermassive black hole growth on star formation}",
      journal = {Nature Astronomy},
         year = 2017,
        month = jul,
       volume = {1},
          eid = {0165},
        pages = {0165},
          doi = {10.1038/s41550-017-0165},
archivePrefix = {arXiv},
       eprint = {1703.06889},
 primaryClass = {astro-ph.GA},
       adsurl = {https://ui.adsabs.harvard.edu/abs/2017NatAs...1E.165H}
}

@ARTICLE{concas22,
       author = {{Concas}, Alice and {Maiolino}, Roberto and {Curti}, Mirko and {Hayden-Pawson}, Connor and {Cirasuolo}, Michele and {Jones}, Gareth C. and {Mercurio}, Amata and {Belfiore}, Francesco and {Cresci}, Giovanni and {Cullen}, Fergus and {Mannucci}, Filippo and {Marconi}, Alessandro and {Cappellari}, Michele and {Cicone}, Claudia and {Peng}, Yingjie and {Troncoso}, Paulina},
        title = "{Being KLEVER at cosmic noon: Ionized gas outflows are inconspicuous in low-mass star-forming galaxies but prominent in massive AGN hosts}",
      journal = {\mnras},
         year = 2022,
        month = jun,
       volume = {513},
       number = {2},
        pages = {2535-2562},
          doi = {10.1093/mnras/stac1026},
archivePrefix = {arXiv},
       eprint = {2203.11958},
 primaryClass = {astro-ph.GA},
       adsurl = {https://ui.adsabs.harvard.edu/abs/2022MNRAS.513.2535C}
}

@ARTICLE{veilleux20,
       author = {{Veilleux}, Sylvain and {Maiolino}, Roberto and {Bolatto}, Alberto D. and {Aalto}, Susanne},
        title = "{Cool outflows in galaxies and their implications}",
      journal = {\aapr},
         year = 2020,
        month = apr,
       volume = {28},
       number = {1},
          eid = {2},
        pages = {2},
          doi = {10.1007/s00159-019-0121-9},
archivePrefix = {arXiv},
       eprint = {2002.07765},
 primaryClass = {astro-ph.GA},
       adsurl = {https://ui.adsabs.harvard.edu/abs/2020A&ARv..28....2V}
}

@ARTICLE{steidel96,
       author = {{Steidel}, Charles C. and {Giavalisco}, Mauro and {Pettini}, Max and {Dickinson}, Mark and {Adelberger}, Kurt L.},
        title = "{Spectroscopic Confirmation of a Population of Normal Star-forming Galaxies at Redshifts Z > 3}",
      journal = {\apjl},
         year = 1996,
        month = may,
       volume = {462},
        pages = {L17},
          doi = {10.1086/310029},
archivePrefix = {arXiv},
       eprint = {astro-ph/9602024},
 primaryClass = {astro-ph},
       adsurl = {https://ui.adsabs.harvard.edu/abs/1996ApJ...462L..17S}
}

@ARTICLE{kriek06,
       author = {{Kriek}, Mariska and {van Dokkum}, Pieter G. and {Franx}, Marijn and {Quadri}, Ryan and {Gawiser}, Eric and {Herrera}, David and {Illingworth}, Garth D. and {Labb{\'e}}, Ivo and {Lira}, Paulina and {Marchesini}, Danilo and {Rix}, Hans-Walter and {Rudnick}, Gregory and {Taylor}, Edward N. and {Toft}, Sune and {Urry}, C. Megan and {Wuyts}, Stijn},
        title = "{Spectroscopic Identification of Massive Galaxies at z \raisebox{-0.5ex}\textasciitilde 2.3 with Strongly Suppressed Star Formation}",
      journal = {\apjl},
         year = 2006,
        month = oct,
       volume = {649},
       number = {2},
        pages = {L71-L74},
          doi = {10.1086/508371},
archivePrefix = {arXiv},
       eprint = {astro-ph/0608446},
 primaryClass = {astro-ph},
       adsurl = {https://ui.adsabs.harvard.edu/abs/2006ApJ...649L..71K}
}

@ARTICLE{vandokkum09,
       author = {{van Dokkum}, Pieter G. and {Kriek}, Mariska and {Franx}, Marijn},
        title = "{A high stellar velocity dispersion for a compact massive galaxy at redshift z = 2.186}",
      journal = {\nat},
         year = 2009,
        month = aug,
       volume = {460},
       number = {7256},
        pages = {717-719},
          doi = {10.1038/nature08220},
archivePrefix = {arXiv},
       eprint = {0906.2778},
 primaryClass = {astro-ph.CO},
       adsurl = {https://ui.adsabs.harvard.edu/abs/2009Natur.460..717V}
}

@ARTICLE{li25,
       author = {{Li}, Yijia and {Leja}, Joel and {Johnson}, Benjamin D. and {Tacchella}, Sandro and {Davies}, Rebecca and {Belli}, Sirio and {Park}, Minjung and {Emami}, Razieh},
        title = "{Cue: A Fast and Flexible Photoionization Emulator for Modeling Nebular Emission Powered by Almost Any Ionizing Source}",
      journal = {\apj},
         year = 2025,
        month = jun,
       volume = {986},
       number = {1},
          eid = {9},
        pages = {9},
          doi = {10.3847/1538-4357/adcab4},
archivePrefix = {arXiv},
       eprint = {2405.04598},
 primaryClass = {astro-ph.GA},
       adsurl = {https://ui.adsabs.harvard.edu/abs/2025ApJ...986....9L}
}

@ARTICLE{kewley19,
       author = {{Kewley}, Lisa J. and {Nicholls}, David C. and {Sutherland}, Ralph S.},
        title = "{Understanding Galaxy Evolution Through Emission Lines}",
      journal = {\araa},
         year = 2019,
        month = aug,
       volume = {57},
        pages = {511-570},
          doi = {10.1146/annurev-astro-081817-051832},
archivePrefix = {arXiv},
       eprint = {1910.09730},
 primaryClass = {astro-ph.GA},
       adsurl = {https://ui.adsabs.harvard.edu/abs/2019ARA&A..57..511K}
}

@ARTICLE{sanders24,
       author = {{Sanders}, Ryan L. and {Shapley}, Alice E. and {Topping}, Michael W. and {Reddy}, Naveen A. and {Brammer}, Gabriel B.},
        title = "{Direct T $_{e}$-based Metallicities of z = 2─9 Galaxies with JWST/NIRSpec: Empirical Metallicity Calibrations Applicable from Reionization to Cosmic Noon}",
      journal = {\apj},
         year = 2024,
        month = feb,
       volume = {962},
       number = {1},
          eid = {24},
        pages = {24},
          doi = {10.3847/1538-4357/ad15fc},
archivePrefix = {arXiv},
       eprint = {2303.08149},
 primaryClass = {astro-ph.GA},
       adsurl = {https://ui.adsabs.harvard.edu/abs/2024ApJ...962...24S}
}

@ARTICLE{reddy20,
       author = {{Reddy}, Naveen A. and {Shapley}, Alice E. and {Kriek}, Mariska and {Steidel}, Charles C. and {Shivaei}, Irene and {Sanders}, Ryan L. and {Mobasher}, Bahram and {Coil}, Alison L. and {Siana}, Brian and {Freeman}, William R. and {Azadi}, Mojegan and {Fetherolf}, Tara and {Leung}, Gene and {Price}, Sedona H. and {Zick}, Tom},
        title = "{The MOSDEF Survey: The First Direct Measurements of the Nebular Dust Attenuation Curve at High Redshift}",
      journal = {\apj},
         year = 2020,
        month = oct,
       volume = {902},
       number = {2},
          eid = {123},
        pages = {123},
          doi = {10.3847/1538-4357/abb674},
archivePrefix = {arXiv},
       eprint = {2009.10085},
 primaryClass = {astro-ph.GA},
       adsurl = {https://ui.adsabs.harvard.edu/abs/2020ApJ...902..123R}
}

@ARTICLE{salim20,
       author = {{Salim}, Samir and {Narayanan}, Desika},
        title = "{The Dust Attenuation Law in Galaxies}",
      journal = {\araa},
         year = 2020,
        month = aug,
       volume = {58},
        pages = {529-575},
          doi = {10.1146/annurev-astro-032620-021933},
archivePrefix = {arXiv},
       eprint = {2001.03181},
 primaryClass = {astro-ph.GA},
       adsurl = {https://ui.adsabs.harvard.edu/abs/2020ARA&A..58..529S}
}

@ARTICLE{simmonds25,
       author = {{Simmonds}, C. and {Tacchella}, S. and {McClymont}, W. and {Curtis-Lake}, E. and {D'Eugenio}, F. and {Hainline}, K. and {Johnson}, B.~D. and {Kravtsov}, A. and {Pusk{\'a}s}, D. and {Robertson}, B. and {Stoffers}, A. and {Willott}, C. and {Baker}, W.~M. and {Belokurov}, V.~A. and {Bhatawdekar}, R. and {Bunker}, A.~J. and {Carniani}, S. and {Chevallard}, J. and {Curti}, M. and {Duan}, Q. and {Helton}, J.~M. and {Ji}, Z. and {Looser}, T.~J. and {Maiolino}, R. and {Maseda}, M.~V. and {Shivaei}, I. and {Williams}, C.~C.},
        title = "{Bursting at the seams: the star-forming main sequence and its scatter at z = 3─9 using NIRCam photometry from JADES}",
      journal = {\mnras},
         year = 2025,
        month = dec,
       volume = {544},
       number = {4},
        pages = {4551-4575},
          doi = {10.1093/mnras/staf1950},
archivePrefix = {arXiv},
       eprint = {2508.04410},
 primaryClass = {astro-ph.GA},
       adsurl = {https://ui.adsabs.harvard.edu/abs/2025MNRAS.544.4551S}
}

@ARTICLE{noeske07,
       author = {{Noeske}, K.~G. and {Weiner}, B.~J. and {Faber}, S.~M. and {Papovich}, C. and {Koo}, D.~C. and {Somerville}, R.~S. and {Bundy}, K. and {Conselice}, C.~J. and {Newman}, J.~A. and {Schiminovich}, D. and {Le Floc'h}, E. and {Coil}, A.~L. and {Rieke}, G.~H. and {Lotz}, J.~M. and {Primack}, J.~R. and {Barmby}, P. and {Cooper}, M.~C. and {Davis}, M. and {Ellis}, R.~S. and {Fazio}, G.~G. and {Guhathakurta}, P. and {Huang}, J. and {Kassin}, S.~A. and {Martin}, D.~C. and {Phillips}, A.~C. and {Rich}, R.~M. and {Small}, T.~A. and {Willmer}, C.~N.~A. and {Wilson}, G.},
        title = "{Star Formation in AEGIS Field Galaxies since z=1.1: The Dominance of Gradually Declining Star Formation, and the Main Sequence of Star-forming Galaxies}",
      journal = {\apjl},
         year = 2007,
        month = may,
       volume = {660},
       number = {1},
        pages = {L43-L46},
          doi = {10.1086/517926},
archivePrefix = {arXiv},
       eprint = {astro-ph/0701924},
 primaryClass = {astro-ph},
       adsurl = {https://ui.adsabs.harvard.edu/abs/2007ApJ...660L..43N}
}

@ARTICLE{tacchella16,
       author = {{Tacchella}, Sandro and {Dekel}, Avishai and {Carollo}, C. Marcella and {Ceverino}, Daniel and {DeGraf}, Colin and {Lapiner}, Sharon and {Mandelker}, Nir and {Primack Joel}, R.},
        title = "{The confinement of star-forming galaxies into a main sequence through episodes of gas compaction, depletion and replenishment}",
      journal = {\mnras},
         year = 2016,
        month = apr,
       volume = {457},
       number = {3},
        pages = {2790-2813},
          doi = {10.1093/mnras/stw131},
archivePrefix = {arXiv},
       eprint = {1509.02529},
 primaryClass = {astro-ph.GA},
       adsurl = {https://ui.adsabs.harvard.edu/abs/2016MNRAS.457.2790T}
}

@ARTICLE{nomoto13,
       author = {{Nomoto}, Ken'ichi and {Kobayashi}, Chiaki and {Tominaga}, Nozomu},
        title = "{Nucleosynthesis in Stars and the Chemical Enrichment of Galaxies}",
      journal = {\araa},
         year = 2013,
        month = aug,
       volume = {51},
       number = {1},
        pages = {457-509},
          doi = {10.1146/annurev-astro-082812-140956},
       adsurl = {https://ui.adsabs.harvard.edu/abs/2013ARA&A..51..457N}
}

@ARTICLE{thomas05,
       author = {{Thomas}, Daniel and {Maraston}, Claudia and {Bender}, Ralf and {Mendes de Oliveira}, Claudia},
        title = "{The Epochs of Early-Type Galaxy Formation as a Function of Environment}",
      journal = {\apj},
         year = 2005,
        month = mar,
       volume = {621},
       number = {2},
        pages = {673-694},
          doi = {10.1086/426932},
archivePrefix = {arXiv},
       eprint = {astro-ph/0410209},
 primaryClass = {astro-ph},
       adsurl = {https://ui.adsabs.harvard.edu/abs/2005ApJ...621..673T}
}

@ARTICLE{vanderwel16,
       author = {{van der Wel}, A. and {Noeske}, K. and {Bezanson}, R. and {Pacifici}, C. and {Gallazzi}, A. and {Franx}, M. and {Mu{\~n}oz-Mateos}, J.~C. and {Bell}, E.~F. and {Brammer}, G. and {Charlot}, S. and {Chauk{\'e}}, P. and {Labb{\'e}}, I. and {Maseda}, M.~V. and {Muzzin}, A. and {Rix}, H.-W. and {Sobral}, D. and {van de Sande}, J. and {van Dokkum}, P.~G. and {Wild}, V. and {Wolf}, C.},
        title = "{The VLT LEGA-C Spectroscopic Survey: The Physics of Galaxies at a Lookback Time of 7 Gyr}",
      journal = {\apjs},
         year = 2016,
        month = apr,
       volume = {223},
       number = {2},
          eid = {29},
        pages = {29},
          doi = {10.3847/0067-0049/223/2/29},
archivePrefix = {arXiv},
       eprint = {1603.05479},
 primaryClass = {astro-ph.GA},
       adsurl = {https://ui.adsabs.harvard.edu/abs/2016ApJS..223...29V}
}

@ARTICLE{strauss02,
       author = {{Strauss}, Michael A. and {Weinberg}, David H. and {Lupton}, Robert H. and {Narayanan}, Vijay K. and {Annis}, James and {Bernardi}, Mariangela and {Blanton}, Michael and {Burles}, Scott and {Connolly}, A.~J. and {Dalcanton}, Julianne and {Doi}, Mamoru and {Eisenstein}, Daniel and {Frieman}, Joshua A. and {Fukugita}, Masataka and {Gunn}, James E. and {Ivezi{\'c}}, {\v{Z}}eljko and {Kent}, Stephen and {Kim}, Rita S.~J. and {Knapp}, G.~R. and {Kron}, Richard G. and {Munn}, Jeffrey A. and {Newberg}, Heidi Jo and {Nichol}, R.~C. and {Okamura}, Sadanori and {Quinn}, Thomas R. and {Richmond}, Michael W. and {Schlegel}, David J. and {Shimasaku}, Kazuhiro and {SubbaRao}, Mark and {Szalay}, Alexander S. and {Vanden Berk}, Dan and {Vogeley}, Michael S. and {Yanny}, Brian and {Yasuda}, Naoki and {York}, Donald G. and {Zehavi}, Idit},
        title = "{Spectroscopic Target Selection in the Sloan Digital Sky Survey: The Main Galaxy Sample}",
      journal = {\aj},
         year = 2002,
        month = sep,
       volume = {124},
       number = {3},
        pages = {1810-1824},
          doi = {10.1086/342343},
archivePrefix = {arXiv},
       eprint = {astro-ph/0206225},
 primaryClass = {astro-ph},
       adsurl = {https://ui.adsabs.harvard.edu/abs/2002AJ....124.1810S}
}

@ARTICLE{vanderwel25,
       author = {{van der Wel}, Arjen and {Meidt}, Sharon E.},
        title = "{Transforming Galaxies with EASE: Widespread structural changes enabled by short-lived spirals}",
      journal = {\aap},
         year = 2025,
        month = dec,
       volume = {704},
          eid = {A147},
        pages = {A147},
          doi = {10.1051/0004-6361/202556758},
archivePrefix = {arXiv},
       eprint = {2509.02847},
 primaryClass = {astro-ph.GA},
       adsurl = {https://ui.adsabs.harvard.edu/abs/2025A&A...704A.147V}
}

%%%%%%%%%%%%%%%%%%%%%%%%%%%%%%%%%%%%%%%%%%

\begin{appendix}

\section{Stellar mass completeness of the parent sample}
\label{sec:Mmin}

\begin{figure*}[]
\includegraphics[width=\textwidth]{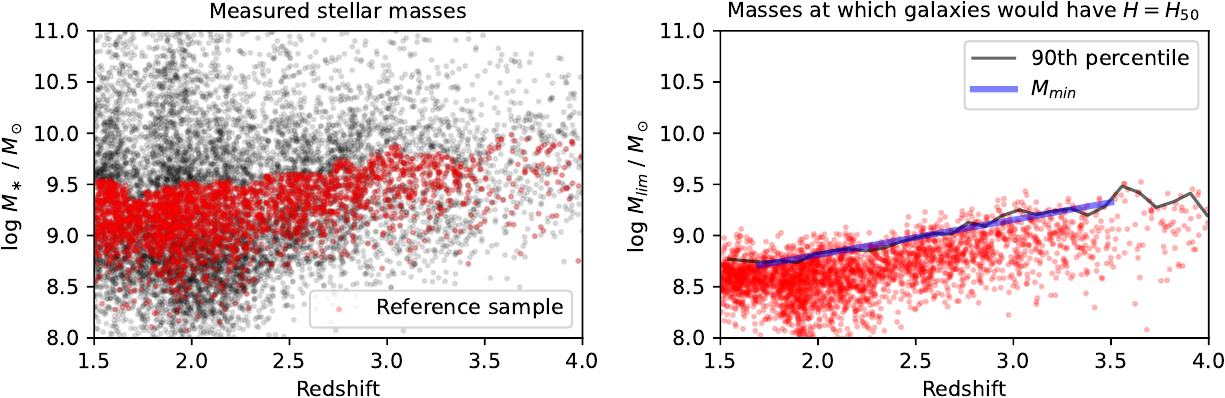}
\caption{Assessing the mass completeness for the parent sample. \textit{Left:} Stellar mass versus redshift for the parent sample. The reference sample, shown in red, is defined as the least massive half of all galaxies that, at a given redshift, are brighter than $H_{90}$. \textit{Right:} $M_\mathrm{lim}$ versus redshift for the reference sample. For each galaxy we define $M_\mathrm{lim}$ as the mass it would need to have in order for its $H$-band magnitude to equal $H_{50}$, assuming that redshift and mass-to-light ratio are unchanged. The blue line marks $M_\mathrm{min}$, which is the top 90th percentile of the $M_\mathrm{lim}$ distribution at each redshift.
}
\label{fig:Mmin}
\end{figure*}

Since we selected targets based on their stellar masses, we need to translate the completeness of the parent catalog from $H$-band magnitude to stellar mass. The mass completeness will depend on redshift and on the mass-to-light ratio distribution of the galaxies that are not observed. We derived this mass-to-light ratio distribution from a reference sample of galaxies selected in the following way: we first excluded objects fainter than $H_{90}$, since they are potentially affected by incompleteness, and then we took, at each redshift, the least massive half of the remaining galaxies. This reference sample, shown in red in the left panel of Fig.~\ref{fig:Mmin}, is bright enough to be unbiased, but also low-mass enough to be representative of the galaxies that are missed by the parent catalog.

Following the method developed by \citet{pozzetti10}, for each galaxy in the reference sample we calculated its limiting mass, defined as the stellar mass it would need to have to be observed exactly at $H_{50}$, assuming the same redshift and mass-to-light ratio. The limiting mass is given by $\log M_\mathrm{lim} = \log M_\ast - 0.4\cdot (H_{50} - H)$. 
The right panel of Fig.~\ref{fig:Mmin} shows the limiting masses for the reference sample: at each redshift they span a range of about one order of magnitude, reflecting the underlying range in mass-to-light ratios. We defined the mass completeness level as the top 90th percentile of the $M_\mathrm{lim}$ distribution as a function of redshift (black line). Within the redshift limits of Blue Jay, the mass completeness is well represented by a linear fit: $\log M_\mathrm{min}/M_\odot = 8.13 + 0.34 z$ (blue line). 

Since we used $H_{50}$ to calculate $M_\mathrm{min}$, the resulting target sample is slightly biased near the mass completeness limit. A more rigorous completeness limit would need to be calculated using $H_{90}$, yielding $\log M^{90}_\mathrm{min} = \log M_\ast - 0.4\cdot (H_{90} - H) = \log M_\mathrm{min} + 0.4 \cdot (H_{50}-H_{90}) = \log M_\mathrm{min} + 0.36$. To assess the importance of incompleteness, we analyzed the properties of galaxies with masses below $M^{90}_\mathrm{min}$ and above the completeness limit. We found that half of these galaxies are brighter than $H_{90}$, and the vast majority of them are much brighter than $H_{50}$. As a result, our parent catalog misses only 15\% of the galaxies with  $M_\mathrm{min} < M_\ast < M^{90}_\mathrm{min}$.

\section{High-redshift filler targets}
\label{sec:hiz}

\begin{figure*}[]
\includegraphics[width=\textwidth]{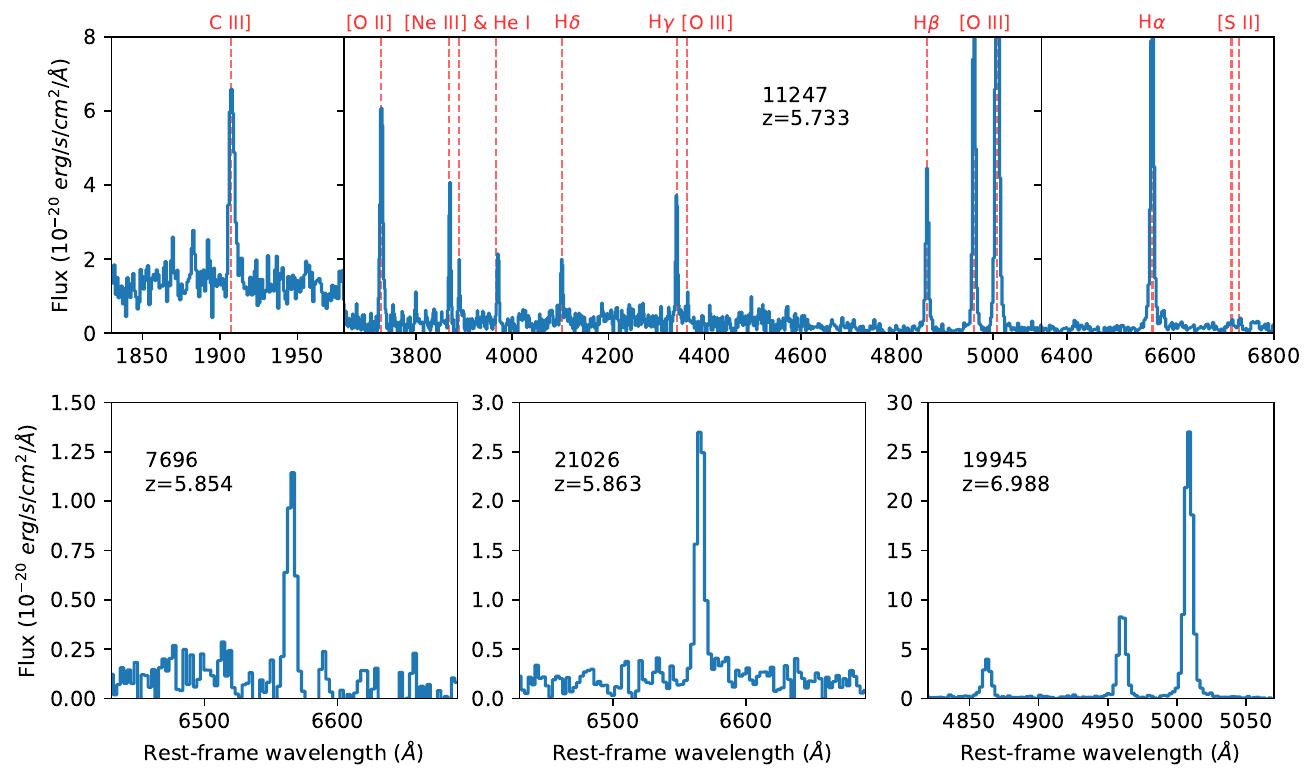}
\caption{Spectra for the four filler targets at high redshifts ($z\sim6-7$). \textit{Top panels}: Spectrum of 11247 with the position of strong emission lines marked, from the rest-UV to the rest-optical. \textit{Bottom panels}: H$\alpha$ emission line (or, in the case of 19945,  H$\beta$ and [O~III]) for each of the other targets.
}
\label{fig:hiz}
\end{figure*}

Four filler targets out of 153 targets were selected with a dropout technique, and were therefore removed from the main Blue Jay sample. Multiple emission lines are detected for each of these four galaxies, yielding spectroscopic redshifts in the range $5.7 < z < 7$, in agreement with the photometric redshifts estimated through the dropout selection. Parts of the spectra of the four fillers are displayed in Fig.~\ref{fig:hiz}. The top panels show a dozen emission lines, from C~III]~1909 to H$\alpha$, detected for galaxy 11247 across the full wavelength range probed by the three NIRSpec gratings.
The bottom panels show the H$\alpha$ emission for 7696 and 21026; and the H$\beta$ and [O~III] emission for 19945, for which H$\alpha$ falls outside the NIRSpec wavelength range due to the slightly higher redshift.

\section{Blue Jay spectra}
\label{sec:spectra}

\begin{figure*}[htbp]
\centering
\includegraphics[width=0.95\textwidth]{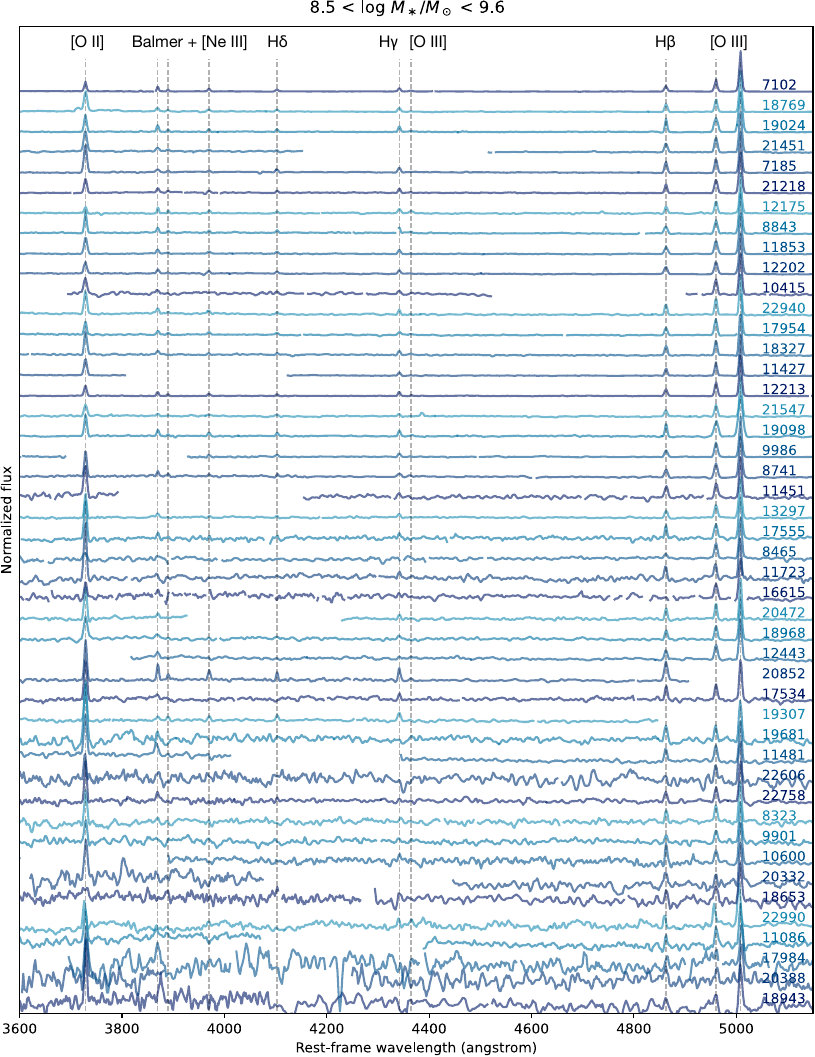}
\caption{Spectra of Blue Jay galaxies with low mass, $8.5 < \log M_\star/M_\odot < 9.6$, sorted by emission line strength. Each spectrum is normalized to have identical peak of the [O~III]$\lambda5008$ emission line. Only part of the rest-frame optical range is shown; some prominent emission lines are marked.
}
\label{fig:spectra1}
\end{figure*}

\begin{figure*}[htbp]
\centering
\includegraphics[width=0.95\textwidth]{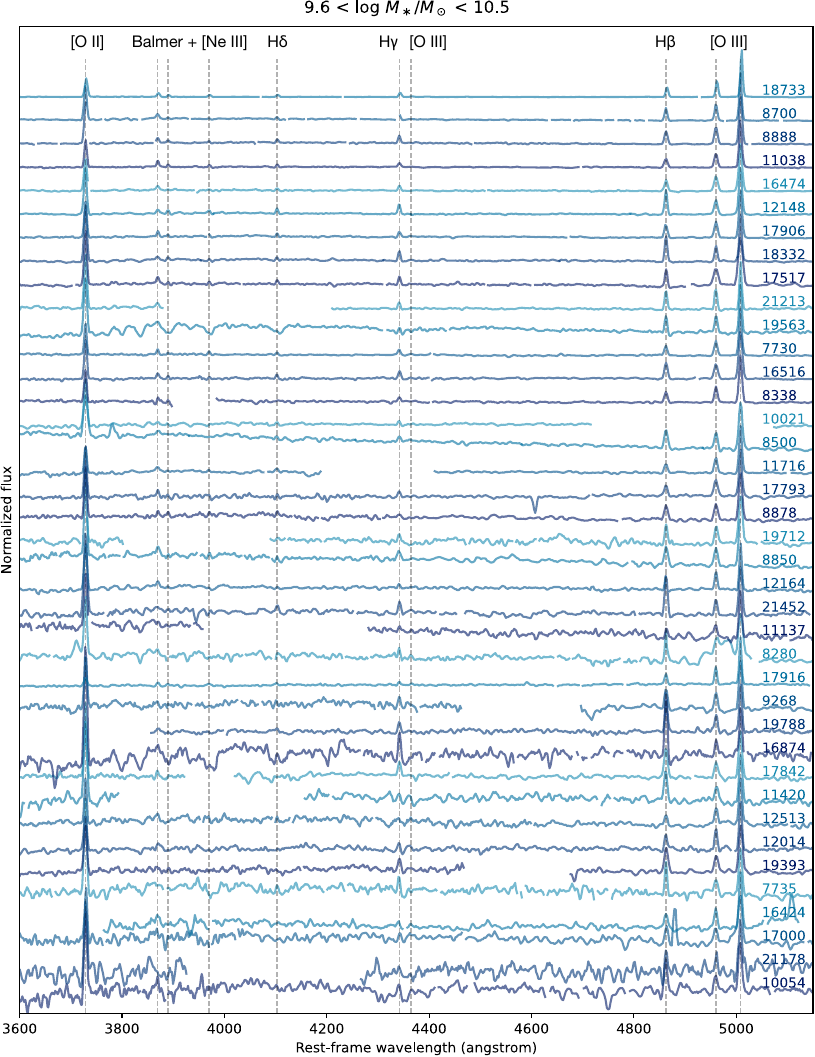}
\caption{Spectra of Blue Jay galaxies with intermediate mass, $9.6 < \log M_\star/M_\odot < 10.5$, sorted by emission line strength. Each spectrum is normalized to have identical peak of the [O~III]$\lambda5008$ emission line. Only part of the rest-frame optical range is shown; some prominent emission lines are marked.
}
\label{fig:spectra2}
\end{figure*}

\begin{figure*}[htbp]
\centering
\includegraphics[width=0.95\textwidth]{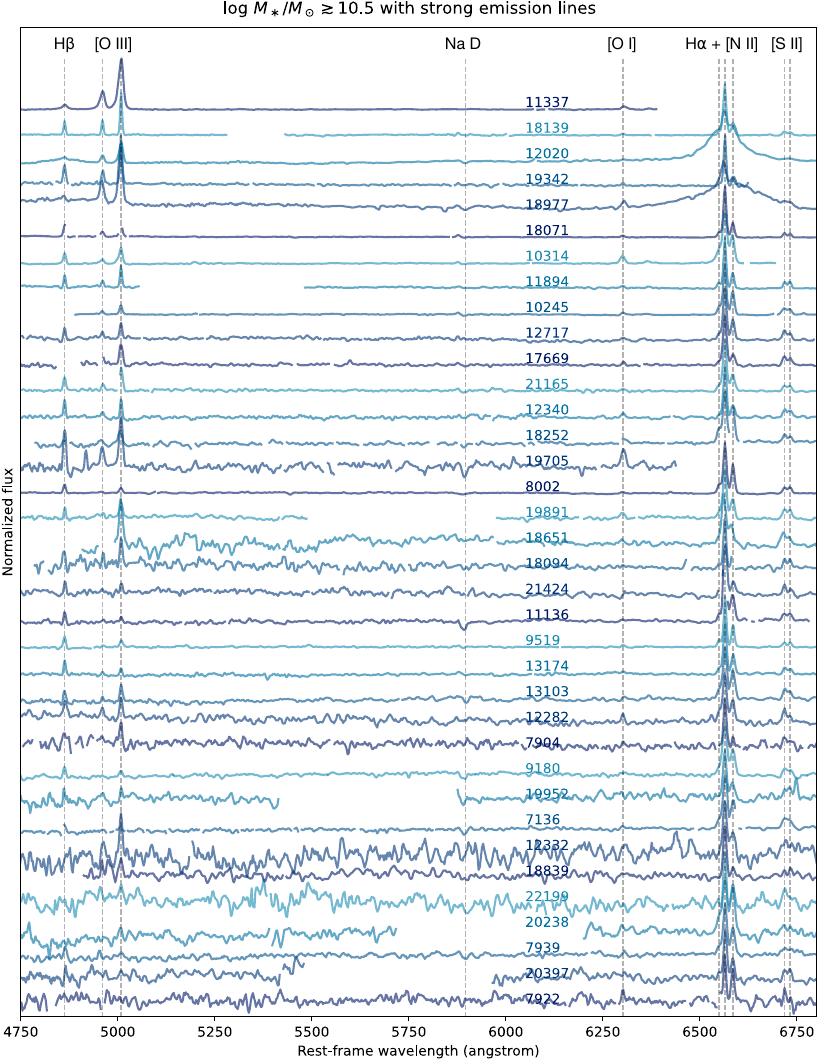}
\caption{Spectra of massive Blue Jay galaxies with strong emission lines, sorted by emission line strength. Each spectrum is normalized to have identical peak of the H$\alpha$ emission line. Only part of the rest-frame optical range is shown; some prominent emission lines are marked.
}
\label{fig:spectra3}
\end{figure*}

\begin{figure*}[htbp]
\centering
\includegraphics[width=0.95\textwidth]{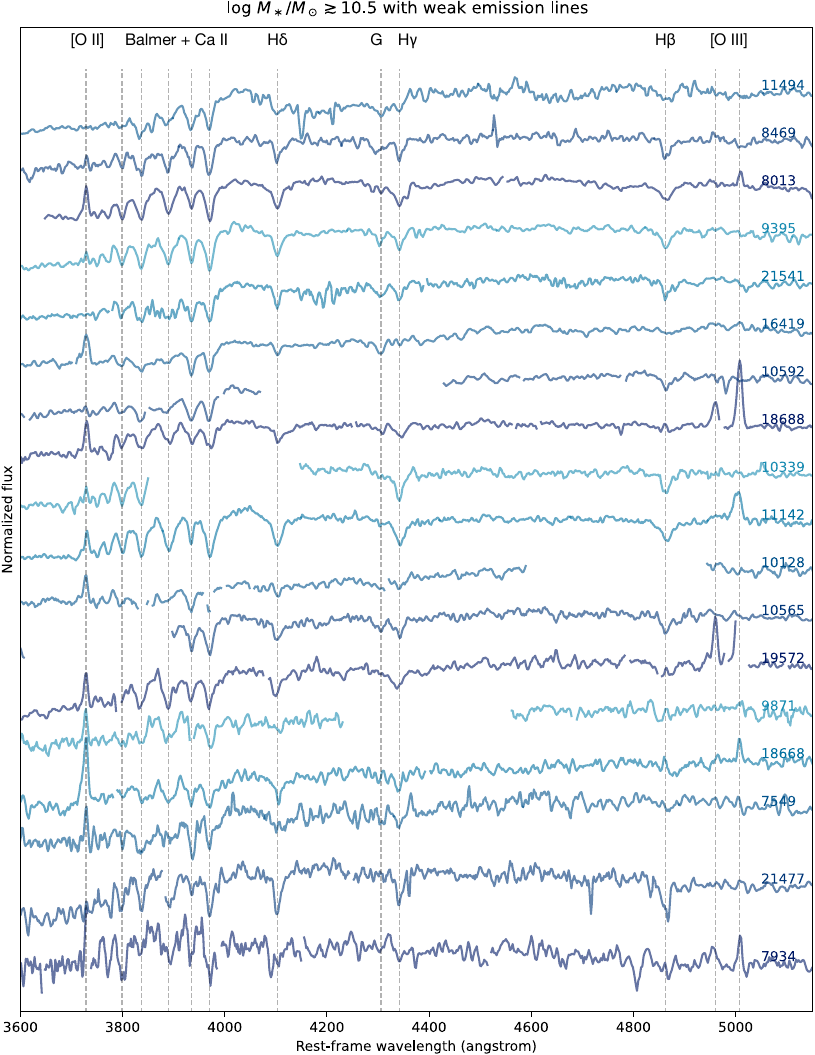}
\caption{Spectra of massive Blue Jay galaxies with weak emission lines, sorted by the S/N of the continuum. Each spectrum is normalized to have identical continuum flux at 5000$\AA$. Only part of the rest-frame optical range is shown; some prominent absorption and emission lines are marked.
}
\label{fig:spectra4}
\end{figure*}

In Figs.~\ref{fig:spectra1} through \ref{fig:spectra4} we show a small portion of each of the 139 spectra with a measured spectroscopic redshift (the four filler targets are excluded; see Appendix~\ref{sec:hiz}). The low- and intermediate-mass galaxies are shown in Figs.~\ref{fig:spectra1} and \ref{fig:spectra2}, respectively. The rest-frame wavelength range shown is between 3600 and 5100$\AA$, containing several important emission lines such as [O~II], [O~III], [Ne~III], and hydrogen Balmer lines. Massive galaxies can be highly attenuated by dust, which makes it harder to detect some of the blue emission lines. For this reason, in Fig.~\ref{fig:spectra3} we show the spectra of massive galaxies using a redder wavelength range, from H$\beta$ to H$\alpha$ and [S~II]; the two broad-line AGNs are clearly recognizable. The normalization needed to display the emission lines prevents a detailed view of the continuum, but many of these spectra also have well-detected absorption lines such as Na~D. Finally, the massive galaxies with weak emission lines, which are mostly quiescent (or on their way to quiescence), are shown in Fig.~\ref{fig:spectra4}. The wavelength range shown is again 3600--5100$\AA$, and the spectra are dominated by absorption lines such as the Balmer series, Ca~II H and K, and the G band. We caution that galaxies were split into different categories corresponding to different figures without using rigorous definitions, but with the sole purpose of displaying the spectra in the clearest possible way.

\end{appendix}

\end{document}